\documentclass[pdflatex,sn-mathphys-num]{sn-jnl}
\usepackage{graphicx}%
\usepackage{multirow}%
\usepackage{amsmath,amssymb,amsfonts,siunitx}%
\usepackage{amsthm}%
\usepackage{mathrsfs}%
\usepackage[title]{appendix}%
\usepackage{xcolor}%
\usepackage{textcomp}%
\usepackage{manyfoot}%
\usepackage{booktabs}%
\usepackage{algorithm}%
\usepackage{algorithmicx}%
\usepackage{algpseudocode}%
\usepackage{listings}%
\usepackage[T1]{fontenc}
\theoremstyle{thmstyleone}%
\usepackage{upgreek}
\usepackage{physics}
\usepackage{hyperref}
\usepackage{bm}
\usepackage[table]{xcolor}   
\usepackage{pdflscape}
\usepackage{changepage}
\usepackage{ragged2e}
\usepackage{tikz}
\usepackage{nicematrix}
\usepackage{capt-of}

\theoremstyle{thmstyletwo}%

\theoremstyle{thmstylethree}%

\usepackage[table]{xcolor}
\usepackage{tabularx}
\usepackage{array}
\newcolumntype{Y}{>{\centering\arraybackslash}X}

\newcommand{\physical}{\text{phys}}
\AtBeginDocument{\RenewCommandCopy\qty\SI} 

\usepackage[automake,acronym]{glossaries}
\makeglossaries

\newacronym{ac}{AC}{Alternating Current}
\newacronym{am}{AM}{amplitude modulation}
\newacronym{bbr}{BBR}{blackbody radiation}
\newacronym{bpsk}{BPSK}{binary phase-shift keying}
\newacronym{cmb}{CMB}{cosmic microwave background}
\newacronym{co}{CO}{carbon monoxide}
\newacronym{dc}{DC}{Direct Current}
\newacronym{decigo}{DECIGO}{DECi-hertz Interferometer Gravitational-wave Observatory}
\newacronym{eia}{EIA}{electromagnetically induced absorption}
\newacronym{eirp}{EIRP}{equivalent isotropically radiated power}
\newacronym{eit}{EIT}{electromagnetically induced transparency}
\newacronym{em}{EM}{electromagnetic}
\newacronym{esa}{ESA}{European Space Agency}
\newacronym{fec}{FEC}{forward error correction}
\newacronym{fm}{FM}{frequency modulation}
\newacronym{gnss}{GNSS}{Global Navigation Satellite System}
\newacronym{gnssro}{GNSS/RO}{GNSS radio occultation}
\newacronym{gw}{GW}{gravitational waves}
\newacronym{hts}{HTS}{high-throughput satellite}
\newacronym{insar}{InSAR}{interferometric synthetic aperture radar}
\newacronym{kagra}{KAGRA}{Kamioka Gravitational Wave Detector}
\newacronym{ligo}{LIGO}{Laser Interferometer Gravitational-Wave Observatory}
\newacronym{lisa}{LISA}{Laser Interferometer Space Antenna}
\newacronym{lna}{LNA}{low-noise amplifier}
\newacronym{lo}{LO}{local oscillator}
\newacronym{mw}{MW}{microwave}
\newacronym{nedt}{NEDT}{noise-equivalent delta temperature}
\newacronym{nef}{NEF}{noise-equivalent field}
\newacronym{nep}{NEP}{noise-equivalent power}
\newacronym{nesz}{NESZ}{noise-equivalent sigma-zero}
\newacronym{polsar}{PolSAR}{polarimetric synthetic aperture radar}
\newacronym{psk}{PSK}{phase-shift keying}
\newacronym{qam}{QAM}{quadrature amplitude modulation}
\newacronym{qpn}{QPN}{quantum projection noise}
\newacronym{qpsk}{QPSK}{quadrature phase-shift keying}
\newacronym{rades}{RADES}{Rydberg Atoms for Dark Evolution Sensing}
\newacronym{rf}{RF}{radio frequency}
\newacronym{rin}{RIN}{relative intensity noise}
\newacronym{rms}{RMS}{root mean square}
\newacronym{ro}{RO}{radio occultation}
\newacronym{sar}{SAR}{synthetic aperture radar}
\newacronym{si}{SI}{International System of Units}
\newacronym{snr}{SNR}{signal-to-noise ratio}
\newacronym{swapc}{SWaP-C}{size, weight, power and cost}
\newacronym{thz}{THz}{terahertz}
\newacronym{ttc}{TT\&C}{Telemetry, Tracking, and Command}
\newacronym{vlbi}{VLBI}{very long baseline interferometry}

\begin{document}

\title[Article Title]{Rydberg Receivers for Space Applications}


\author[1]{\fnm{Gianluca} \sur{Allinson}}\email{gianluca.allinson@durham.ac.uk}

\author[2]{\fnm{Mark} \sur{Bason}}\email{mark.bason@stfc.ac.uk}

\author[3]{\fnm{Alexis} \sur{Bonnin}}\email{alexis.bonnin@onera.fr}

\author[4,5]{\fnm{Sebastian} \sur{Borówka}}\email{s.borowka@cent.uw.edu.pl}

\author*[6]{\fnm{Petronilo} \sur{Martin-Iglesias}}\email{petronilo.martin.iglesias@esa.int}

\author[6]{\fnm{Manuel} \sur{Martin Neira}}\email{manuel.martin-neira@esa.int}

\author[4]{\fnm{Mateusz} \sur{Mazelanik}}\email{m.mazelanik@cent.uw.edu.pl}

\author[6]{\fnm{Richard} \sur{Murchie}}\email{richard.murchie@esa.int}

\author[4,5]{\fnm{Michał} \sur{Parniak}}\email{m.parniak@cent.uw.edu.pl}

\author[6]{\fnm{Sophio} \sur{Pataraia}}\email{sophio.pataraia@esa.int}

\author[7]{\fnm{Thibaud} \sur{Ruelle}}\email{thibaud.ruelle@csem.ch}

\author[3]{\fnm{Sylvain} \sur{Schwartz}}\email{sylvain.schwartz@onera.fr}

\author[6]{\fnm{Aaron} \sur{Strangfeld}}\email{aaron.strangfeld@esa.int}


\affil[1]{\orgdiv{Department of Physics}, \orgname{Durham University}, %
           \orgaddress{\city{Durham}, \country{United Kingdom}}}

\affil[2]{\orgdiv{UKRI-STFC}, \orgname{Science and Technology Facilities Council}, %
          \orgaddress{\city{Didcot}, \country{United Kingdom}}}

\affil[3]{\orgname{DPHY, ONERA, Université Paris-Saclay}, %
          \orgaddress{\city{Palaiseau}, \country{France}}}
          
\affil[4]{\orgdiv{Centre for Quantum Optical Technologies, Centre of New Technologies}, \orgname{University of Warsaw}, %
          \orgaddress{\city{Warsaw}, \country{Poland}}}

\affil[5]{\orgdiv{Faculty of Physics}, \orgname{University of Warsaw}, %
\orgaddress{\city{Warsaw}, \country{Poland}}}

\affil[6]{\orgname{European Space Agency (ESA)}, %
          \orgaddress{\city{Noordwijk}, \country{Netherlands}}}

\affil[7]{\orgname{CSEM SA}, \orgaddress{\city{Neuchâtel}, \country{Switzerland}}}

\abstract{Rydberg-atom sensors convert radiofrequency, microwave and terahertz fields into optical signals with SI-traceable calibration, high sensitivity, and broad tunability. This review assesses their potential for space applications by comparing five general architectures (Autler-Townes, AC-Stark, superheterodyne, radiofrequency-to-optical conversion, and fluorescence) against space application needs. We identify promising roles in radiometry, radar, terahertz sensing, and in-orbit calibration, and outline key limitations, including shot noise, sparse terahertz transitions, and currently large Size, Weight, Power and Cost. A staged roadmap highlights which uncertainties should be resolved first and how research organisations, industry and space agencies could take the lead for the different aspects.}

\maketitle
\section{Introduction}\label{sec:introduction}

In the last decade, the growth of quantum technologies has led to the development of a variety of sensors that exploit quantum effects, including clocks, gravimeters, accelerometers, magnetometers, gyroscopes and quantum imaging sensors~\cite{dowling2003quantum,acin2018quantum,krelina2021quantum}, with some pointing to space applications~\cite{alonso2022coldatoms,Abend2023}. In addition, a new class of sensors, based upon Rydberg atoms, has emerged, covering a wide range of \gls{em} fields~\cite{adams2019rydberg}.

\begin{figure}[ht]
\centering
\includegraphics[width=0.9\textwidth]{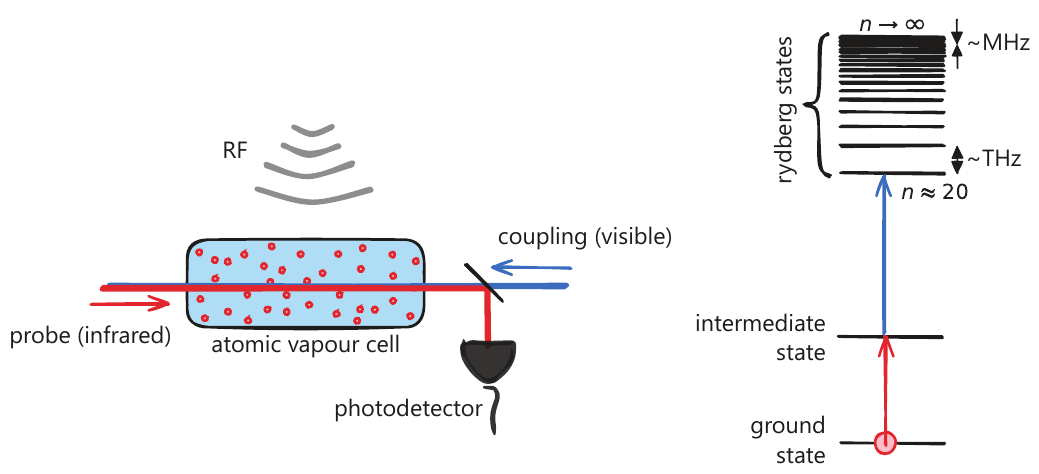}
\caption{Rydberg excitation process using two optical fields. \textit{Left:} A common experiment showing the atoms contained within a glass cell, which is excited using two lasers, a `probe' and a `coupling' laser. The probe laser is read out on a photodetector, separated from the coupling using a dichroic mirror, and the presence of the \gls{rf} field strength can be determined. \textit{Right}: An illustration of the atomic energy levels in an atom. The probe laser allows the coupling to an intermediate atomic state, from which the coupling laser further couples to Rydberg states. The energy separation between Rydberg states, on the order of microwaves, then allows the \gls{rf} field to be coupled to the quantum system.}
\label{fig:rydberg_excitation}
\end{figure}

These Rydberg sensors rely upon the laser excitation of an atom such that one (or more) of its electrons occupies an atomic state with a high principal quantum number ($n$), i.e., to a relatively high energy. This excitation process is shown in Figure~\ref{fig:rydberg_excitation} (right) using two optical fields, as is typically the case for the alkali atoms chosen for Rydberg sensors.

The use of Rydberg atoms in atomic physics has a long and successful history, including in tests of quantum electrodynamics~\cite{raimond2001manipulating,meschede1985one,gleyzes2007quantum}, precision spectroscopy~\cite{gallagher2006rydberg,li2003millimeter,jones2020probing}, many-body physics~\cite{browaeys2020many,zeiher2016many}, simulations of condensed matter systems~\cite{scholl2021simulation} and quantum computing~\cite{saffman2010quantum,adams2019rydberg}. However, it is the ability of Rydberg atoms to sense \gls{em} radiation that is perhaps the most promising development for potential deployment for space applications.

When compared to a classical antenna, the uniqueness of Rydberg sensing emerges. Instead of an \gls{em} field inducing current in a metal antenna, Rydberg sensors exploit atom-light interactions in an atomic vapour (hot or cold atoms) to change the state of laser light in the presence of the \gls{em} field. Thus, the task of sensing the \gls{em} field is reduced to that of measuring visible or infrared light.

This novel sensing method has a number of unique features. For example, the sensing volume can be made electrically small~\cite{cox2018electricallysmall,backes2024performance,mao2024resonator, elgee2025electrically}, tuned over a very wide range~\cite{meyer2021waveguide,elgee2023satellite,meyer2020assessment}, detect the angle of arrival~\cite{robinson2021determining,richardson2025study,schlossberger2025angle,yan2023three}, phase~\cite{meyer2018digital,holloway2019phase,anderson2020rydberg} and \gls{em} field polarisation~\cite{wang2023precise,elgee2024complete}. Sensitivity to \gls{rf} and \gls{mw} fields is also improving rapidly~\cite{schlossberger2024rydberg,cai2023sensitivity, borowka2024conversion,legaie2024mmwave,sandidge2024structures} and starting to compete with conventional approaches. In principle, the frequency range of the sensed \gls{em} fields can span from \gls{dc} to \gls{thz}. Performance characteristics differ, however, depending on the use of discretely distributed, narrow atomic electron transitions or off-resonant methods.

The interested reader is encouraged to study a number of recent review articles~\cite{adams2019rydberg,meyer2020assessment,fancher2021comm,simons2021review,yuan2023quantum,liu2023electric,zhang2024rydberg,schlossberger2024rydberg}.

This paper examines the needs and trends for \gls{rf}/\gls{mw} sensing in space and how Rydberg sensing may help. It is intended for two audiences:
\begin{itemize}
  \item \textbf{Developers of Rydberg sensors:} researchers and engineers developing Rydberg atomic sensors, seeking space-relevant requirements and application-driven performance targets.
  \item \textbf{Developers of space systems:} scientists and engineers evaluating Rydberg sensors for radiometry, radar, communications, and calibration.
\end{itemize}
The paper proceeds from space applications (Section~\ref{sec:space-applications}) to sensor principles and architectures (Section~\ref{sec:principles-architectures}), then to performance specification (Section~\ref{sec:performance-spec}) and system matching (Section~\ref{sec:architecture_application}),  unique capabilities (Section~\ref{sec:unique_capabilities}) and current limitations (Section~\ref{sec:current_limitations}) and closes with a roadmap (Section~\ref{sec:roadmap}).

\section{Radiofrequency and Microwave Space Applications}\label{sec:space-applications}

The objective of this section is to outline the current and future areas of \gls{rf}/\gls{mw} space applications, in which Rydberg sensors may find their use. We identify these areas as passive radiometry, spectroscopy, \gls{thz} imaging, radar, active sounding, communication, gravitational wave detection, pulsar timing and a range of other future applications.

\subsection{Passive Radiometry, Spectroscopy, and Terahertz Imaging}

\subsubsection{Overview of Microwave Radiometers}

An \gls{mw} radiometer is a passive remote sensing instrument that measures 
thermal radiation naturally emitted by surfaces or the atmosphere at \gls{mw} frequencies. 
They are widely used in meteorology, oceanography, and Earth observation to retrieve 
parameters such as atmospheric temperature, water vapour amount, and cloud liquid water amount.

\gls{mw} radiometers are implemented in several architectures, including the 
Total-Power Radiometer, Dicke Radiometer~\cite{dicke1946thermal}, 
Noise-Adding Radiometer~\cite{goggins1967feedback}, and 
pseudo-Correlation Radiometer~\cite{mennella2003plancklfi}. 
The Total-Power Radiometer, the simplest design, directly measures the power of 
incoming radiation. It offers high sensitivity but requires careful calibration to 
correct for gain fluctuations. The antenna delivers broadband thermal noise, extending 
over a bandwidth larger than the receiver bandwidth $B_w$.

Radiometer sensitivity is commonly expressed as the \gls{nedt}, 
the \gls{rms} brightness-temperature fluctuation equivalent to the instrument noise. For a Total-Power Radiometer, it can be written as
\begin{equation}
\text{NEDT} = (T_A + T_\text{Rx}) \,
\sqrt{\frac{1}{B_w \tau} + \left(\frac{\Delta G}{G}\right)^2},
\end{equation}
where $T_A$ is the antenna temperature, $T_\text{Rx}$ the receiver system noise temperature, $B_w$ the 
detection bandwidth, $\tau$ the integration time, and $\Delta G/G$ the fractional gain fluctuation within the integration time. We call
\begin{equation}
T_\text{sys}=T_A+T_\text{Rx}
\end{equation}
the system noise temperature.
The $1/(B_w \tau)$ term reflects thermal (radiometric) noise, while $(\Delta G/G)^2$ quantifies 
instability of the receiver gain. The system noise temperature combines several contributions. Any passive element adds thermal noise according to its physical temperature and loss, while active elements 
amplify input fluctuations and add their own internal noise. In a cascaded system, the 
dominant contribution comes from the first high-gain stage, typically a \gls{lna}.

In practice, $T_\text{Rx}$ and $G$ are determined from power measurements with reference 
loads of known physical temperature. Using the radiometer equation
\begin{equation*}
P_\text{out} = G k_\mathrm{B} (T_A + T_\text{Rx}) B_w,
\end{equation*}
where $k_\mathrm{B}$ is the Boltzmann's constant, one measures $P_\text{out}$ for two (or more) known $T_A$ values, then extracts both the 
receiver gain $G$ and system noise temperature $T_\text{Rx}$ by linear regression. This 
hot-cold (or multi-load) calibration is standard in radiometry and links the noise power 
sensitivity directly to $T_\text{Rx}$ and $G$.

\subsubsection{Spectrometry and Terahertz Spectral Line Analysis}

A special case of passive radiometry is spectrometry, used in two main contexts: 
hyperspectral radiometry and high-resolution spectral line analysis. Calibration 
remains essential to correct gain fluctuations ($\Delta G / G$), ensuring accurate 
temperature measurements and reducing uncertainty in \gls{nedt}. 

For illustrative purposes, Table~\ref{tab:nedt_example} shows a typical 183~GHz radiometer channel with its \gls{nedt} calculation.

\begin{table}[ht]
\centering
\caption{Example \gls{nedt} calculation for a 183 GHz radiometer channel.}
\begin{tabular}{cccccc}
\toprule
$T_A$ (K) & $T_\text{Rx}$ (K) & $B_w$ (MHz) & $\tau$ (ms) & $\Delta G/G$ & NEDT (K) \\
\midrule
250.00 & 600.00 & 1000.00 & 15.00 & $1.5 \times 10^{-5}$ & 0.22 \\
\bottomrule
\end{tabular}

\label{tab:nedt_example}
\end{table}

Hyperspectral \gls{mw} radiometers provide broad spectral coverage for retrieving 
temperature and humidity, while high-resolution spectral 
line analysis targets narrow molecular transitions for trace gas detection 
(e.g.\ ozone, \gls{co}, water vapour). This latter approach is widely used in atmospheric 
chemistry, planetary exploration, and deep-space sensing. At \gls{thz} frequencies, specific lines, such as the 4.7448~THz oxygen transition, have been studied with MHz 
resolution to capture line shape, broadening, and Doppler shifts, yielding 
insight into composition, temperature, and wind dynamics~\cite{hansen2025spectrometer}.

\gls{mw} spectrometry is typically implemented with digital spectrometers~\cite{pradhan2024radiometer} 
or \gls{mw} photonic techniques~\cite{pett2018radiometer}, both enabling high spectral 
resolution and flexibility to address scene variability.

\subsection{Radar and Active Sounding}\label{sec:radar_and_active_sounding}

By transmitting a well-defined waveform and measuring timing and its modification upon return, active \gls{rf}/\gls{mw} sensing can range and 
probe atmospheric, surface, and subsurface properties.

Well-established techniques illustrate the diversity of radar applications. \gls{sar} synthesises a large aperture by using platform motion and coherent 
summation of successive pulses~\cite{brown67sar, stewart17sar}. \gls{insar} 
compares repeat-pass acquisitions to detect changes in surface topography~\cite{griffiths95isar}, 
while \gls{polsar} exploits polarisation diversity to identify scattering 
mechanisms~\cite{verbout92sar}. \gls{esa}’s Sentinel-1 mission combines \gls{sar} with \gls{insar} and partial 
\gls{polsar}~\cite{torres12sentinel}. Scatterometers infer surface roughness, often used as a 
proxy for ocean winds. Altimeters measure elevation or sea level from signal delay, and 
subsurface sounders exploit the penetration of low-frequency waves. Broadband radar 
spectroscopy uses frequency-dependent scattering, absorption, and Doppler shifts to probe 
atmospheric composition and dynamics.

From a physical perspective, radar performance is set by the balance between the received 
signal power and the noise floor of the receiver. The expected signal power is given by the 
radar equation
\begin{equation}
P_r = \frac{P_t G_t G_r \lambda^2 \sigma}{(4\pi)^3 R^4 L_s L_p},
\label{eq:radar}
\end{equation}
where \(P_t\) is the transmitted power, \(G_t\) and \(G_r\) the transmitter and receiver 
gains, \(\lambda\) the wavelength, \(\sigma\) the radar cross section of the target, \(R\) 
the range, and \(L_s, L_p\) system and propagation losses.

For a fundamental point-target analysis, \(\sigma\) is simply the target radar cross 
section. In imaging radar, however, the relevant quantity is the normalised radar cross 
section \(\sigma_0\), with the effective cross section of a resolution cell written as 
\(\sigma = \sigma_0 A_\text{res}\), where \(A_\text{res}\) is the ground-projected 
resolution cell area. Finally, when waveform processing and pulse integration are 
included, the received power is effectively enhanced by a processing gain 
\(G_\text{proc}\), so that
\begin{equation*}
P_{r,\text{proc}} =
\frac{P_t G_t G_r \lambda^2 \sigma_0 A_\text{res}}{(4\pi)^3 R^4 L_s L_p}\,
G_\text{proc}.
\end{equation*}
Detectability requires comparison with 
the receiver noise power,
\begin{equation*}
P_n = k_\mathrm{B} T_\text{sys} B,
\end{equation*}
where $k_\mathrm{B}$ is the Boltzmann’s constant, $T_\text{sys}$ the effective system noise temperature, 
and $B$ the receiver bandwidth. The fundamental figure of merit is the
\gls{snr},
\begin{equation*}
\mathrm{SNR} = \frac{P_r}{P_n}.
\end{equation*}

Receiver figures of merit such as $T_\text{sys}$ and $B$ enter through \gls{snr} into higher-level 
radar metrics. The maximum range scales as $R_\text{max} \propto (1/T_\text{sys})^{1/4}$. 
The \gls{nesz}, i.e.\ the smallest detectable backscatter 
coefficient, is
\begin{equation}
\mathrm{NESZ} =
\frac{\sigma_0}
     {\mathrm{SNR}},
\label{eq:nesz}
\end{equation}
thus, lowering $T_\text{sys}$ 
or narrowing $B$ directly improves sensitivity. Conversely, increasing bandwidth benefits other radar figures of merit: 
the range resolution improves as $\delta R = c/(2B)$, and the 
time-bandwidth product $B\tau_p$ defines the pulse-compression or processing gain 
$G_\text{proc}$ that enhances effective \gls{snr} in coherent radars. 
Broadband operation also enables spectral discrimination of targets and 
frequency-dependent scattering analysis.

Recent radar developments emphasise improved algorithms (\gls{sar} motion compensation, real-time 
processing)~\cite{jmse13020355}, multi-frequency and multi-polarisation capability~\cite{ZHOU2025133513}, and machine learning for 
data interpretation~\cite{rs15153742}. Key challenges include managing the rapidly growing data volume and 
achieving tighter integration with other sensing modalities.
\subsection{Communication}

\subsubsection{Telemetry, Tracking, and Command}

\gls{ttc} systems are vital for monitoring satellite health and controlling its operations. Telemetry involves collecting real-time spacecraft data (e.g., attitude control, power levels, thermal conditions). Tracking determines the satellite's orbital parameters using ranging signals, and Command operations enable ground control to send operational instructions. These links operate primarily in S-band and X-band, utilising low-data-rate modulation schemes for robustness against interference and signal degradation.

\subsubsection{Data Downlink}

Scientific and Earth observation missions require high-speed data downlinks due to the large volumes of data generated by onboard sensors, imaging systems, and scientific instruments.

X-band and Ka-band are widely used for high-throughput payload data transmission. Optical communications using laser-based systems (e.g., low Earth orbit to geostationary orbit, intersatellite links or satellite-to-ground optical links) are being developed to overcome \gls{rf} spectrum limitations.

\subsubsection{Telecommunications}

Satellite-based telecommunications provide broadband internet, voice, and video services across vast geographical regions, particularly in remote or underserved areas. The \gls{rf} link consists of uplink (Earth-to-satellite) and downlink (satellite-to-Earth) transmissions, typically operating from L-band to Ka-band. Ongoing advancements are pushing operations into even higher frequency ranges to support greater throughput.

Modern high-throughput satellites maximise bandwidth efficiency and capacity through frequency reuse, spot beam technology, and advanced beamforming techniques. Next-generation satellites incorporate a large number of independently controlled beams, dynamically adjusting power and frequency allocation based on real-time demand. Beam hopping further enhances efficiency by allowing satellites to shift resources as needed, optimising coverage in high-traffic areas.

Traditional analogue payloads are being replaced by fully digital beamforming systems, enabling greater flexibility and adaptability in coverage. Large-scale low Earth orbit constellations, such as Starlink and OneWeb, are transforming satellite communications by delivering global, low-latency broadband. These networks leverage inter-satellite laser links to enhance connectivity and reduce dependence on ground-based infrastructure.

\subsubsection{Link Budget and Performance}

The link budget is a standard method for evaluating the performance of a communication system. All of the introduced parameters and their symbols are defined in Table \ref{Ka-band-budget}, along with exemplary values.

Consider a transmitting antenna located in space and a receiving antenna on the ground. 
The effectiveness of the transmitting system is represented by the \gls{eirp}, defined as
\begin{equation}
\text{EIRP} = P_T + G_T - L_\text{FTx},
\end{equation}
where \(P_T\) is the transmitter output power, \(G_T\) is the transmit antenna gain, and \(L_\text{FTx}\) is the feeder loss on the transmit side.

The receiving system is commonly characterised by the ratio \(G/T\), referred to as the receiver figure of merit. This quantity represents the ratio of the receive antenna gain to the total noise temperature of the receiving system. It is defined as
\begin{equation}
\frac{G}{T} = G_R - 10 \log_{10}\left(T_{\text{sys}}\right),
\end{equation}
where \(G_R\) is the receive antenna gain and \(T_{\text{sys}}\) is the system noise temperature at the antenna input. A higher value of \(G/T\) indicates a more sensitive receiving system.

The received carrier-to-noise spectral density ratio is
\begin{equation}
\frac{C}{N_0} = \text{EIRP} - L + \frac{G}{T} + 228.6,
\end{equation}
where \(C\) is the received carrier power, \(N_0\) is the noise spectral density, and the constant \(228.6\) accounts for the Boltzmann constant and the conversion of temperature to dB units. The total propagation loss \(L\) is
\begin{equation}
L = L_\text{FSL} + L_\text{atm} + L_\text{rain} + L_\text{scint} + L_\text{pol} + L_\text{ion},
\end{equation}
where
\(L_\text{FSL}\) is free space path loss,
\(L_\text{atm}\) is atmospheric absorption,
\(L_\text{rain}\) is rain attenuation,
\(L_\text{scint}\) is scintillation fade,
\(L_\text{pol}\) is polarisation mismatch loss,
and \(L_\text{ion}\) accounts for ionospheric effects. The free space path loss is given by
\begin{equation}
L_\text{FSL} = 20 \log_{10}\left(\frac{4 \pi d}{\lambda}\right),
\end{equation}
where \(d\) is the distance between transmitter and receiver and \(\lambda\) is the carrier wavelength.

Finally, the ratio of energy per bit to noise spectral density is
\begin{equation}
\frac{E_b}{N_0} = \frac{C}{N_0} - 10 \log_{10}(R),
\end{equation}
where \(R\) is the data rate (bit/s). Representative required values of energy per bit to noise spectral density ratio are included for common modulation and coding schemes, following standard link-budget practice and used here as indicative thresholds for relative performance comparison.


\begin{table}[ht]
\centering
\caption{Ka-band communication link budget example (\gls{psk}, \gls{qam}, \gls{fec}, \gls{bpsk}, \gls{qpsk})}
\begin{tabular}{|l|l|l|l|}
\hline
\textbf{Parameter} & \textbf{Symbol} & \textbf{Value} & \textbf{Units / Notes} \\
\hline
Transmit Power & $P_T$ & 20 & dBW (100 W) \\
Transmit Antenna Gain & $G_T$ & 45 & dBi \\
Feeder Loss (Tx side) & $L_\text{FTx}$ & 2 & dB \\
Effective Isotropic Radiated Power & $\text{EIRP}$ & 63 & dBW \\
Free-Space Path Loss & $L_\text{FSL}$ & 206.5 & dB (20 GHz, GEO $\approx 36 000$ km) \\
Atmospheric Absorption Loss & $L_\text{atm}$ & 2 & dB \\
Rain Attenuation Loss & $L_\text{rain}$ & 3 & dB \\
Other Losses & $L_\text{others}$ & 1 & dB \\
Total Propagation Loss & $L$ & 212.5 & dB \\
Receive Antenna Gain & $G_R$ & 50 & dBi \\
Antenna Noise Temperature & $T_a$ & 100 & K \\
Receiver Noise Temperature & $T_R$ & 100 & K \\
Feeder Loss (linear) & $L_F$ & 1.5 & ($\approx$ 1.76 dB) \\
System Noise Temperature & $T_\text{sys}$ & 395 & K \\
Figure of Merit & $G/T$ & 24.03 & dB/K \\
Carrier-to-Noise Density Ratio & $C/N_0$ & 103.13 & dBHz \\
Data Rate & $R$ & $10^8$ & bps \\
Energy per Bit to Noise Density Ratio & $E_b/N_0$ & 23.13 & dB \\
Required $E_b/N_0$ for BPSK (FEC 1/2) & & 3 & dB \\
Required $E_b/N_0$ for QPSK (FEC 1/2) & & 4 & dB \\
Required $E_b/N_0$ for 8PSK (FEC 3/4) & & 7.5 & dB \\
Required $E_b/N_0$ for 16QAM (FEC 3/4) & & 11 & dB \\
\hline
\end{tabular}
\label{Ka-band-budget}
\end{table}

\subsection{Future Applications}

\subsubsection{Terahertz Detection}
The modern frontier of \gls{rf} sensing lies between 300 GHz and 30 \gls{thz}, known as the \gls{thz} gap. A region of the \gls{em} spectrum between electronics and optics, this range is too high to easily generate with semiconductors and too low to generate with laser optics. The \gls{thz} gap is being driven by numerous high-profile applications, e.g., development of 6G wireless communication~\cite{mao2024terahertz}, radio astronomy~\cite{li2025terahertz}, and navigation systems~\cite{kanhere2021position}. Another promising approach for future systems includes \gls{thz} radar~\cite{lu2016thzradar}, which would create opportunities to exploit measurements in frequency ranges between traditional radar and lidar, Light Detection and Ranging. New \gls{thz} sources and receivers development, such as Rydberg receivers, might address the demand for such \gls{thz} applications.

\subsubsection{Highly Selective Radiofrequency Detection}
Modern applications of \gls{rf} sensors frequently suffer from interference from other \gls{rf} devices. There is a demand for \gls{rf} sensors that are robust against interference. Systems with Rydberg sensors can be a promising architecture to address this issue due to their selectivity, especially for higher frequencies.

\subsubsection{Search for Dark Matter and Axions}
Furthermore, on the frontiers of astroparticle physics, Rydberg atoms have been identified as a promising candidate to detect hypothetical particles, such as axions, to test the Standard Model and explain various dark matter hypotheses~\cite{gue2023darkmatter,graham2024axion,engelhardt2024axion}.

\subsubsection{Gravitational Wave Detection and Pulsar Timing}

Since the first direct detection of \gls{gw} by the \gls{ligo}-Virgo collaboration~\cite{abbott2016gravitation}, \gls{gw} astronomy has become an active research field. Existing experiments target different frequency bands, determined by both expected astrophysical sources and detector sensitivity limits. The operating ranges of \gls{ligo}, Virgo, \gls{kagra}, \gls{lisa}, \gls{decigo} and pulsar timing arrays are summarised in Table~\ref{tab:sensitivity}.

\begin{table}[ht]
\centering
\caption{Sensitivity range of \gls{gw} detectors.}
\begin{tabular}{|l|c|}
\hline
\textbf{Experiment} & \textbf{\gls{gw} Sensitivity} \\
\hline
\gls{ligo}/Virgo/\gls{kagra}~\cite{abbott2018gravitation}     & $10 \sim 10^{2}\,\text{Hz}$ \\
\gls{lisa}~\cite{amaro2017laserantenna} \& \gls{decigo}~\cite{yagi2011neutron} & $10^{-4} \sim 10^{1}\,\text{Hz}$ \\
Pulsar timing arrays~\cite{agazie2023nanograv,reardon2023gravitation} & $10^{-9} \sim 10^{-7}\,\text{Hz}$ \\
\hline
\end{tabular}
\label{tab:sensitivity}
\end{table}

Several theoretical models, including scenarios involving primordial black holes and early universe dynamics, suggest the possible existence of high-frequency \gls{gw}. Rydberg atom sensors offer sensitivity over a wide frequency range, from kHz~\cite{jau2020vapor,li2023super} to \gls{thz}~\cite{chen2022terahertz}, and have therefore been proposed for detecting such signals. In particular, \gls{gw} propagation through a static magnetic field can generate a weak oscillating electric field that may be measured using heterodyne Rydberg detection in the 0.3 to 16 GHz range~\cite{jing2020superhet}. Achieving this requires very high sensor sensitivity~\cite{kanno2025gravitation}, enabled, for example, by the superheterodyne approach discussed in Section~\ref{subsubsec:Superheterodyne}.

\subsection{Summary}

\begin{figure}[ht]
  \centering
  \includegraphics[width=\linewidth]{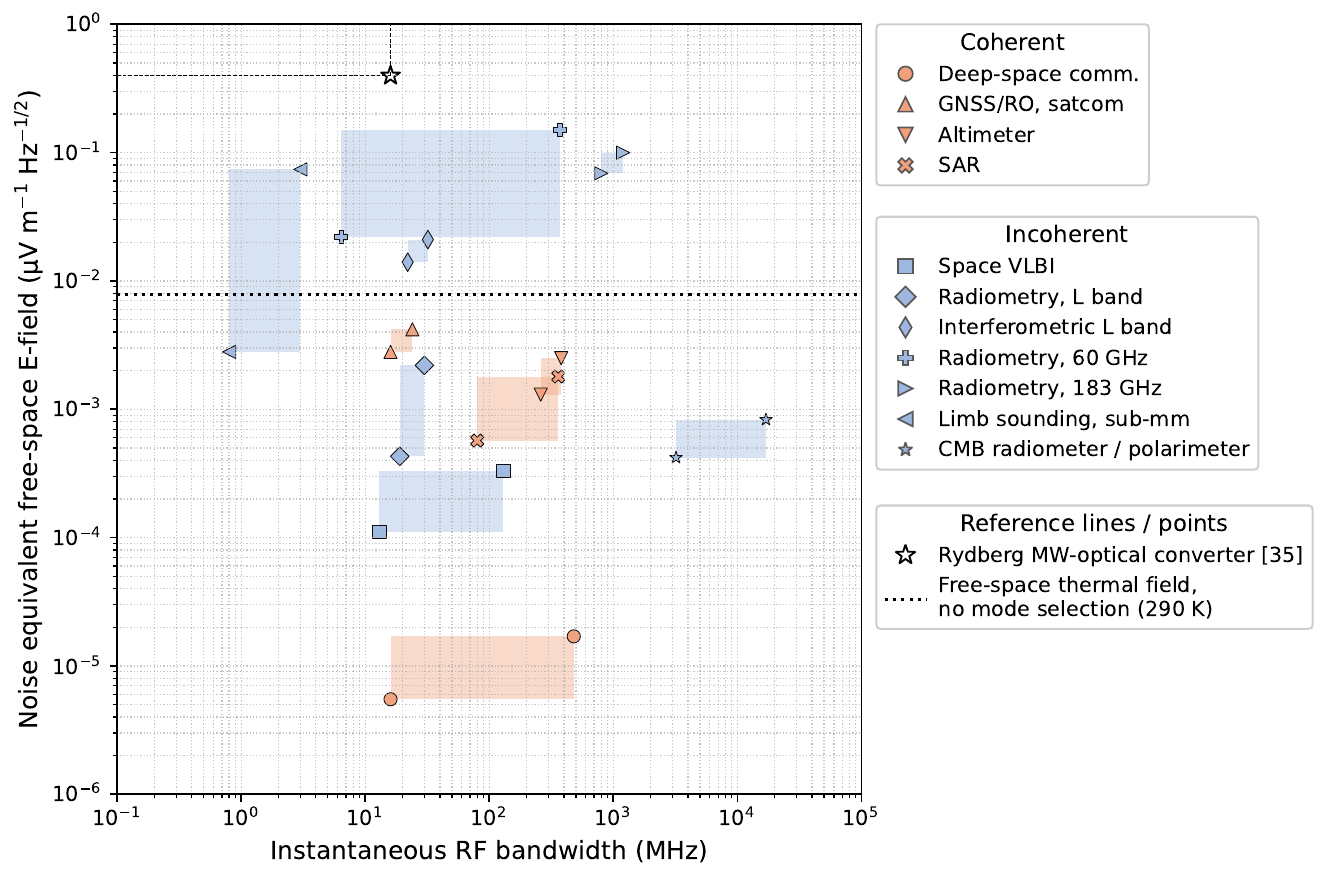}
  \caption{Instantaneous \gls{rf} bandwidth versus noise equivalent free-space electric-field for exemplary spaceborne receiver classes. Shaded rectangles indicate synthesised ranges of instantaneous bandwidth and input-referred free-space electric-field amplitude spectral density per $\sqrt{\mathrm{Hz}}$, derived from the instrument dataset; the construction of these ranges is described in the Annex. Marker shapes distinguish receiver classes and colour encodes coherence (coherent vs.\ incoherent), as indicated in the legend. Sensitivities are expressed as an equivalent free-space field, $\tilde{E}_{\mathrm{free}}=\sqrt{k_\mathrm{B} T_\mathrm{sys}\eta_0/(\rho^2 A_e)}$, corresponding to the electric-field spectral density that would deliver the system noise power through the effective aperture $A_e$ for a single receiving polarisation. Coherent, polarisation-matched systems use $\rho^2=1$, while single-channel incoherent receivers use $\rho^2=1/2$. The horizontal dotted line marks the global free-space thermal electric-field amplitude at 290~K, obtained without spatial or polarisation mode selection. The starred point highlights a Rydberg \gls{mw}--optical converter, with dashed guide lines indicating its bandwidth and sensitivity coordinates.}

  \label{fig:RydbergSpaceApplications}
\end{figure}

Figure~\ref{fig:RydbergSpaceApplications} summarises the parameter space spanned by representative spaceborne \gls{rf} and \gls{mw} receiver classes by comparing instantaneous \gls{rf} bandwidth with an equivalent free-space electric-field sensitivity per~$\sqrt{\mathrm{Hz}}$. This metric combines system noise temperature and effective aperture into a single input-referred quantity, enabling heterogeneous instruments to be compared on a common physical basis.

Deep-space communication systems occupy the lower region of the diagram, reaching the smallest noise-equivalent free-space field levels, down to $\sim10^{-5}\,\mu\mathrm{V\,m^{-1}\,Hz^{-1/2}}$. These values primarily reflect very large effective apertures and high directivity, which strongly couple a desired plane-wave mode while restricting the number of admitted spatial and polarisation modes, thereby suppressing coupled thermal background. Radar and active sounding instruments exhibit higher noise-equivalent free-space fields; in practice, their sensitivity is recovered through the coherent processing gains.

Atmospheric radiometers populate intermediate to higher noise ranges, typically between $\sim10^{-3}$ and $10^{-1}\,\mu\mathrm{V\,m^{-1}\,Hz^{-1/2}}$, while spanning several orders of magnitude in bandwidth. In these instruments, effective aperture is constrained by footprint requirements, and system noise temperature generally increases with frequency. For incoherent radiometric systems in particular, aperture primarily sets spatial resolution and scene selection, whereas sensitivity is governed by system noise temperature and calibration stability.

The free-space thermal field reference indicates the electric-field amplitude of thermal radiation integrated over all spatial modes and polarisations in free space, in the absence of mode selection, and does not represent a fundamental detection limit. Operation below this level requires restricting the number of coupled modes through effective aperture, antenna directivity, polarisation selectivity, and, where applicable, temporal or spatial correlation and coherent processing. In this context, the placement of the Rydberg \gls{mw}--optical converter \cite{borowka2024conversion} highlights a scaling challenge. For Rydberg vapour-cell receivers operated in their standard configuration, reaching comparable noise-equivalent free-space fields while suppressing thermal background depends critically on achieving substantial local field enhancement together with controlled coupling to a limited set of spatial modes. Assessing whether such enhancement and mode selectivity can be realised in a scalable, system-compatible manner is therefore central to evaluating the applicability of Rydberg-based receivers to low-field, background-limited space sensing.

Beyond these classical applications, the chapter explored \gls{gw} detection at high frequencies using Rydberg atom sensors and other future applications, such as \gls{thz} detection, interference-free, highly selective \gls{rf} receiving, and detection of new particles.
Rydberg atoms also promise other breakthroughs in quantum communication, quantum computing, and advanced sensing, beyond the presently defined applications. Their large electric dipole moments and strong interactions may enable a platform for harnessing entanglement and precise single-photon-level control.

\newpage

\section{Principles and Architectures of Rydberg Sensors}\label{sec:principles-architectures}
\subsection{Principles}
Let us begin with a brief outline of the fundamental principles and scaling laws underlying the operation of atomic Rydberg sensors~\cite{schlossberger2024rydberg,fancher2021comm,yuan2023quantum,liu2023electric,zhang2024rydberg}, starting with the basics of electronic atomic transitions.

Atoms have discrete energy levels, which arise from solutions to the Schrödinger equation for the electron bound to the nucleus. These quantised energy states -- called eigenstates -- form the basis for understanding atomic structure and transitions. 

Semi-classically, an external electric field perturbs the electron distribution in the atom, inducing a dipole moment that in turn interacts with the field. The strength of this interaction is governed by dipole transition moments -- matrix elements of the dipole operator -- which depend on the spatial structure of the electron wavefunctions involved, and on the polarisation of the applied field.

\begin{figure}[ht]
    \centering
    \includegraphics[width=1.0\textwidth]{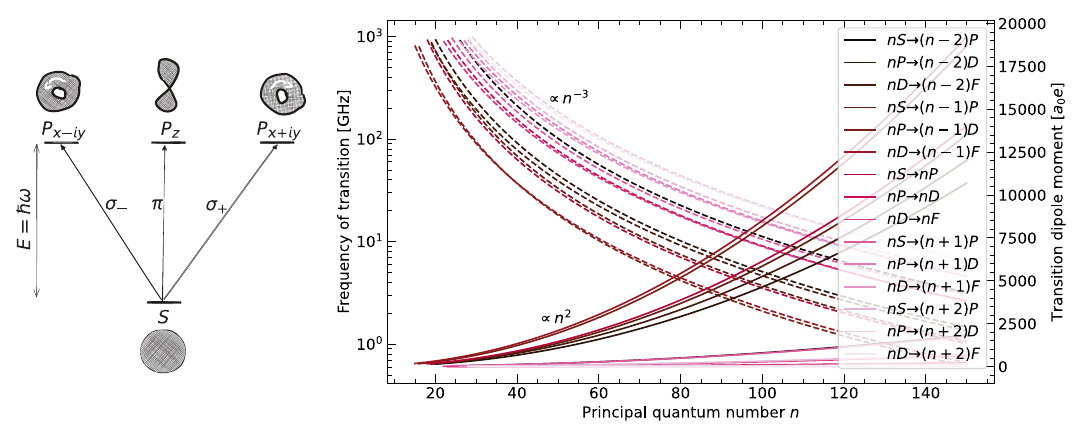}
    \caption{Principles of Rydberg atomic transitions. (left) Illustration of three possible electronic transitions involving the electric field of three orthogonal polarisations -- $\sigma_\pm$ - left-hand ($-$) and right-hand ($+$) circular polarisations in $x$-$y$ plane, and $\pi$ -- linear polarisation in $z$ direction. The orbitals are drawn for the simplest case of 1S to 2P transitions, and the arrows represent the interaction-free evolution. (right) Examples of possible transitions between Rydberg states in $^{87}\mathrm{Rb}$. The solid lines represent the transition dipole moment (right axis) and the dashed lines the transition frequency (left axis)~\cite{sibalic2017arc}.}
    \label{fig:rydberg-branches}
\end{figure}

In atoms, the fundamental dipole modes include one linear dipole along the $z$-axis and two circular dipoles rotating in opposite directions within the $x$-$y$ plane. These modes correspond to different polarisation components: linear polarisation along the $z$-axis, and two circular polarisations in the $x$-$y$ plane. Depending on the specific quantum states involved, each mode can exhibit a different transition strength. A schematic representation of these three possible modes, in the simplest case of a transition between s and p orbitals, is shown in Figure~\ref{fig:rydberg-branches}.

Importantly, only transitions driven by \gls{em} waves near resonance significantly contribute to the atomic response. This behaviour mirrors that of a driven harmonic oscillator or antenna, where the response follows a Lorentzian profile and scales inversely with the detuning from resonance ($\propto 1/\Delta$). 

Transitions between highly excited states -- Rydberg states -- are of particular interest here. These transitions occur at frequencies ranging from the MHz to the \gls{thz} regime, depending on the principal quantum number $n$ and angular momentum $\ell$, of the states involved. The energy spacing between adjacent Rydberg states scales approximately as $n^{-3}$~\cite{adams2019rydberg}: transitions in the high-$n$ regime fall in the high MHz to low GHz domain, while those at lower $n$ can reach the \gls{thz} range~\cite{fan2015rfsensing}.

As illustrated in Figure \ref{fig:rydberg-branches}, there are numerous possible transition branches between Rydberg states. However, only the strongest transitions reachable by existing laser technology are considered in practice. These typically can be reached via one to three intermediate transitions from the ground state. The transition strengths characterised by the dipole moment for Rydberg-Rydberg transitions scale as $n^2$. Consequently, the sensitivity to external fields is greatly enhanced for high-$n$ states.

Atomic species from group 1 -- such as rubidium (Rb) and caesium (Cs) -- are preferred for sensing due to their single valence electron, which results in a hydrogen-like level structure and a well-established both theoretical and experimental frameworks. While the fundamental Rydberg physics is similar across these species, differences arise in the practical implementation. These include variations in the laser wavelengths needed to reach Rydberg states~\cite{low2012guide}, as well as differences in vapour pressure at room temperature, which determines whether the vapour cell needs to be heated for operation.

Finally, while most implementations of Rydberg sensors use thermal atoms, cold-atom systems offer excellent quantum control, narrow linewidths, and the potential for long-term frequency standards and calibration. These properties make cold-atom systems particularly promising for high-precision measurements and applications requiring long-term frequency stability or calibration~\cite{duverger2024metrology}. However, they require complex, often bulky apparatuses and typically operate in a pulsed mode (low duty cycle). In contrast, warm-atom systems, based on thermal vapours, are more robust, compact, and capable of continuous operation, albeit with distinct additional noise sources and typically broader linewidths due to atomic motion.

\subsection{Architectures}
\begin{figure}[ht]
    \centering
    \includegraphics[width=1.0\textwidth]{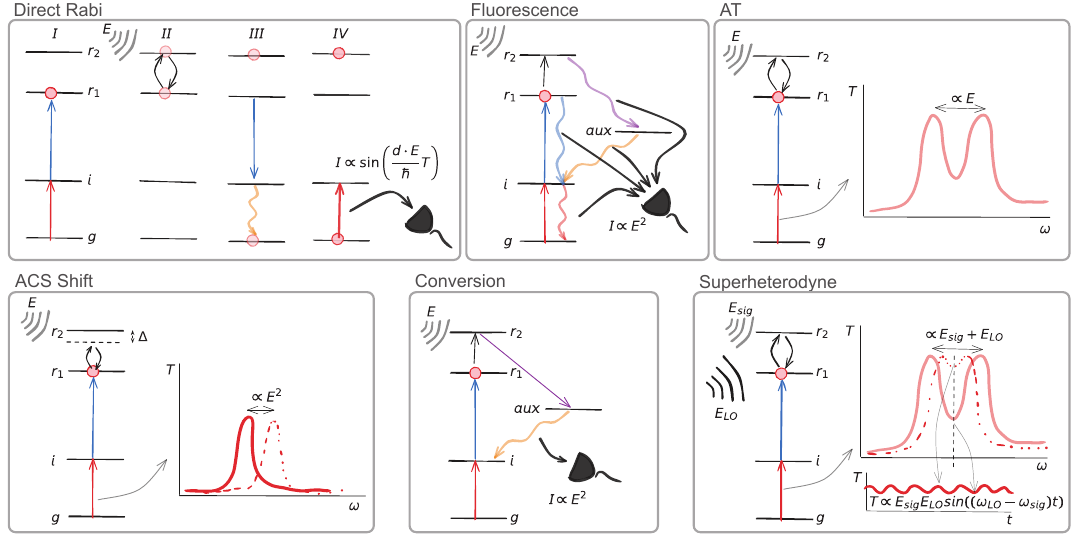}
    \caption{Different architectures of Rydberg sensors. In each architecture, atoms are excited from the ground state ($g$) to the Rydberg state ($r_1$) via the intermediate state ($i$). Depending on the architecture, the \gls{rf} electric field to be detected ($E$) couples the two Rydberg states ($r_1$ and $r_2$) in resonance or off-resonantly. The auxiliary state ($aux$) is additionally employed in Fluorescence and Conversion schemes. In the case of superheterodyne, an additional (\gls{lo}) \gls{rf} field ($E_\mathrm{LO}$) is employed to facilitate the detection of a weak signal \gls{rf} field $E_\mathrm{sig}$.} 
    \label{fig:rydberg-architectures}
\end{figure}

\renewcommand{\arraystretch}{1.3} 

\makeatletter
\tikzset{
  double color fill/.code={%
    \def\tempA##1,##2,\relax{%
      \def\first{##1}%
      \def\second{##2}%
    }%
    \expandafter\tempA#1,\relax
    \pgfdeclareverticalshading{diagonalfill}{100bp}{
      color(0bp)=(\first);
      color(50bp)=(\first);
      color(50bp)=(\second);
      color(100bp)=(\second)
    }
    \tikzset{shade, shading=diagonalfill, shading angle=-45}
  }
}

\tikzset{
  triple color fill/.code={%
    \def\tempA##1,##2,##3,\relax{%
      \def\first{##1}%
      \def\second{##2}%
      \def\third{##3}%
    }%
    \expandafter\tempA#1,\relax
    \pgfdeclareverticalshading{triplefill}{100bp}{%
      color(0bp)=(\first);
      color(40bp)=(\first);
      color(40bp)=(\second);
      color(60bp)=(\second);
      color(60bp)=(\third);
      color(100bp)=(\third)
    }%
    \tikzset{shade, shading=triplefill, shading angle=-45}
  }
}
\makeatother

\tikzset{%
  diagonal fill/.style={%
      double color fill={red,blue},
      shading angle=45,
      opacity=0.8},
  tdiagonal fill/.style={%
      triple color fill={red,green,blue},
      shading angle=45,
      opacity=0.8}
}

\begin{table}[ht]

\footnotesize
\centering

\setlength{\tabcolsep}{5pt}
\renewcommand{\arraystretch}{1.2}

\definecolor{demonstrated}{RGB}{102,194,165}
\definecolor{promising}{RGB}{141,160,203}
\definecolor{limited}{RGB}{252,141,98}
\definecolor{auxiliary}{RGB}{231,212,124}
\definecolor{none}{RGB}{220,220,220}
\NiceMatrixOptions{caption-above}
\begin{minipage}{\linewidth} 
\begin{tabular}{c}
\end{tabular}
\end{minipage}

\begin{NiceTabular}{@{}p{2.2cm}
                >{\centering\arraybackslash}p{1.4cm}
                >{\centering\arraybackslash}p{1.4cm}
                >{\centering\arraybackslash}p{1.4cm}
                >{\centering\arraybackslash}p{1.4cm}
                >{\centering\arraybackslash}p{1.4cm}
                >{\centering\arraybackslash}p{1.4cm}@{}}[
caption = Comparison of Rydberg-based measurement techniques across multiple sensing modalities. Colour encodes qualitative capability.
]
\toprule
Method & Power & Phase & Polarisation & Spatial (Imaging / AoA) & BBR detection & Off-Resonant Sensing \\
\midrule
Direct Rabi &
\cellcolor{demonstrated}\cite{romalis2024rades}, SI-traceable &
\cellcolor{promising} &
\cellcolor{promising} &
\cellcolor{promising} &
\Block[
  tikz={double color fill={promising,auxiliary}},
]{1-1}{ }With cavity &
\cellcolor{limited}
\\

Fluorescence &
\cellcolor{demonstrated}\cite{downes2020imaging} &
\Block[
  tikz={double color fill={promising,auxiliary}, inner sep=3em},
]{1-1}{ }With ext. \gls{lo} &
\cellcolor{promising} &
\cellcolor{demonstrated}\cite{wade2016imaging,schlossberger2025angle} &
\cellcolor{promising} &
\cellcolor{limited}
\\

Autler-Townes &
\cellcolor{demonstrated}\cite{sedlacek2012microwave}, SI-traceable &
\cellcolor{auxiliary}With ext. \gls{lo} -- see superhet. row &
\cellcolor{demonstrated}\cite{sedlacek2013polarization} &
\cellcolor{auxiliary}\cite{fan2014subwavelength,holloway2014sub}, With laser scanning &
\cellcolor{limited}\cite{kaur2025impact}, Line shape modifications  &
\cellcolor{limited}\cite{simons2016detuning}
\\

AC-Stark Shift &
\cellcolor{demonstrated}\cite{jiao2016spectroscopy}, SI-traceable &
\cellcolor{auxiliary}\cite{meyer2021waveguide}, With ext. \gls{lo}  &
\cellcolor{promising} &
\cellcolor{promising} &
\cellcolor{promising} &
\cellcolor{demonstrated}\cite{jiao2016spectroscopy}
\\

Conversion with Photon Counting &
\cellcolor{demonstrated}\cite{borowka2024conversion} &
\cellcolor{none} &
\cellcolor{promising} &
\cellcolor{promising} &
\cellcolor{demonstrated}\cite{borowka2024conversion} &
\cellcolor{limited}\cite{krokosz2025comb}
\\

Superheterodyne &
\cellcolor{demonstrated}\cite{gordon2019weak,jing2020superhet} &
\cellcolor{demonstrated}\cite{simons2019rydberg} &
\cellcolor{demonstrated}\cite{elgee2024complete} &
\cellcolor{demonstrated}\cite{robinson2021determining} &
\Block[
  tikz={double color fill={promising,auxiliary}},
]{1-1}{ }\cite{santamaria2022comparison}, With cavity  &
\cellcolor{limited}\cite{yao2022detuned}
\\
\bottomrule
\end{NiceTabular}

\vspace{0.7em}
\begin{minipage}{\linewidth}\footnotesize
\textbf{Legend:} 
\textcolor{demonstrated}{\rule{0.6em}{0.6em}} Experimentally demonstrated \quad
\textcolor{promising}{\rule{0.6em}{0.6em}} Theoretically promising or possible in principle \quad
\textcolor{limited}{\rule{0.6em}{0.6em}} Limited capability \quad
\textcolor{auxiliary}{\rule{0.6em}{0.6em}} Conditional requirement (cavity, scanning, or external \gls{lo}) \quad
\textcolor{none}{\rule{0.6em}{0.6em}} No demonstrated capability
\end{minipage}

\label{tab:rydberg_techniques_colored}
\end{table}

\subsubsection{Direct Observation of Rabi Oscillations}

Rydberg atoms have been interfaced with \gls{mw} fields for many years~\cite{gallagher1979interactions, figger1980photon,agarwal1984vacuum,gallagher2006rydberg}, well before their modern role in quantum sensing. A major milestone was the construction of a single-Rydberg-atom maser in 1985~\cite{meschede1985one}. These developments culminated in the 2012 Nobel Prize co-awarded to S.~Haroche (shared with D.J.~Wineland) for the control and detection of single \gls{mw} photons using Rydberg atoms~\cite{haroche2013nobel}.

Of special relevance to current sensing applications is the recently introduced \gls{rades} method developed at Princeton~\cite{romalis2024rades}. In this protocol, atoms are first optically excited in a vapour cell. The subsequent Rabi rotation is induced by the measured \gls{mw} field in the absence of any optical fields. Final readout is performed optically. Though inherently pulsed, this method offers a promising alternative to directly observing the \gls{mw} field's effect on Rydberg atoms.

\subsubsection{Fluorescence Detection}

A simple method introduced in~\cite{wade2016imaging} relies on detecting fluorescence emitted from Rydberg atoms as they decay to the ground state. Atoms are first excited to a Rydberg state via one or more laser fields. The applied \gls{rf} field then promotes them to a second Rydberg state, from which decay occurs, emitting light at specific wavelengths characteristic of intermediate transitions and final relaxation to the ground state.

The fluorescence intensity is proportional to the second Rydberg state's population and hence to the intensity of the \gls{rf} field. For optimal contrast and noise suppression, fluorescence is typically detected along decay paths that are distinct from the excitation and primary decay paths. Spectrometers or wavelength-selective detectors can be used to simultaneously monitor multiple decay channels.

Additionally, this approach is suitable for spatial imaging of \gls{rf} fields using a camera~\cite{downes2020imaging, downes2022rapid, downes2023practical,li2024dual}. However, the incoherent nature of fluorescence means that it lacks phase information despite some promise of better sensitivity~\cite{prajapati2024fluorescence}.

\subsubsection{Autler-Townes Method}

Another simple and historically the first method for Rydberg atoms-based electrometry is based on Autler-Townes (A-T) splitting~\cite{sedlacek2012microwave,holloway2014broadband}. This method relies on \gls{eit}~\cite{boller1991eit} in a ladder-type three-level system. A probe laser is scanned across the transition between the ground state and an intermediate state, while a second, typically counter-propagating, coupling laser drives the transition from the intermediate state to the first Rydberg state, creating a narrow \gls{eit} window in the probe beam transmission~\cite{mohapatra2007coherent}.

When an \gls{rf} field is applied and resonant with a transition from the Rydberg state to another Rydberg state, it induces a splitting of the \gls{eit} peak -- known as A-T splitting. The separation between the split peaks is proportional to the Rabi frequency $\Omega$ of the \gls{rf} transition and therefore to the electric field $\mathbf{E}$ and the transition dipole moment $\mathbf{d}$:
\begin{equation}\label{eqn:ATsplitting}
\Omega = \frac{\mathbf{d} \cdot \mathbf{E}}{\hbar},
\end{equation}
where $\hbar$ is the reduced Planck constant.

Because dipole moments can be calculated with high precision, the \gls{rf} field amplitude can be determined directly without external calibration. This property makes the A-T method especially useful as a calibration tool for more sensitive, but not directly \gls{si}-traceable, detection methods.

If the \gls{rf} field is off-resonant, it does not cause splitting, but instead induces a Stark shift of the Rydberg levels, resulting in a shift of the \gls{eit} line. However, this off-resonant regime is significantly less sensitive, as the shift scales with the square of the field amplitude and inversely with the detuning, as we explain in the following subsection.

Interestingly, the intermediate regime -- where the \gls{rf} is near-resonant but not fully on resonance -- can also be exploited to extract the \gls{rf} field detuning, as demonstrated in~\cite{duverger2024metrology,borowka2024automotive}.

Several enhancements to the basic A-T method have been developed. For instance, exploiting two-photon \gls{rf} transitions by selecting three-level ladders of Rydberg states with comparable energy separations or employing additional \gls{rf} fields allows extending the set of feasible transitions, thereby broadening the frequency range over which detection is possible~\cite{anderson2014multiphoton, anderson2017highintensity, xue2021twophoton}. 

Furthermore, the sensitivity of the method can be improved through modulation techniques, such as using lock-in detection by modulating the probe laser and demodulating the transmitted signal, which enhances the \gls{snr}~\cite{kumar2017lockin}.

The performance of the A-T method is ultimately limited by the linewidth of the \gls{eit} resonance, which limits the minimal resolvable splitting. Although the natural linewidth can be extremely narrow (on the order of kHz), atomic motion broadens the observed line through the Doppler effect, and, to a lesser extent, through the transit time broadening effect due to atoms entering and exiting the interaction beam and thus interacting only through a limited time. In typical configurations using rubidium or caesium atoms, the counter-propagating probe and coupling beams reduce the Doppler broadening (from 200 MHz down to about 10 MHz in rubidium), yet residual broadening remains due to their wavelength mismatch, often exceeding the natural linewidth by four orders of magnitude. This limitation can be mitigated by employing a three-photon excitation scheme with carefully selected intermediate states to match laser wavelengths, maximising Doppler cancellation. This has been recently experimentally demonstrated~\cite{bohaichuk2023three,glick2025warm}.

The A-T method is also sensitive to time-varying fields because it responds to both the amplitude and frequency of the \gls{rf} field. Hence, it can be used to detect \gls{am} or \gls{fm} signals. In such cases, the modulation of the \gls{rf} field is imprinted onto the probe transmission, which can be detected using high-speed photodetectors and subsequently demodulated for signal analysis~\cite{borowka2022mod}.

Finally, by selecting intermediate and Rydberg states that allow transitions coupled by different polarisations -- each characterised by distinct dipole moments -- the A-T method can also be used to detect the polarisation of the \gls{rf} field. This is achieved by analysing how the splitting or lineshape depends on the \gls{rf} field's polarisation and the orientation of the atomic dipoles~\cite{sedlacek2013polarization}. Polarisation selectivity can be further enhanced by incorporating the atomic medium into a resonant structure that amplifies the \gls{rf} field for a chosen polarisation~\cite{anderson2018resonant}.

\subsubsection{Off-resonant AC-Stark Detection}

The previously described methods rely on the \gls{rf} field frequency being closely resonant with atomic transitions. While this enables high sensitivity by maximising atomic response, it limits the detectable frequency bands to those matched to atomic resonances. However, detection of fields whose frequency is off-resonance is possible by measuring the \gls{ac}-Stark shift induced by the \gls{rf} field~\cite{mohapatra2008giant, hu2022continuously}.

This method is conceptually related to the far-detuned A-T regime mentioned in the previous section. When the \gls{rf} field is far-detuned from any Rydberg-Rydberg transition, it no longer causes resolvable \gls{eit} splitting but instead shifts energy levels via the \gls{ac}-Stark effect. This shift manifests as a measurable displacement of the \gls{eit} transparency window, with a magnitude proportional to the square of the Rabi frequency and inversely proportional to the detuning:
\begin{equation}
    \Delta_{\mathrm{acS}} \propto \frac{|\Omega|^2}{\Delta}.
\end{equation}

This method comes at the price of a reduced intrinsic sensitivity~\cite{liu2022highly}. To enhance interaction strength and improve detection sensitivity, experimental setups often incorporate metal electrodes to concentrate the \gls{rf} electric field within the vapour cell. However, these electrodes introduce a capacitive load that limits the operational frequency range, typically below 100 MHz, although demonstrations up to 500 MHz have been reported~\cite{paradis2019atomic}.

Alternatively, a waveguide integrated into the glass cell can enable detection of higher-frequency fields (e.g., up to 20~GHz~\cite{meyer2021waveguide}). The electrodes can also be embedded in an \gls{rf} resonator to enhance the field strength further and improve detection sensitivity. Nevertheless, at lower frequencies, performance is constrained by screening effects due to the alkali-metal atoms adsorbed into the inner surface of the vapour cell glass, which reduce \gls{rf} penetration into the atomic medium~\cite{jau2020vapor}.

Furthermore, an \gls{eit}-based method for off-resonant \gls{rf} field sensing of weak fields has recently been experimentally demonstrated~\cite{Trinh2024}. In this scheme, phase modulation applied to the coupling laser is transferred to the probe transmission via nonlinearity in the atomic medium, which provides the observable response to the \gls{rf} field.

\subsubsection{Conversion}

A fundamentally different approach to \gls{rf} field detection leverages coherent photon-to-photon conversion, wherein the \gls{rf} field is converted into an optical signal via multi-wave mixing in the atomic medium. In this method, atoms are first excited from the ground state to a Rydberg state using a sequence of laser fields. The \gls{rf} field then couples this state to a second Rydberg level, effectively being absorbed in the process. A second set of laser beams completes the excitation-deexcitation cycle, resulting in the emission of a coherent optical field that closes the interferometric loop. 

In this loop, the \gls{rf} signal is coherently translated into an optical signal -- the phase and amplitude of the \gls{rf} field are mapped onto the emitted optical field. This technique was initially demonstrated in ultracold atomic ensembles~\cite{han2018coherent, vogt2019efficient, kumar2023quantum} and has more recently been realised in warm atomic vapours~\cite{borowka2024conversion, li2024conversion}, marking a substantial step toward real-world applications. 

Notably, this is the only method utilising warm atomic media that enables single-photon counting and is sensitive enough to detect thermal \gls{mw} radiation with an inferred sensitivity of $4\ \mathrm{nV}/\mathrm{cm}/\sqrt{\mathrm{Hz}}$~\cite{borowka2024conversion}. While offering unique advantages -- coherent detection, high sensitivity, and compatibility with optical quantum technologies -- the technique requires a complex experimental setup involving multiple laser fields to define excitation and de-excitation pathways.

A critical requirement for efficient conversion is phase matching, which ensures momentum conservation among the interacting optical and \gls{rf} photons. This condition leads to an antenna-like reception pattern for the \gls{rf} field. By adjusting the angles of the interacting laser beams, the antenna pattern -- and hence the directionality of the receiver -- can be tailored.

\subsubsection{Superheterodyne} \label{subsubsec:Superheterodyne}

Another advanced technique, based on using the atomic medium as a free-space \gls{rf} mixer with optical readout, is the superheterodyne method~\cite{jing2020superhet,simons2019rydberg, gordon2019weak}. Due to its similarity with classical radio receivers, it is often described as a quantum analogue of superheterodyne detection. 

Conceptually, the method can be understood in two complementary ways. In one picture, it is an extension of the A-T method. Here, a strong \gls{lo} \gls{rf} field at frequency $f_{\mathrm{LO}}$ and a weaker signal \gls{rf} field at frequency $f_{\mathrm{\gls{rf}}}$ produce a beat frequency $f_s = |f_{\mathrm{LO}} - f_{\mathrm{\gls{rf}}}|$ that modulates the A-T splitting. The probe laser, stabilised at the point of highest sensitivity on the A-T split \gls{eit} resonance peak, detects this modulation as an oscillating transmission signal. This signal constitutes the optical readout from the mixer, enabling phase-sensitive detection of \gls{rf} fields. 

In addition to providing the phase reference, the \gls{lo} effectively biases the atomic medium, shifting its operating point to the region of maximal response, which allows the system to reach its highest sensitivity~\cite{jing2020superhet}. However, as in the A-T method, both \gls{rf} fields must also be close in frequency to the atomic resonance (typically within a few MHz).

Alternatively, the scheme can be interpreted as a specific implementation of a frequency-conversion mechanism, in which the \gls{lo} field and the coupling laser form a de-excitation path. This process leads to the emission of photons at the probe beam wavelength but with a slight frequency offset corresponding to the beat frequency $f_s$. The resulting frequency-shifted optical field interferes with the original probe beam, creating an optical beat signal that is detectable using a photodiode.

As with the previously discussed wave-mixing conversion schemes, phase matching plays a critical role. In the frequency-conversion picture, it stems from momentum conservation across the interacting fields. In the A-T modulation picture, constructive interference occurs only when the \gls{rf} modulation is in-phase along the probe beam propagation direction. Otherwise, destructive interference reduces the signal due to spatial averaging. Just like in the conversion scheme, this phase matching can be exploited to shape the sensor's antenna pattern.

Various extensions of this scheme exist. For instance, by using multiple \gls{lo} fields with different polarisations, it is possible to infer the polarisation state of the incoming signal \gls{rf} field~\cite{wang2023polarisation,elgee2024complete}. Another extension involves incorporating more intermediate Rydberg states into \gls{mw} transition loops~\cite{anderson2022interferometer,berweger2023loop}. In this configuration, the \gls{lo} field is split into multiple \gls{rf} components that form a closed excitation path through the atomic levels, similar to loops in optical wave-mixing schemes. This architecture decouples the signal and \gls{lo} frequencies, allowing for less intrusive measurements and expanding the range of addressable Rydberg-Rydberg transitions. In a related development, fully optical versions of the superheterodyne method have been proposed~\cite{borowka2024alloptical,schmidt2025alloptical}, where the \gls{rf} \gls{lo} is replaced by additional laser fields to implement an all-optical detection scheme.

\subsubsection{Cold Atom Architectures}\label{sec:cold_atom_architectures}

Cold atom systems are an interesting alternative to hot vapour systems as they circumvent some fundamental limitations of the latter. Typically, atoms from a room-temperature vapour are cooled by a combination of three pairs of counterpropagating lasers and a gradient of magnetic field, allowing to reach sub-mK temperatures for a sample of about $10^8$ atoms. Noticeably, the physical mechanism responsible for the atomic cooling is the radiation pressure, and no cryogenic systems are required.

A cold atomic sample exhibits, by definition, a lower Doppler broadening, and an increase in response of the atomic medium caused by the reduction of thermal motion. As in the case of the 3-photon excitation scheme in caesium, this leads to a smaller minimal resolvable splitting and thus to a larger measurement range of electric fields, e.g., in~\cite{zhou2023improving}, where \gls{eit} linewidths down to 500~kHz have been reported. Moreover, the Doppler mismatch factor in the A-T splitting is cancelled with such cold vapours~\cite{zhou2022effect}. Smaller linewidths may also lead to an improvement in sensitivity, but cold atom sensors generally suffer from smaller instantaneous bandwidths and sequential/pulsed operation modes, although a continuous time detection (at a bandwidth of 4.7~kHz) was recently demonstrated~\cite{jamieson2025uhf}.

In addition to a much lower Doppler broadening, the high degree of control combined with the high isolation from their environments of cold atom samples make them promising for metrology, as is the case for cold atom clocks. Such an aspect is pointed out in~\cite{duverger2024metrology}, where no noticeable drift on Rydberg state spectroscopy in a magneto-optical trap is observed over a few hours. These characteristics, together with a strongly reduced transit time broadening, offer a longer coherence time that could be harnessed to enhance the sensor sensitivity~\cite{fancher2021comm}.

Cold atom-based sensors also allow the implementation of a direct 2-photon excitation scheme from ground to Rydberg states. In that case, the population of the intermediate (first-excited) state remains negligible thanks to a large single photon-detuning which approaches a quasi-ideal A-T configuration, resulting in a better linearity of the sensor scale factor~\cite{ duverger2024metrology}. In the so-called \gls{eia} regime of~\cite{liao2020electrometry}, this linearity is accompanied by narrow linewidths down to 400~kHz corresponding to a minimal resolvable field of about 100~$\upmu$V/cm. Cold atom-based sensor also demonstrated the possibility of measuring non-resonant \gls{mw} fields over a frequency range on the order of 100 MHz by looking are the centre position of the Autler-Townes doublet~\cite{duverger2024metrology}.

In a recent paper~\cite{tu2024approaching}, a first demonstration of heterodyne detection in a cold atom Rydberg sensor was carried out. The obtained sensitivity of $10~\mathrm{nV/cm/\sqrt{Hz}}$ is comparable to their ``warm vapour counterparts'' and is only 2.6-fold above the associated \gls{qpn} for $5.2\times10^5$ atoms, which allowed investigation into the different underlying noise mechanisms. This sensor also exhibits a resolution of $540~\mathrm{pV/cm}$ for an integration time of 420~s and a bandwidth of 2.3~MHz.

\section{Performance Specification}\label{sec:performance-spec}

Rydberg techniques infer the signal electric field from a measured Rabi frequency, using either optical power readout of a probe beam (methods such as A-T or heterodyne) or photon counting from the atomic vapour (fluorescence, frequency conversion, wave mixing). The primary figure of merit is the \gls{nef},
\begin{equation}
  \mathrm{NEF} \equiv \frac{E_{\min}}{\sqrt{\Delta f}},
\end{equation}
with units such as $\mathrm{V\,m^{-1}\,Hz^{-1/2}}$ or $\mathrm{nV\,cm^{-1}\,Hz^{-1/2}}$, representing the minimum detectable field $E_{\min}$ that yields $\mathrm{SNR}=1$ in measurement bandwidth $\Delta f$. To compare heterogeneous receivers in field units rather than received power, we use an input-referred equivalent free-space electric-field spectral density for a plane wave. Using the effective aperture and a polarisation-coupling factor, the equivalent free-space field at SNR \(=1\) for a single receiving polarisation is
\begin{equation}
  \tilde{E}_{\mathrm{req,free}}\equiv\sqrt{\frac{k_\mathrm{B}\,T_\mathrm{sys}\,\eta_0}{\rho^{2}\,A_e}},
  \label{eq:eref_Ae}
\end{equation}
where \(k_\mathrm{B}\) is Boltzmann’s constant, \(\eta_0 \approx 377~\Omega\) is the free-space wave impedance, \(T_\mathrm{sys}\) is the system noise temperature, and \(A_e\) is the effective collecting area coupling external radiation into the receiver chain. The factor \(\rho\) accounts for polarisation alignment between the incident field and the receive channel \((0\le\rho\le 1)\). Common choices are \(\rho^{2}=1\) for a polarisation-matched coherent signal, and \(\rho^{2}=1/2\) for unpolarised incoherent emission observed with a single linear polarisation channel.

In this form, an existing instrument with known \(T_\mathrm{sys}\) and aperture properties can be mapped to an equivalent free-space field level for like-for-like comparison across receiver classes. Eq.~\eqref{eq:eref_Ae} can be written in the gain-based form used in~\cite{santamaria2022comparison} by substituting $A_e=\lambda^{2}G/(4\pi)=c^{2}G/(4\pi f^{2})$, where $G$ is the equivalent gain, $\lambda$ the wavelength, $f$ the corresponding frequency and $c$ the speed of light. Furthermore, substituting $\eta_0=1/(\varepsilon_0 c)$ gives
\begin{equation}
\mathrm{NEF}=\sqrt{\frac{4\pi f^{2}}{\varepsilon_0 c^{3}G}}\sqrt{k_\mathrm{B} T_\mathrm{sys}}\ \ (\rho^{2}=1), 
\qquad
\mathrm{NEF}=\sqrt{\frac{8\pi f^{2}}{\varepsilon_0 c^{3}G}}\sqrt{k_\mathrm{B} T_\mathrm{sys}}\ \ (\rho^{2}=\tfrac{1}{2}),
\label{eqn:NEF2Tn}
\end{equation}
so the \gls{nef} can be interpreted as the electric field spectral density noise equivalent to a receiving system with noise temperature $T_\mathrm{sys}$ and antenna gain $G$. The inverse direction, using a field sensitivity to infer an implied \(T_\mathrm{sys}\) and \(A_e\), is generally not unique without additional modelling, because both the effective mode coupling (captured only crudely by \(\rho^{2}A_e\)) and the relevant noise behaviour depend on the specific receiver architecture, detection bandwidth, and measurement observable. It is typically discussed for electrically small detectors~\cite{backes2024performance}, can, in some cases, be measured from the sensing process~\cite{borowka2024conversion}, and has been treated theoretically in several works~\cite{santamaria2022comparison,weichman2024doppler}. In practice, if one wanted to report a noise temperature, one should report \gls{nef} together with the assumed $(A_e,\rho)$ so that the free-space requirement Eq.~\eqref{eq:eref_Ae} is explicit and coupling architectures can be designed to enhance the local field. Further discussion, including a more general route from traditional power sensitivities to field sensitivities, is given in the Annex. To ensure consistency of terminology in the context of considering these requirements, we advocate for using the noise-equivalent electric field as the primary sensitivity criterion for assessing the suitability of a Rydberg receiver.

In the following sections, we will give an overview of noise sources in Rydberg detectors. As there are a variety of architectures, the noise contributions will be described in the context of the most relevant architecture. We classify two types of noise:\\
\textbf{Intrinsic noise}, which is inherent to the Rydberg sensor and measurement;\\
\textbf{Extrinsic noise}, which originates from sources external to the sensor.

\subsection{Intrinsic Noise}
\subsubsection{Quantum Projection Noise}
\gls{qpn} is the main noise source that separates Rydberg -- and most quantum -- sensors from their classical counterparts. \gls{qpn} is an inherently quantum noise source that arises due to the measurement of a quantum state~\cite{kitching2011atomic}. This is common across all forms of quantum sensors or instruments and, as such, it is often seen as the analogue to Johnson-Nyquist noise in traditional detectors~\cite{bussey2022shotnoise}. These comparisons often favour Rydberg systems under the ``perfect'' condition of no other noise sources, especially when compared to electrically-small antennas~\cite{backes2024performance}.

In all electrical systems, there is some resistance which, at some temperature $T$, generates a noise power. Similarly, in all quantum sensing systems, a measurement requires the collapse of a quantum state, which generates \gls{qpn}. For Rydberg receivers, this noise is given by~\cite{fan2015rfsensing}:

\begin{equation} \label{eq:QPN}
\text{NEF}_{\mathrm{qpn}} = \frac{h}{|\mathbf{d}|} \cdot \frac{1}{\sqrt{N \tau_{\mathrm{coh}}}},
\end{equation}
where $N$ is the number of atoms contributing to the signal, $\mathbf{d}$ is the dipole moment of the Rydberg transition, and $\tau_{\mathrm{coh}}$ is the coherence time, the duration over which the quantum state maintains a well-defined phase relationship. In this expression, the measurement (integration) time is assumed to be greater than the coherence time.

\gls{qpn} represents the ``ultimate'' noise floor in Rydberg detectors as it constitutes a lower limit on their sensitivity. In a Rydberg superheterodyne study~\cite{jing2020superhet}, the \gls{qpn} was estimated to be $\sim700$ pV/cm/$\sqrt{\text{Hz}}$ (corresponding approximately to $T_n \sim 100$ K), which was a factor of 80 lower than the total observed noise. While this kind of noise represents the theoretical limit, in practice, other noise sources often dominate. Moreover, estimating the correct value of $N$ that actually describes the \gls{qpn} in Eq.~\eqref{eq:QPN} can be a difficult problem, depending on the details of the implementation.

\subsubsection{Photon Shot Noise}
Photon shot noise is the counting statistics associated with the probe power measurement. The associated noise equivalent power spectral density is given by~\cite{haus2000electromagnetic}:
\begin{equation}
P_{\mathrm{SN}} = \sqrt{\bar{P}_p h \nu_p},
\end{equation}
where $\bar{P}_p$ is the detected (average, adjusted for detection efficiency) probe power, and $\nu_p$ is the frequency of the probe laser. 

The amount of shot noise is dependent on the sensing architecture. In typical Rydberg heterodyne systems, photon shot noise often represents the first practical sensitivity barrier. Its magnitude depends on two main factors: the atomic vapour's response to the \gls{mw} field, that is, the resulting change in transmission from a change in the signal \gls{mw} field, and the power of the lasers used~\cite{santamaria2022comparison,prajapati2023comparison}. For conversion/wave-mixing schemes or fluorescence detection, photon shot noise is influenced by various parameters, such as laser powers and the overall efficiency of the photon conversion process~\cite{vogt2019efficient,borowka2024conversion,li2024conversion}.

\subsubsection{Laser Frequency Noise}
Laser frequency noise refers to fluctuations in the stability of the laser frequency (or wavelength). To interact resonantly with atomic energy levels, the laser frequency must be stabilised. A measure of this stability is the laser linewidth, $\gamma$, which represents the width of its optical spectrum. In \gls{eit}-based Rydberg sensors, even small fluctuations in frequency can affect vapour transmission or introduce amplitude noise in the probe laser measurement~\cite{jing2020superhet,yang2024noise}. \par 
More generally, laser frequency noise influences how the atomic medium responds to the \gls{mw} signal field, e.g., changing the transmission response for a given \gls{mw} field strength in \gls{eit}-based applications or the efficiency of photon-conversion processes.

\subsubsection{Transit Noise}
Transit noise arises due to the thermal motion of atoms across the laser beams in room-temperature or heated vapour cells. Atoms that leave the laser path no longer contribute to the signal, while atoms entering the beam, without prior excitation, can introduce noise. Additionally, the power distribution across a laser beam waist is not constant and varies depending on the position of the atom across the laser beam, which can cause transit noise. This transit effect causes signal variability dependent on the beam waist and average atomic velocity~\cite{aoki2016random}. 

Brownian motion of the atoms in the vapour cell can introduce amplitude noise in vapour-cell frequency standards~\cite{micalizio2020brownian}. Experimental demonstrations show that increasing the laser beam waist increases transit noise in \gls{eit}-based Rydberg receivers~\cite{wang2023noise}.

\subsubsection{Photodetector Noise}
Photodetector noise arises in the signal readout of a Rydberg sensor. In an \gls{eit}-based system, the probe laser power is measured with a photodiode whose responsivity dictates the resulting electrical signal. Photodiodes also exhibit intrinsic noise, quantified as a \gls{nep}, even without incident light. The \gls{nep} includes Johnson-Nyquist noise due to electrical resistance, as well as dark current. Typical values range from 10$^{-15}$ to 10$^{-12}$ W/$\sqrt{\mathrm{Hz}}$. Optical homodyne and heterodyne techniques for detection of the probe laser or physically cooling the photodiode could decrease the noise from the \gls{nep} of a photodiode in the case where this is limiting~\cite{santamaria2022comparison}. Common-mode rejection of the probe laser using balanced photodetectors is used in recent \gls{eit}-based Rydberg literature to further improve sensitivity and reduce amplitude noise on the probe laser~\cite{sandidge2024structures}.

In photon-counting setups for conversion or wave-mixing architectures, additional photodetector noise is tied to the quantum efficiency of the single-photon counting.

\subsubsection{Laser Amplitude Noise}
Laser amplitude noise refers to fluctuations (``jitter'') in the optical power of the probe or coupling lasers. In \gls{eit}-based sensors, probe laser amplitude noise is particularly important because it appears directly in the power measurement on a photodiode. This noise is typically expressed as a power spectral density in dBc/Hz reported as \gls{rin}. Balanced photodetection techniques can mitigate \gls{rin}. 

Just like frequency noise, amplitude fluctuations cause time-varying changes in atomic response. In \gls{eit}-based sensors, this alters probe transmission; in conversion-based schemes, it may change the number of converted photons for a given \gls{mw} field strength.

\subsection{Extrinsic Noise}
Noise is introduced by the finite temperature thermal background due to the \gls{bbr} sensed by the Rydberg receiver. We refer to this as thermal noise to distinguish it from Johnson-Nyquist noise in traditional detectors. Unlike conventional receivers, most Rydberg atom sensors (whether based on cold or hot atoms) do not employ an antenna or directive element. Instead, the direct coupling of the atoms to the surrounding fluctuating radiation field leads to a noise contribution akin to an antenna temperature~\cite{santamaria2022comparison,zhu2025general}. Crucially, this thermal field is not limited to the ambient environment as any nearby surface at finite temperature contributes, including the vapour cell glass itself.

In hot atomic sensors, the vapour cell must be heated to 300~K or higher to ensure a sufficient number of atoms participate in the measurement. This heating sets a lower bound on the thermal noise, as the hot glass may contribute to the background \gls{bbr}. Although cold-atom-based sensors are not themselves heated, they are still susceptible to thermal radiation from nearby hot components outside the intended signal path. Thus, both vapour-cell and cold-atom sensors could exhibit a limiting \gls{nef}, due to the physical temperature of their surroundings. Only the use of a cold, confining \gls{mw} structure, leveraging the gain pattern of these receivers, and/or radiative cooling techniques, could mitigate this extrinsic thermal contribution in both hot and cold experimental systems~\cite{santamaria2018sensitivity,santamaria2022comparison}.

For a free-space Rydberg receiver, the thermal noise arising from the total external blackbody field, including vacuum contributions, could be limited by the hot vapour cell itself at 300~K~\cite{santamaria2022comparison,sandidge2024structures}. Direct measurements of the noise introduced by the local blackbody using heterodyne techniques have yet to be performed, whereas conversion techniques have measured it directly~\cite{borowka2024conversion}. Other Rydberg architectures, in the context of thermometry, have measured specific \gls{mw} frequency components of the local blackbody field in a cold system~\cite{schlossberger2025thermometry} and at longer wavelengths in hot vapour cells~\cite{lamantia2025sensor} through other processes.

\subsection{Bandwidth and Detection Frequencies}

There are several interpretations of bandwidth in the context of Rydberg atom detectors, similar to terminology used in traditional \gls{rf} sensing.

\subsubsection{Instantaneous Bandwidth}
The instantaneous bandwidth refers to the maximum frequency range over which a signal can be modulated around a carrier frequency at a given time. Equivalently, it is the bandwidth of a modulated signal that can be received or transmitted without retuning the system. While the architecture influences this range, the ultimate limitation arises from atomic properties, such as the decay rate of the Rydberg state and transit broadening due to atomic motion.

Typical instantaneous bandwidths are on the order of 5--10~MHz, but values as low as 10--100~kHz~\cite{prajapati2022video,jing2020superhet,legaie2024mmwave} and as high as tens of MHz~\cite{borowka2024conversion,yan2025bandwidth} have been reported. This depends on parameters like the beam waists or laser powers~\cite{prajapati2024rydberg}. In optical-to-\gls{mw} photon conversion architectures, the bandwidth is referred to as conversion bandwidth -- the frequency range over which the \gls{mw} field is converted to an optical photon.

\subsubsection{Carrier, Sensing and Tuning Bandwidth}
These terms refer to the overall \gls{rf} frequency range over which the system is capable of detecting signals.

The detection frequency can be tuned across a range of centre frequencies, allowing the system to adapt to different operating bands over time. This tuning bandwidth is typically in the range of 10--100~MHz. This can be achieved through applying external \gls{dc} electric or magnetic fields to tune the Rydberg energy levels~\cite{schlossberger2024zeeman}, by modifying the laser frequencies~\cite{liu2024stable} or the addition of \gls{rf} fields~\cite{berweger2023engineering}.

More broadly, the sensing bandwidth is determined by the large number of Rydberg transitions that can be addressed with suitable laser wavelengths. Depending on the architecture, a Rydberg detector is capable of detecting \gls{dc}--\gls{thz} frequencies. Resonant \gls{eit}-based and wave-mixing architectures require a transition between Rydberg states (shown in Figure \ref{fig:rydberg-branches}), which results in carrier bandwidths ranging from tens of MHz to hundreds of GHz, with the highest demonstrated detection reported at 1.1~THz~\cite{chen2022terahertz}. Non-resonant methods rely on a less-sensitive mechanism but are capable of sensing bandwidths from \gls{dc} to \gls{thz}.

A generic Rydberg detector can then be said to only be \emph{sensitive} to a large number of discrete windows of \gls{mw} frequencies, defined by the Rydberg state being addressed by the laser. This stands in contrast to traditional radio receivers, which typically exhibit broad, continuous sensing bandwidths set mainly by the properties of the \gls{lna} or receiving antenna. To bridge that gap, some efforts have been made to cover frequencies up to 20 GHz~\cite{meyer2021waveguide}. However, higher frequencies still present a challenge due to wider gaps between possible Rydberg transitions.

Additionally, the sensing bandwidth in Rydberg detectors may also be limited by the front-end hardware used, such as the antenna, if used in combination, or any band-pass filters that precede the atomic medium.

\subsection{Experimental Demonstrations and Performance}

\begin{figure}[ht]
  \centering
  \includegraphics[width=\textwidth]{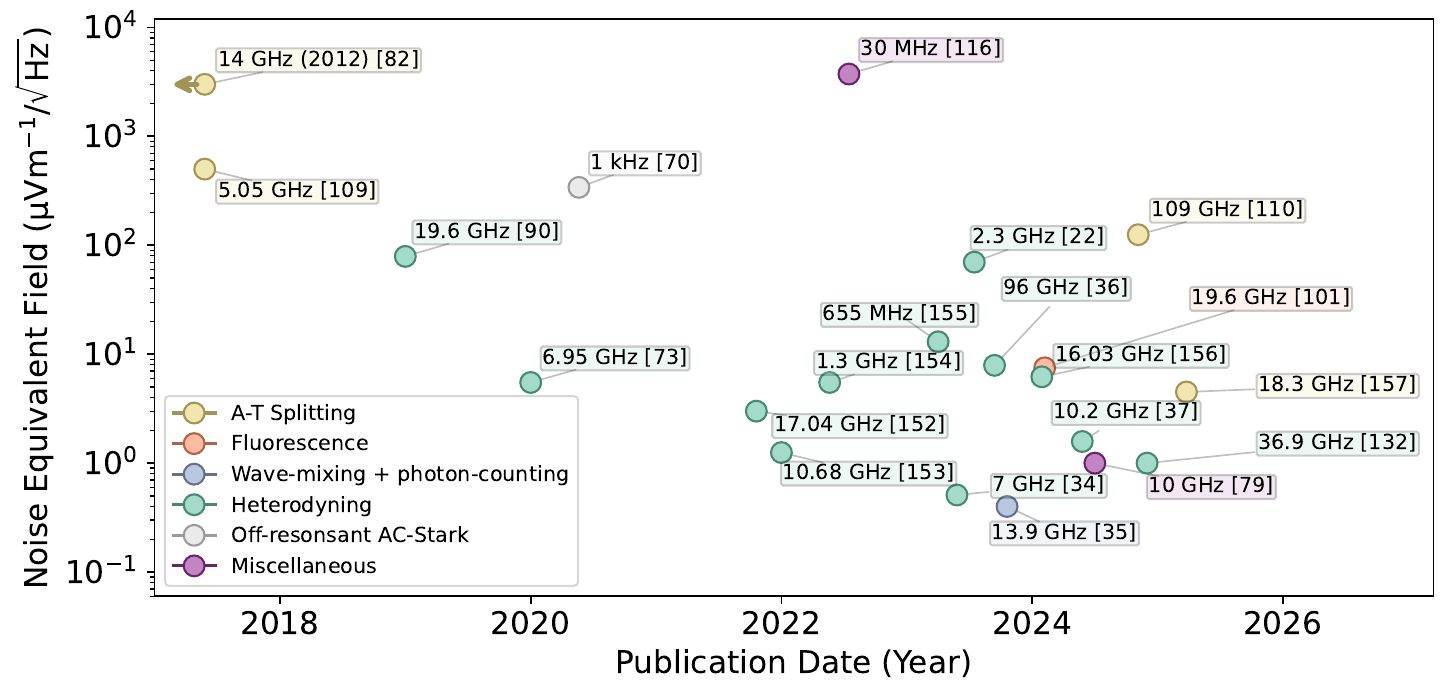}
  \caption{\gls{nef} sensitivities reported for assorted
  Rydberg receivers. Points are labelled by the carrier
  frequency and the reference. The references for each data point are~\cite{sedlacek2012microwave, kumar2017lockin, gordon2019weak, jing2020superhet, jau2020vapor, prajapati2021enhancement, cai2022sensitivity,holloway2022rydberg,liu2022highly,brown2023very,cai2023sensitivity,elgee2023satellite,legaie2024mmwave, borowka2024conversion, yang2024highly, prajapati2024fluorescence,sandidge2024structures,romalis2024rades,bohaichuk2023three,tu2024approaching,venu2025three} are provided in the plot. }
  \label{fig:neftrend}
\end{figure}

The general trend in Figure \ref{fig:neftrend} shows continuous improvement of
sensitivity over the last decade; most state-of-the-art results near
$10\,$GHz report NEFs of $\sim\!1~\upmu\mathrm{V\,m^{-1}\sqrt{Hz}}$. Recent
wave-mixing (up-conversion) experiments presently hold the record, while other techniques, such as fluorescence detection, require more benchmarking.

Because \gls{nef} is convertible to a NET (noise-equivalent temperature), these points
can be benchmarked against the performance of low-noise electronic LNAs.
For example, the $7.9~\upmu\mathrm{Vm^{-1}/\sqrt{Hz}}$ result of
Legaie \emph{et al.}~\cite{legaie2024mmwave} (with free-space coupling) implies a noise
temperature of $T_n\!\approx\!7{,}000\,$K when a simple dipole gain
$G\!=\!3/2$ is assumed. By contrast, Sandidge \emph{et al.}~
\cite{sandidge2024structures} couple the \gls{mw} field to an antenna and cavity,
yielding $1.58~\upmu\mathrm{Vm^{-1}/\sqrt{Hz}}$; as the antenna had a gain
$G\!\approx\!5$ this maps to $T_n\!\approx\!80{,}000\,$K.

Off-resonant \gls{ac}-Stark techniques deliver poorer \gls{nef}, yet extend detection
down to the sub-MHz regime, rivalling ideal $1$-cm dipoles. A-T
splitting measurements remain essential in many architectures as \emph{in-situ} calibration, even where its sensitivity is not as good as traditional receivers (when comparing to ideally matched dipole antennas with LNAs).

Photon shot noise currently limits most demonstrations
\cite{prajapati2023comparison,santamaria2022comparison,wang2023noise}. Using larger laser beams and simultaneously using larger powers, or using
\gls{mw} enhancement structures (cavities, split-ring resonators) may push
this limit downward by increasing the \gls{snr}. A number of enhancement-based receivers have already
been reported
\cite{meyer2021waveguide,legaie2024mmwave,mao2024resonator,prajapati2021enhancement,sandidge2024structures}.

Up-conversion experiments at higher carrier frequencies continue to expand
the Rydberg toolkit. Li \emph{et al.}~report a noise-equivalent intensity of approx. $2.2\times10^{-9}\,$W m$^{-2}$ Hz$^{-1/2}$ at $0.107\,$\gls{thz}
\cite{li2024conversion} while Krokosz \emph{et al.}~have indicated $4.5\times10^{-10}\,$W m$^{-2}$ Hz$^{-1/2}$ at $0.125$ \gls{thz}~\cite{krokosz2025comb}.

\section{Matching of Architecture with Application}
\label{sec:architecture_application}

\begin{table}[ht]
\footnotesize
\centering
\caption{Suitability of Rydberg architectures by application. Colour intensity reflects the current maturity or suitability for each sensing or communication modality.}
\setlength{\tabcolsep}{5pt}
\renewcommand{\arraystretch}{1.2}

\definecolor{demonstrated}{RGB}{102,194,165}   
\definecolor{promising}{RGB}{141,160,203}      
\definecolor{limited}{RGB}{252,141,98}         
\definecolor{auxiliary}{RGB}{231,212,124}      
\definecolor{none}{RGB}{220,220,220}           

\begin{tabular}{@{}l
                >{\centering\arraybackslash}p{1.4cm}
                >{\centering\arraybackslash}p{1.4cm}
                >{\centering\arraybackslash}p{1.4cm}
                >{\centering\arraybackslash}p{1.4cm}
                >{\centering\arraybackslash}p{1.4cm}
                >{\centering\arraybackslash}p{1.4cm}@{}}
\toprule
Architecture & Radiometry & \gls{thz} Imaging/Sensing & Radar / Sounding & Communi-cations & Calibration / Metrology \\
\midrule
Fluorescence &
\cellcolor{limited} &
\cellcolor{demonstrated} \cite{wade2016imaging} &
\cellcolor{none} &
\cellcolor{limited} &
\cellcolor{auxiliary} \\

Autler-Townes &
\cellcolor{auxiliary} &
\cellcolor{demonstrated} \cite{holloway2014sub,chen2022terahertz} &
\cellcolor{demonstrated} \cite{watterson2025radar} &
\cellcolor{auxiliary} \cite{holloway2021amfm,anderson2021amfm,borowka2022mod} \gls{am}/\gls{fm} &
\cellcolor{demonstrated} \cite{sedlacek2012microwave} \\

\gls{ac}-Stark &
\cellcolor{limited} &
\cellcolor{limited} &
\cellcolor{none} &
\cellcolor{demonstrated} \cite{prajapati2024high} &
\cellcolor{auxiliary} \\

Conversion &
\cellcolor{promising} \cite{borowka2024conversion} &
\cellcolor{demonstrated} \cite{krokosz2025comb} &
\cellcolor{none} &
\cellcolor{demonstrated} \cite{zuo2025digital} &
\cellcolor{auxiliary} \\

Superheterodyne &
\cellcolor{promising} &
\cellcolor{limited} &
\cellcolor{demonstrated} \cite{chen2025radar} &
\cellcolor{demonstrated} \cite{holloway2019phase,nowosielski2024qam} &
\cellcolor{auxiliary} \\
\bottomrule
\end{tabular}

\vspace{0.7em}
\begin{minipage}{\linewidth}\footnotesize
\textbf{Legend:} 
\textcolor{demonstrated}{\rule{0.6em}{0.6em}} Experimentally demonstrated \quad
\textcolor{promising}{\rule{0.6em}{0.6em}} Theoretically promising \quad
\textcolor{limited}{\rule{0.6em}{0.6em}} Limited sensitivity or bandwidth \quad
\textcolor{auxiliary}{\rule{0.6em}{0.6em}} Auxiliary or indirect role \quad
\textcolor{none}{\rule{0.6em}{0.6em}} No reported demonstration
\end{minipage}

\label{tab:rydberg_applications_colored}
\end{table}

\subsection{Radiometry}
For radiometry, the most promising Rydberg-based architectures are \gls{rf}-to-optical conversion and superheterodyne \gls{eit}-based methods. \gls{rf}-to-optical conversion has demonstrated the ability to detect thermal radiation, including in room-temperature setups, with noise-equivalent temperatures below 4~K~\cite{borowka2024conversion}. Such performance meets the requirements of radiometric missions, such as directional gain and low noise temperature. The ability to tune the detection frequency of a Rydberg sensor and detect in high-frequency regimes provides further advantages. The optical photon counting and conversion efficiency intrinsic to this method currently limit the \gls{nedt}. 

In contrast, superheterodyne architectures have not yet experimentally demonstrated sufficient sensitivity for thermal background detection and remain limited by other noise sources, principally the shot noise associated with measuring the power of the probe laser~\cite{prajapati2023comparison}. Theoretical studies show that hard sensitivity limits have yet to be reached experimentally and can be surpassed by the introduction of resonant \gls{mw} structures, achieving very low noise temperatures ($<300$~K)~\cite{santamaria2022comparison}. This architecture also has the advantage of integrating into present \gls{rf} systems, where the receiving antenna can be connected directly to the Rydberg sensor~\cite{sandidge2024structures}. Such an integration could feasibly be added to \gls{rf}-to-optical conversion methods. However, this specific architecture then limits the frequency tunability often seen as an advantage in Rydberg systems. Fluorescence-based approaches show potential but have not matched the sensitivity needed for practical radiometry~\cite{prajapati2024fluorescence}. Meanwhile, A-T detection, while useful for calibration, lacks the sensitivity and dynamic range for stand-alone radiometric use. Off-resonant \gls{ac}-Stark architectures are similarly disadvantaged, limited by low sensitivity and offer advantages, when compared to traditional radio detection, in electrically small, low-frequency ($<$ 100~MHz) contexts.

Generally, the fixed low bandwidth of Rydberg systems means that the measurement bandwidth cannot be arbitrarily increased, as opposed to in traditional systems, to improve the \gls{nedt}. This means they may be better suited in contexts where high selectivity is needed, such as in sounding or in a congested spectrum with large RFI.

There are several opportunities to further study the Rydberg detectors for radiometry application; the gain pattern, both for wave-mixing processes and heterodyne techniques, could be studied for different configurations of the vapour cell (that with different beam geometry and cell manufacturing) and detection frequency. This would further clarify its use for wideband detection and as a different detection architecture which requires no \gls{rf} frontend.

\subsection{Imaging and Terahertz Sensing}

In the case of \gls{thz} imaging and sensing, fluorescence detection and superheterodyne detection are currently the most viable approaches. Fluorescence detection offers spatially resolved field detection and has demonstrated two-dimensional \gls{thz} imaging with high frame rates~\cite{downes2020imaging}. However, this system has mostly been studied in active sensing contexts such as material discrimination and non-destructive testing. The sensitivity of this method has not been compared using similar metrics to conventional \gls{thz} detectors used in passive sensing contexts (e.g., radiometric missions), but it gives a minimum detectable power of 190 fW Hz$^{-1/2}$.

Other architectures, such as superheterodyne detection, lack spatial resolution, making them less suitable for imaging. However, these architectures can be used for general \gls{thz} (100+~GHz) sensing with recent work achieving very high sensitivities measuring a 96~GHz field~\cite{legaie2024mmwave}. More study should be undertaken to demonstrate good sensitivities with heterodyne architectures in much higher frequency regimes, 100+~GHz. Although \gls{rf}-to-optical conversion holds theoretical promise -- especially through low-noise and coherence-preserving techniques -- it has not yet been experimentally established in the \gls{thz} frequency regime. An A-T measurement has been demonstrated as a use-case for calibrating \gls{thz} fields acting as a ``standard candle''~\cite{chen2022terahertz}. A-T measurements have also been used for sub-wavelength imaging of microwaves within the vapour cell~\cite{fan2014subwavelength, holloway2014sub}. Off-resonant \gls{ac}-Stark methods are not currently developed for use at \gls{thz} frequencies and likely do not offer sufficient sensitivity.

\subsection{Radar and Sounding}
For radar and active sounding applications, superheterodyne detection is the most suitable Rydberg architecture. Its ability to resolve signal phase, measure electric field amplitude with high spectral resolution, and detect modulated waveforms aligns well with the requirements of radar systems~\cite{anderson2020rydberg,arumugam2024soil}. Demonstrations of pulsed signal detection and multifrequency operation further confirm its practical viability~\cite{anderson2020rydberg}. While A-T detection has shown some capability for pulsed signal analysis~\cite{bohaichuk2022transient}, its lack of phase sensitivity and generally lower sensitivity overall make it less competitive. Fluorescence-based and off-resonant \gls{ac}-Stark methods lack the temporal resolution and field sensitivity necessary for radar detection. \gls{rf}-to-optical conversion has not yet been deployed in this context and remains limited by bandwidth and system complexity.

\subsection{Communications}
In the context of communications, the superheterodyne, and more generally, resonant-\gls{eit} architectures have shown the most promise and have been demonstrated across a range of modulation formats, including \gls{am}, \gls{fm}, phase modulation (PM)~\cite{holloway2019phase}, and \gls{qam}~\cite{fancher2021comm}. The architecture's phase sensitivity and high spectral resolution are crucial for handling modern communication signals. Experimental results include the detection of communication signals from satellites~\cite{elgee2023satellite} and multi-band signals~\cite{liu2022multifrequency,meyer2023simultaneous,allinson2024multiband} with practical SNRs~\cite{nowosielski2024qam}. While A-T detection can, in principle, detect \gls{am}/\gls{fm} signals, its lack of phase sensitivity limits its broader applicability. \gls{rf}-to-optical conversion, although capable of detecting modulated signals~\cite{zuo2025digital}, requires precise optical LOs and remains technically complex for most communication applications. Fluorescence and \gls{ac}-Stark methods again lack the phase sensitivity or resolution required for practical use in this domain. Furthermore, similarly to radiometry, the gain pattern of a generic Rydberg receiver has not been well studied. Such a demonstration of an arbitrary gain pattern for the detection frequency, beam geometry and vapour cell design would help draw better comparisons between traditional \gls{rf} communications architectures and Rydberg \gls{rf} detectors.

\subsection{Calibration and Metrology}
Finally, in calibration and metrology, the A-T splitting method is uniquely well-suited, and cold-atom architectures may be more suited due to their longer coherence times. Although not an effective detection method on its own, it offers an \gls{si}-traceable calibration of electric field amplitudes via the well-known transition dipole moments of Rydberg states~\cite{holloway2014broadband}. This makes it ideal for establishing or maintaining calibrations within more sensitive detection schemes, such as superheterodyne, conversion-based architectures or even as an auxiliary to traditional detectors. It provides a direct link between atomic-scale interactions and macroscopic measurement units without requiring external references like hot/cold loads or amplified noise sources. No other architecture currently offers this level of fundamental calibration capability, making A-T a valuable auxiliary in many experimental contexts, both for Rydberg systems and traditional receivers.

\section{Unique Capabilities}
\label{sec:unique_capabilities}

\subsection{Self-calibration / SI traceability}

Rydberg receivers typically measure the strength of the coupling between two Rydberg states induced by the \gls{em} field of interest, given in the case of a resonant coupling by the Eq.~\eqref{eqn:ATsplitting}. In that formula, \( \mathbf{E} \) is the amplitude of the \gls{em} field to be measured, and \( \mathbf{d} \) is the matrix element of the dipole operator connecting the two Rydberg states involved in the transition. Because those matrix elements can be calculated very accurately for rubidium and caesium, it is in principle possible to link the measured signal \( \Omega \) to the electric field of interest using fundamental constants only. This provides Rydberg receivers with the unique capability of being self-calibrated, or \gls{si}-traceable~\cite{holloway2014broadband}.

This property is reminiscent of other atomic sensors, such as atomic clocks or cold-atom gravimeters, which provide self-calibrated measurements using atomic frequencies as a reference. It should be pointed out, however, that the value of the electric field seen by the atoms, which can indeed be measured in an \gls{si}-traceable way, is, on the other hand, affected by the presence of the glass cell containing the atoms, and also possibly by other elements needed to operate the Rydberg receiver, such as collimators or detectors. It should also be pointed out that a direct self-calibrating measurement is lost when using a \gls{lo} to increase the sensitivity or measure the phase, as performed for example in~\cite{jing2020superhet}, but the measurement can still be traced back to an earlier A-T measurement.

\subsection{Dielectric aspect}

Classical antennas typically rely on measuring an electric current induced in a piece of conductor by the \gls{em} field of interest. Conversely, Rydberg receivers measure electric fields using an atomic vapour, which is a dielectric medium. Provided the other elements required to operate the sensor are also made of dielectric materials (or deported far away, for example, by using optical fibres), the sensor head can, in principle, be made fully dielectric, providing the Rydberg receiver with the unique capability of making measurements of the \gls{em} field with minimal disturbance to the latter~\cite{amarloo2025photonic}.

It should be pointed out that a disturbance is, however, still present, depending on the permittivity of the materials used for the vapour cell and other elements (see previous paragraph). Furthermore, while the gaseous alkali metals do not introduce reflections and ohmic losses, the deposits condensed in the vapour cell may act in such a way. This can be mitigated by filling the vapour cells with precise amounts of alkali medium (so that the solid/liquid phase is not present at the operating temperature). It should also be noted that sensors based on electro-optic materials~\cite{gaborit2014probe} share the dielectric aspect with Rydberg receivers.

\subsection{Tunability, size being independent of the wavelength}

Rydberg receivers typically have an instantaneous bandwidth of several MHz, with a frequency of maximum sensitivity that can be tuned over a broad frequency range (typically from MHz to \gls{thz}) by changing the value of the principal quantum number, which can, for example, be done by tuning the wavelength of the coupling laser. This provides Rydberg receivers with the unique capability to perform sensitive measurements at very different frequencies of interest using one single sensor.

A related unique characteristic of these sensors is the fact that their overall size is independent of the wavelength of the signal to be measured, while the size of a classical antenna will typically be on the order of the latter wavelength, which can be very large for some applications.

\subsection{Other capabilities:}

\subsubsection{Dynamic range}
Superheterodyne detection offers linear dynamic ranges on the order of $80$--$90\ \mathrm{dB}$~\cite{jing2020superhet,Wu2023,borowka2024alloptical}. These values compete with the state of the art of classical receivers at higher frequencies (above $70\ \mathrm{GHz}$).

\subsubsection{High selectivity}
The bandwidth of Rydberg receivers remains relatively constant ($10$--$20\ \mathrm{MHz}$), while the receiving frequency can be increased by addressing lower Rydberg transitions. This results in proportionally higher selectivity at higher frequencies (for example, $Q > 10{,}000$ above $100\ \mathrm{GHz}$).

\subsubsection{All-optical measurement}
An all-optical reception of \gls{qam} signals has been demonstrated in an architecture similar to the superheterodyne configuration~\cite{borowka2024alloptical}. This points to the prospect of fully optical, phase-sensitive reception even without the use of a \gls{lo}.

\subsubsection{Perspectives for sensor arrays}
Rydberg receivers are well suited for operation in sensing arrays, where additional data capacity gains have been demonstrated~\cite{otto2021data}.

\subsubsection{High refresh rate in terahertz imaging}
The demonstrated \gls{thz} imaging capabilities of Rydberg sensors represent the state of the art, achieving frame rates of $3000\ \mathrm{fps}$~\cite{downes2020imaging}.

\section{Current Limitations}
\label{sec:current_limitations}

The technological transfer of Rydberg atomic sensing faces some obstacles and challenges that need to be addressed before considering practical applications. These limitations can be divided into three main categories: technical (T), which can be overcome when better equipment and subsystem solutions become widely available, fundamental (F), dictated by the properties of atoms, and definitional (D), based on the formulation of proper benchmarks and key parameters. The limitations with their indicated criticality for space development are presented in Table \ref{tab:rydberg_limitations_concise} and explained in more detail below.

\begin{table}[ht]
\footnotesize
\centering
\caption{Condensed summary of limitations and enabling factors for Rydberg atomic sensing, grouped by technical (T), fundamental (F), and definitional (D) nature.}
\setlength{\tabcolsep}{5pt}
\renewcommand{\arraystretch}{1.25}

\begin{tabular}{@{}p{3.6cm}p{4.3cm}p{4.3cm}@{}}
\toprule
\textbf{Aspect / Limitation} & \textbf{Description} & \textbf{Implication for Space Development} \\
\midrule

\multicolumn{3}{l}{\textbf{Technical (T)}} \\[2pt]

Baseline transduction efficiency &
Atom-field coupling limits signal conversion without geometric or resonant enhancement. &
Beam shaping and field confinement might raise sensitivity to usable levels. \\

Instrument integration and miniaturisation &
Integration of optical, \gls{rf}, and control subsystems remains fragmented. &
System-level packaging and modular interfaces are needed for space-qualified instruments. \\

Lack of compact laser systems &
Few robust, low-mass lasers exist at the required wavelengths (480--510 nm). Current sources are bulky. &
Compact excitation lasers are key to reducing \gls{swapc} and enabling integrated payloads. \\

Vapour-cell performance &
Surface charges, dielectric losses, and alkali adsorption degrade stability and calibration. &
Requires improved materials, coatings, and temperature control for reproducible behaviour. \\

\midrule
\multicolumn{3}{l}{\textbf{Fundamental (F)}} \\[2pt]

Nonlinear and collective effects &
At higher atomic densities, interactions cause nonlinearity and spatially varying response. The nature and degree of these interactions are still topics in ongoing research. &
These effects might degrade overall performance and calibration. \\

Atomic transition constraints &
Instantaneous bandwidth (MHz level) and dynamic range follow atomic properties and are difficult to enhance; \gls{thz} access is limited by sparse transitions. &
Architectures must target specific transitions. \\

Shot-noise floor &
Many practical implementations, especially in the superheterodyne scheme, are shot-noise limited; squeezed-light methods remain challenging. &
Fundamental research focus should be to improve readout and detection schemes. \\

\midrule
\multicolumn{3}{l}{\textbf{Definitional (D)}} \\[2pt]

Lack of common benchmarks &
Architectures use incompatible performance metrics. &
Unified reference parameters are needed for cross-comparison and qualification. \\

Proof-of-concept maturity &
Several key functions remain at PoC stage (tuning, \gls{thz} detection, quantum limits). &
Application-driven studies and demonstrations are essential for raising TRL. \\

\bottomrule
\end{tabular}

\label{tab:rydberg_limitations_concise}
\end{table}

To better understand the limiting and enabling factors in the development of Rydberg atomic sensing, it may be helpful to briefly analyse the historical progress in this field. A form of Rydberg \gls{rf} sensing appeared in the atomic beam era of experiments~\cite{figger1980photon}, but the crucial developments~\cite{mohapatra2007coherent,mohapatra2008giant,bason2010enhanced} that led to the modern realisations were enabled by the commercial availability of lasers based on second harmonic generation (such as a 480~nm coupling laser for rubidium). The application of Rydberg atoms for \gls{rf} sensing was shown in~\cite{sedlacek2012microwave} and as a calibration tool in~\cite{holloway2014broadband}. Sensitivity was limited by the Doppler effect, which was later overcome by using the atoms as an \gls{rf} mixer with a \gls{lo} field~\cite{simons2019rydberg,gordon2019weak,jing2020superhet}. Later, it was demonstrated that some architectures allow for near-complete cancellation of the Doppler effect even without using the \gls{lo} field~\cite{bohaichuk2023three}. On the other hand, some significant advances have been made to miniaturise and integrate Rydberg sensors~\cite{simons2018fiber,anderson2021self,amarloo2025photonic}.

\subsection{Baseline transduction efficiency (T)}

While Rydberg receivers have demonstrated promising results in terms of achieved sensitivities, there is still some room for improvement. One of the most straightforward methods to increase the sensitivity is to enhance the efficiency of signal reception. While no comprehensive study has been conducted, several results indicate that controlling laser beam shapes and vapour cell temperature can increase the number of Rydberg atoms participating in the detection process, particularly in the frequency-conversion regime~\cite{borowka2024conversion,li2024conversion,weichman2024doppler,santamaria2022comparison}. An additional enhancement route is the use of resonant \gls{rf} structures coupled to classical antennas~\cite{holloway2022rydberg,santamaria2022comparison,mao2024resonator,yang2025local}. This can improve transduction efficiency, although at the expense of wideband tuning operation.

\subsection{Instrument integration and miniaturisation (T)}

While sensor-level architectures have matured considerably, integration of optical, \gls{rf} and electronic subsystems remains fragmented. Current implementations often rely on laboratory-scale optical tables or modular racks. For deployment in space or compact terrestrial platforms, system-level packaging, thermally robust enclosures and standardised interfaces are needed. Integrated photonic platforms and fibre-based architectures~\cite{simons2018fiber,anderson2021self,amarloo2025photonic} provide promising directions, but these approaches have not yet matured to full space-qualified readiness.

\subsection{Lack of compact laser systems (T)}

The main factor limiting the \gls{swapc} of Rydberg sensors is the lack of compact and resilient lasers at some of the required wavelengths. While a significant effort was put into the integration of the sensor part, the laser systems of sufficient powers, linewidth and reliability still weigh several kilograms and come at a significant cost in case of typical 480~nm and 510~nm coupling lasers in Rydberg excitation via two lasers. At present, this could constitute the majority of the Rydberg sensing system's weight and size. What is more, most of the time, tuning the lasers' wavelengths to various Rydberg levels needs to be done manually, as automated systems are not always readily available.

The biggest hope lies in the development of compact laser solutions for these wavelengths. The scientific development of such systems is progressing and very near reaching the requirements~\cite{holguinlerma2020laser,trageser2024blue}, but these solutions are not yet commercially available. Another route to consider could be some of the all-infrared Rydberg excitations via three lasers (i.e., 780~nm + 776~nm + 1260~nm in rubidium and 852~nm + 917~nm + 1145~nm in caesium), where the powers needed are significantly lower (e.g., when the same transitions are considered, 100~mW of 480~nm laser corresponds to only 9~mW of 1260~nm laser in rubidium, and 100~mW of 510~nm laser corresponds to 30~mW of 1145~nm laser in caesium, both cases in terms of Rabi frequencies), and compact commercial solutions are appearing.

\subsection{Vapour-cell performance (T)}

Despite intrinsic atomic calibration advantages, the vapour cell introduces reflections, dielectric losses, and long-term stability challenges. Early work examined \gls{rf} propagation through vapour cells~\cite{fan2015rfsensing,fan2015cell}, and later studies identified multiple mechanisms affecting stability: alkali adsorption on internal surfaces~\cite{ma2025study,sandidge2025impact}, charge accumulation on cell walls~\cite{ma2020dc}, and frequency- and temperature-dependent dielectric responses of glass~\cite{hulst1981light,westphal1972dielectric,strom1977temperature}. These effects complicate reproducible calibration. Material improvements, surface treatments and temperature stabilisation methods are required to mitigate these issues. Another approach is to calibrate out the effect of the cell by measuring it with other sensing methods, such as thermofluorescence imaging~\cite{prost2025microwave}.

\subsection{Nonlinear and collective effects (F)}

At higher atomic densities, dipole-dipole interactions between Rydberg atoms lead to nonlinear and spatially varying response characteristics. These phenomena can produce strong localised enhancement in transduction efficiency, but they reduce overall predictability and calibration repeatability~\cite{carr2013nonequilibrium,weichman2025resonant}. Stable operation therefore requires balancing density against interaction-induced distortions or developing operational regimes that exploit these effects in a controlled manner.

\subsection{Atomic transition constraints (F)}

Key sensing properties, including bandwidth and dynamic range, are dictated by atomic structure. Enhancing one performance dimension often requires concessions in another. For instance, using a strong \gls{lo} can increase instantaneous bandwidth, but at the cost of reduced sensitivity~\cite{hu2022continuously}. While Rydberg transitions are sufficiently dense in the \gls{mw} region to allow near-continuous tuning, transitions in the \gls{thz} domain are sparse, leading to inaccessible frequency gaps~\cite{yuan2023quantum,downes2023practical}. These constraints must be accounted for during system-level frequency plan design.

\subsection{Shot-noise floor (F)}

In superheterodyne readout schemes, the dominant sensitivity limit arises from optical shot noise in the probe beam. Although squeezed light can reduce this noise and has been implemented in large-scale interferometers such as \gls{ligo}~\cite{jia2024squeezing}, equivalent improvements in compact Rydberg architectures are currently unrealistic. Achieving the \gls{qpn} limit would require more than an order of magnitude improvement~\cite{brown2023very}, making the shot-noise floor effectively fundamental in present architectures.

\subsection{Lack of common benchmarks (D)}

Because of the diversity of demonstrated Rydberg detection schemes, reported performance often uses incompatible metrics, complicating meaningful comparison. Additionally, comparison to classical antenna receivers requires carefully chosen equivalent reference quantities~\cite{santamaria2022comparison,weichman2024doppler}. Establishing shared figures of merit and reference measurement procedures will be needed to evaluate maturity and readiness in a system context.

\subsection{Proof-of-concept maturity (D)}

Several sensor capabilities still require dedicated proof-of-concept demonstrations. This applies to fast and reliable tuning between Rydberg transitions, stable mmWave and \gls{thz} operation, approaches for operation near quantum-noise limits, and demonstrations of single-photon-level detection. Raising the technology readiness level requires application-driven demonstrations performed in operationally relevant environments.

\section{Proposed Roadmap and Recommendations} \label{sec:roadmap}

Sections \ref{sec:architecture_application} and \ref{sec:unique_capabilities} identified superheterodyne and \gls{rf}-to-optical conversion architectures as leading candidates for early deployment. Their tunability, \gls{si}-traceability, and phase sensitivity make them particularly well-suited for testing in operationally relevant applications such as calibration sources, mmWave passive radiometry, and \gls{thz} radar receivers.

This section outlines a strategic roadmap for advancing Rydberg atomic sensor technologies from laboratory prototypes to operational space systems. It supports the paper’s cross-disciplinary vision and emphasises actionable steps for stakeholders. The graphical roadmap (Table~\ref{tab:roadmap}) illustrates time-staged goals across technology maturation and application readiness.

\subsection*{Application Readiness Ranking}

Table~\ref{tab:DRL} ranks architecture-application pairs by estimated time to deployment, aligning with the roadmap categories in Table~\ref{tab:roadmap}.

\begin{table}[htbp]
\centering
\caption{Estimated deployment readiness of Rydberg sensor applications}

\small

\definecolor{early}{RGB}{179,225,210}    
\definecolor{mid}{RGB}{243,234,190}      
\definecolor{late}{RGB}{254,198,177}     

\begin{tabular}{|p{3.cm}|p{3.cm}|p{4.cm}|p{2cm}|}
\hline
\textbf{Application Area} & \textbf{Architecture(s)} & \textbf{Key Advantages} & \textbf{Deployment Horizon} \\
\hline
\hline
\rowcolor{early}
Radiometric Calibration & Autler-Townes & \gls{si}-traceable, stable & short-term \\
\rowcolor{mid}
\gls{thz} Imaging & Fluorescence & High refresh rate, spatial resolution & medium-term \\
\rowcolor{mid}
Radiometric Sensing & Conversion, Superheterodyne + Resonator & Cryo-free, high sensitivity & medium-term \\
\rowcolor{mid}
Radar & Superheterodyne & Phase resolution, bandwidth, SWaP & medium-term \\
\rowcolor{mid}
Communications & Superheterodyne & Stealth, modulation flexibility & medium-term \\
\rowcolor{late}
Quantum Communication & Conversion & New capability, entanglement-preserving & long-term \\
\rowcolor{late}
Quantum Networking & Conversion & Entanglement transfer and routing & long-term \\
\rowcolor{late}
Gravitational Wave Detection & Superheterodyne, Conversion & kHz--THz access, quantum limits & very long-term \\
\hline
\end{tabular}

\label{tab:DRL}
\end{table}

\subsection{Technology Roadmap}

The development of Rydberg receivers for space can be structured as a multi-stream, multi-timescale roadmap. The near-term emphasis remains on improving sensitivity, bandwidth, and stability, yet the technology should be regarded as a platform rather than a single-purpose detector. While laboratory work has already demonstrated proof-of-concept field measurements and on-ground electric-field probes exploiting \gls{si}-traceability and dielectric sensing, the exploration and exploitation of the wider set of Rydberg-based capabilities for space applications remain open and essential tasks.

\subsubsection{Sensitivity and Bandwidth-Focused Progression}

Current demonstrations of Rydberg atomic receivers exhibit \gls{nef} sensitivities in the range of \(10^{-6}\)--\(10^{-4}\,\mathrm{V\,m^{-1}\,Hz^{-1/2}}\) with instantaneous bandwidths near \(1\text{--}10~\mathrm{MHz}\). Most space-borne \gls{em} and radiometric applications, however, require field sensitivities beyond \(10^{-7}\) \(\mathrm{V\,m^{-1}\,Hz^{-1/2}}\) and bandwidths of \(10\text{--}10^{4}~\mathrm{MHz}\). Bridging this gap defines the principal long-term challenge and motivates the structure of the roadmap.

\subsubsection{Technology Platform-Level Exploration.}

Running in parallel to the sensitivity and bandwidth track, an exploratory stream should investigate capabilities that are not directly sensitivity-limited but may yield disruptive advantages. Such features position Rydberg receivers as a flexible atomic-sensor platform. Even partial realisation of these properties could open roles beyond the canonical receiver function, such as field calibration references, diagnostic payloads, or hybrid \gls{em}-optical instruments.

\begin{table}[htbp]
\centering
\caption{Community roadmap for Rydberg receiver development, showing priority activities and primary responsibility over time.}

\renewcommand{\arraystretch}{1.4}
\setlength{\tabcolsep}{3pt}

\definecolor{research}{RGB}{198,208,229}
\definecolor{industry}{RGB}{179,225,210}
\definecolor{space}{RGB}{254,198,177}

\begin{tabularx}{\textwidth}{p{3.cm}|YYY}
 & \textbf{0--4 years} & \textbf{4--8 years} & \textbf{8--12 years} \\
\hline

\textbf{Benchmarking and performance definition}
& \cellcolor{research} Establish sensitivity, bandwidth and dynamic range metrics. Perform inter-lab comparisons and report reproducibility.
& \cellcolor{research} Develop shared characterisation protocols and publish reference datasets.
& \cellcolor{space} Adopt traceable standards and integrate calibration frameworks into mission planning. \\

\hline

\textbf{Application and use-case identification}
& \cellcolor{space} Define candidate mission requirements and provide access to testing environments.
& \cellcolor{space}Conduct structured field or airborne trials and identify thresholds for transition to orbit.
& \cellcolor{space} Complete accommodation studies and specify thermal, power and interface requirements.\\

\hline

\textbf{Integration and packaging}
& \cellcolor{industry} Design and test \gls{rf} coupling structures and compact optical layouts.
& \cellcolor{industry} Miniaturise vapour cells and detectors toward MEMS or PIC platforms, and perform subsystem-level environmental testing.
& \cellcolor{space} Complete flight qualification of compact physics packages. \\

\hline

\textbf{Measurement and readout techniques}
& \cellcolor{research} Improve optical readout sensitivity and bandwidth.
& \cellcolor{research} Demonstrate real-time noise-limited operation with tunable and wide input bandwidth.
& \cellcolor{space} Consider multi-channel or array readout architectures for payload designs. \\

\hline

\textbf{Core enabling technologies}
& \cellcolor{industry} Develop compact, low-noise lasers and vapour-cell fabrication approaches suitable for scaling.
& \cellcolor{industry} Engineer radiation-tolerant and temperature-stable laser systems.
& \cellcolor{space} Qualify integrated photonic-laser-cell modules for long-duration missions. \\

\hline

\textbf{Exploratory platform capabilities (parallel stream)} & \multicolumn{3}{p{9.62cm}}{%
  \cellcolor{research}%
  \vspace{-0.5em}
  \parbox{\dimexpr\linewidth-2\tabcolsep\relax}{%
    \centering
    Explore and evaluate novel sensing modalities, including sensitive \gls{si}-traceable field measurements, dielectric sensor heads, frequency-agile schemes and fully optical or distributed architectures. Publish performance characterisation and operational limits.
  }%
  \vspace{0.4em}
} \\

& \multicolumn{3}{p{9.62cm}}{%
  \cellcolor{industry}%
  \vspace{-0.5em}
  \parbox{\dimexpr\linewidth-2\tabcolsep\relax}{%
    \centering
    Assess manufacturability and integration pathways for emerging concepts. Develop engineering designs for multi-band tunability, wavelength-independent scaling and compact implementations suitable for scalable production.
  }%
  \vspace{0.4em}
} \\

& \multicolumn{3}{p{9.62cm}}{%
  \cellcolor{space}%
  \vspace{-0.5em}
  \parbox{\dimexpr\linewidth-2\tabcolsep\relax}{%
    \centering
    Identify mission roles where these capabilities provide system-level benefit. Conduct environment evaluations and incorporate viable concepts into future payload studies.
  }%
  \vspace{0.4em}
} \\

\hline
\end{tabularx}

\vspace{0.7em}
\begin{minipage}{\linewidth}\footnotesize
\textbf{Legend:} 
\textcolor{research}{\rule{0.6em}{0.6em}} Research institutes \quad
\textcolor{industry}{\rule{0.6em}{0.6em}} Industry \quad
\textcolor{space}{\rule{0.6em}{0.6em}} Space agencies
\end{minipage}

\label{tab:roadmap}

\end{table}

\subsection{Recommendations for Stakeholders}

A coordinated effort across research institutes, industry, and space agencies is essential to unlock the full potential of Rydberg-based sensing technologies for space applications. The advancement of this field requires a balance between fundamental research, applied engineering, and strategic programme support, ensuring that progress in sensitivity and platform capabilities converges toward credible, mission-ready systems.

\subsubsection{Academia}

Academic institutions should continue to advance the theoretical and experimental foundations of Rydberg physics, expanding the understanding of atomic interactions, transduction mechanisms, and noise processes that ultimately define achievable sensitivity. Research groups are encouraged to explore novel receiver architectures, hybrid readout schemes, and new operational regimes such as all-optical configurations. Parallel efforts should focus on establishing standardised measurement protocols and reproducible benchmarks so that results from different laboratories can be directly compared and integrated into community-wide datasets supporting roadmap validation.

\subsubsection{Industry}

Industrial partners play a critical role in translating laboratory advances into manufacturable and reliable technologies. Development efforts should prioritise key enabling components, such as compact laser sources, integrated control electronics, and vapour-cell assemblies, whose performance and robustness determine overall system maturity. Particular attention should be given to the ruggedisation and miniaturisation of physics packages, aiming for low-mass, vibration-tolerant subsystems that meet space qualification requirements. Close collaboration with academic researchers and system integrators is essential to ensure that designs remain compatible with emerging measurement techniques and applications and can be scaled for both terrestrial and space markets.

\subsubsection{Space Agencies}

Space agencies should provide the strategic and programmatic framework that allows Rydberg technology to mature from laboratory demonstrations to operational payloads. This includes identifying and articulating specific mission scenarios where atomic receivers can provide unique benefits, such as self-calibrated field references, low-disturbance probes, or frequency-agile diagnostics, and defining the associated system requirements. Agencies should fund comparative trade-off studies and pathfinder missions to validate performance in relevant environments, while also embedding Rydberg-based systems within broader quantum technology and metrology roadmaps. Long-term institutional support will be necessary to integrate Rydberg receivers into standard development cycles for future space instrumentation.

\section{Conclusion}

Rydberg atomic receivers are emerging as a transformative technology for \gls{em} field sensing, bridging quantum metrology and classical radio engineering. By converting \gls{rf}, \gls{mw}, and \gls{thz} radiation into optical signals with \gls{si}-traceable calibration, they provide a fundamentally new measurement paradigm that complements traditional electronic architectures. The analysis presented in this paper shows that Rydberg sensors could serve multiple roles in space applications, ranging from passive radiometry and \gls{thz} imaging to radar, communications, and calibration references.

Across the architectures reviewed (Autler-Townes, \gls{ac}-Stark, fluorescence, superheterodyne, and \gls{rf}-to-optical conversion), each exhibits distinct advantages in sensitivity, bandwidth, or frequency coverage. Laboratory demonstrations already approach \gls{nef} levels in the microvolt-per-centimetre range, with continuous improvement observed over the last decade. However, sensitivity still falls several orders of magnitude short of the requirements for radiometry or low-noise radar, establishing a clear focus for near-term research. Achieving the necessary performance will depend on progress in core enabling technologies, including low-noise and radiation-tolerant lasers, robust vapour-cell materials, and integrated optical-electronic control architectures.

At the same time, Rydberg receivers should not be viewed solely as alternatives to antennas or amplifiers, but as a platform with unique properties. Their intrinsic \gls{si} traceability, dielectric operation, wavelength-independent scaling, and all-optical readout provide a foundation for entirely new instrument classes, such as self-calibrating field probes or hybrid optical-\gls{em} sensors. These capabilities may ultimately prove as impactful as sensitivity itself.

The proposed roadmap integrates both a sensitivity-driven maturation track and an exploratory technology platform-development stream. The first addresses measurable metrics such as \gls{nef}, dynamic range, and \gls{swapc}; the second targets longer-term, high-payoff functionalities including self-calibration, tunability, and array architectures.

Realising this vision will require close coordination between academia, industry, and space agencies. If these efforts converge, Rydberg atomic receivers could evolve within the next decade from laboratory prototypes to mission-ready instruments. Their development represents not only an incremental improvement in sensitivity but the establishment of a new measurement paradigm for spaceborne electrometry and radiometry.

\printglossary[type=\acronymtype,title={List of abbreviations}]

\section*{Declarations}
\bmhead{Availability of data and materials}
The datasets used and/or analysed during the current study are available from the corresponding author on reasonable request.
\bmhead{Competing interests}
The authors declare that they have no competing interests.
\bmhead{Funding}
Members of the University of Warsaw acknowledge the support of the "Quantum Optical Technologies" project (FENG.02.01-IP.05-0017/23; carried out within the Measure 2.1 International Research Agendas programme of the Foundation for Polish Science, co-financed by the European Union under the European Funds for Smart Economy), as well as the PRELUDIUM (2024/53/N/ST7/02730) and SONATA (2021/43/D/ST2/03114) grants from the National Science Centre, Poland. S.B.~is a recipient of the Foundation for Polish Science START 2025 scholarship. The team at ONERA acknowledges funding from the French government through Agence Nationale
de la Recherche within the project ANR-22-CE47-0009-
03 and within France 2030 with the reference ANR-23-PETQ-0004. CSEM acknowledges funding for R\&D activities by the canton of Neuchâtel.

\newcommand{\secref}[1]{Section~\ref{#1} (\nameref{#1})}

\bmhead{Authors' contributions}
The authors are listed in alphabetical order to reflect the collaborative and collective nature of this community-based paper. All authors contributed to the conception during a community meeting on 7th of November, 2024 and to the development of the manuscript. Specific focus areas are presented below:\\

\textbf{Gianluca Allinson}: Contributed to \secref{sec:performance-spec}, and matching of architectures with applications in \secref{sec:architecture_application}.\\

\textbf{Mark Bason}: Drafted \secref{sec:introduction} and contributed to \secref{sec:current_limitations}.\\

\textbf{Alexis Bonnin}: Drafted most of section \secref{sec:cold_atom_architectures}.\\

\textbf{Sebastian Borówka}: Gathered and evaluated data on \secref{sec:current_limitations}, and on \secref{sec:architecture_application}. Gathered and organised references.\\

\textbf{Petronilo Martin-Iglesias}: Gathered data for and contributed to \secref{sec:space-applications}, with emphasis on microwave radiometers and telecommunication systems.\\

\textbf{Manuel Martin Neira}: Contributed considerations on the application of Rydberg sensor technology to microwave radiometry across the document.\\

\textbf{Mateusz Mazelanik}: Gathered, evaluated and illustrated data for the principles of Rydberg receivers in chapter \secref{sec:principles-architectures}. Prepared figures and contributed to tables across the document.\\

\textbf{Richard Murchie}: Gathered information for \secref{sec:radar_and_active_sounding}. Evaluated and illustrated aspects of \secref{sec:from_classical}.\\

\textbf{Michał Parniak}: Outlined \secref{sec:principles-architectures}, analysed properties, and prepared figures and tables.\\

\textbf{Sophio Pataraia}: Gathered information for \secref{sec:space-applications}, especially for future applications.\\

\textbf{Thibaud Ruelle}: Drafted \secref{sec:roadmap}.\\

\textbf{Sylvain Schwartz}: Wrote most of \secref{sec:space-applications} and contributed to \secref{sec:cold_atom_architectures}.\\

\textbf{Aaron Strangfeld}: Led overall project management, including organisation of meetings, coordination of timelines and tasks, and alignment of contributions. Gathered, evaluated, and illustrated data used for comparison with classical receiver technologies presented in Figure \ref{fig:RydbergSpaceApplications} and \secref{sec:methodology}.\\

All authors reviewed, discussed, and approved the final manuscript.

\bmhead{Acknowledgements}
The authors thank colleagues and collaborators for valuable discussions and technical insights, and acknowledge contributions from research institutes, industrial partners, and space agencies involved in quantum sensing and metrology. This document is intended as a collaborative reference for future research on Rydberg-based sensing for space applications, and benefits from the broader scientific community’s ongoing exchange of ideas.

\bmhead{Disclaimer}
\gls{esa} does not formally endorse the specific recommendations, roadmap, or conclusions presented in this publication.
\newpage
\appendix
\section{Annex}
\label{sec:annex}

\subsection{Size of interaction region vs. effective aperture}

One way of interconnecting power sensitivity with electric field sensitivity in Rydberg detection would be to consider the standard relation between detected power $P$ and electric field amplitude $E$
\begin{equation}\label{eq:power-field}
    P = \frac{A E^2}{2 \eta_0},
\end{equation}
where $A$ is the aperture, through which the electric field is integrated, and $\eta_0 = 377\ \mathrm{Ohm}$ is the impedance of free space.

It may be tempting to consider the unknown parameter $A$ as simply the (optical) aperture of the interaction region. For a simplified example of laser beams of diameter $D$ interacting with atomic vapours through length $L$, this could mean that $A = \pi D^2 /4$ for collinear \gls{rf} detection or $A = D L$ for perpendicular detection. This approach, however, leads to several inconsistencies.

Firstly, as the atomic interaction region is usually subwavelength to the detected \gls{rf}, diffraction needs to be taken into account, thus increasing the effective aperture. This is critical in the collinear detection, where for \gls{mw} wavelengths $\lambda$ usually the $D < \lambda < L$ relation holds true, and the efficiency of detection may be defined more through the length of the medium rather than laser beam diameters.

Secondly, some of the Rydberg sensor architectures (superheterodyne and conversion) require geometrical phase matching between the optical and \gls{rf} fields, which may lead to reception patterns differing from a dipole antenna. The proper relation between the detected electric field and power needs to include this consideration, which is lost in assuming that the interaction region can be taken as the effective aperture.

Thirdly, due to effects such as transit noise and Rydberg interactions, the electric field sensitivity may be considered nonlocal (in the sense that it cannot be defined point by point in space, without considering the whole sensor). This leads to invalidity of the Eq.~\eqref{eq:power-field} relation when applied to sensitivities, in regimes where these effects are prominent.

These arguments point to the need for more careful formulation of the relation between the electric field sensitivity (easier to define in atomic measurements) and power sensitivity (more applicable in practical sensing).

\subsection{From Classical Power Sensitivity to Field Sensitivity}
\label{sec:from_classical}
Classical \gls{rf} receivers are specified by noise temperature or power sensitivity referred to a 
\SI{50}{\ohm} port, for example \SI{-140}{dBm/Hz}.
A Rydberg field probe, however, responds directly to the local electric 
field amplitude rather than to a matched electrical port.
To compare such systems consistently, the classical receiver is first expressed as an equivalent 
free-space field requirement, which can then be compared with the probe’s intrinsic field sensitivity at the vapour.

\subsubsection{Step 1. Classical receiver reference}

A conventional \gls{rf} receiving system is characterised by its system noise temperature $T_\mathrm{sys}$ and effective collecting area $A_e$, which together define the power sensitivity referred to free space. For unity \gls{snr} in \SI{1}{Hz} bandwidth, the corresponding power flux density is the system equivalent flux density (SEFD):
\begin{equation*}
  \mathrm{SEFD} = \frac{k_\mathrm{B} T_\mathrm{sys}}{\rho^2 A_e}
  \quad\left[\mathrm{W\,m^{-2}\,Hz^{-1}}\right],
\end{equation*}
where $k_\mathrm{B}$ is Boltzmann’s constant and $\rho$ ($0\!\le\!\rho\!\le\!1$) quantifies polarisation alignment between the incident field and the receiver feed. For an unpolarised signal received by one linear polarisation channel, $\rho^2=1/2$, giving the common form $\mathrm{SEFD}=2k_\mathrm{B} T_\mathrm{sys}/A_e$. Using the free-space impedance $\eta_0=377~\mathrm{Ohm}$, the equivalent free-space electric-field spectral density of a plane wave is
\begin{equation}
  \mathrm{NEF}\equiv\tilde{E}_{\mathrm{req,free}}\equiv
  \sqrt{\frac{k_\mathrm{B} T_\mathrm{sys}\,\eta_0}{\rho^2 A_e}}
  \quad\left[\mathrm{V\,m^{-1}\,Hz^{-1/2}}\right].
  \label{eq:Ereq_free_simple}
\end{equation}
Equation~\eqref{eq:Ereq_free_simple} is interpreted as an \gls{nef} given the constraints of the classical receiver reference. In other words, this is the minimum incident free-space field per $\sqrt{\mathrm{Hz}}$ (we denote the spectral density relation by the tilde $\tilde{\cdot}$) that yields \gls{snr}~1 for polarisation coupling $\rho$.
We note that in incoherent radiometry, the antenna effective area 
$A_e$ and the antenna beam solid angle $\Omega_A$ determines spatial resolution via $A_e\Omega_A \approx \lambda^2$, where $\Omega_A$ is the antenna beam solid angle. The radiometric temperature sensitivity (also known as \gls{nedt}) 
$\Delta T_B = T_\mathrm{sys}/\sqrt{B\tau}$ 
is therefore independent of aperture size, where $\tau$ is integration time (in seconds). 
In the present context, however, $A_e$ is not treated as a design parameter 
but as a proportionality factor that converts between the incident free-space field 
and received power. It serves only to express the equivalent free-space 
electric-field requirement at the antenna aperture, rather than implying 
a change in radiometric performance.

\subsubsection{Step 2. Field enhancement}

Any field-shaping or resonant structure is summarised by a single, dimensionless enhancement factor $\beta_\mathrm{enh}\!\ge\!1$ that maps the incident field at the aperture or reflector to the local field at the sensor:
\begin{equation}
  E_\mathrm{loc} = \beta_\mathrm{enh}\,E_\mathrm{free} \quad\left[\mathrm{V\,m^{-1}}\right],
  \label{eq:Eloc_placeholder}
\end{equation}
where $E_\mathrm{free}$ is a real incoming, free-space field. The value of $\beta_\mathrm{enh}$ is left unspecified here. It may depend on frequency, bandwidth, geometry, coupling, and the degree of coherence, and can be set later once a specific design is chosen.

\subsubsection{Step 3. Local field requirement and comparison}

\begin{figure}[ht]
\centering
\includegraphics[width=1\textwidth]{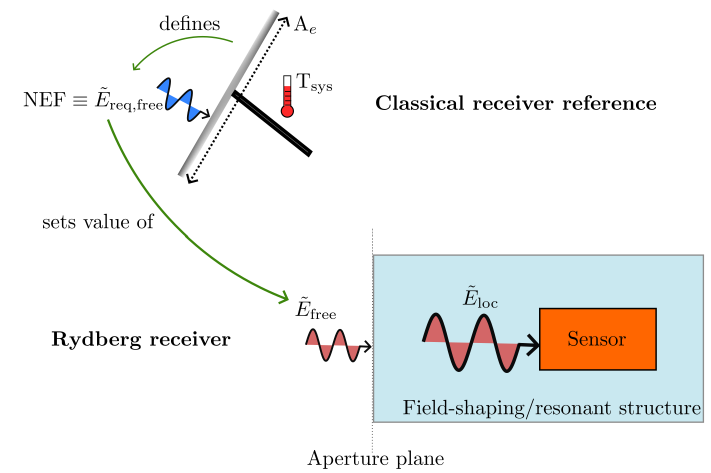}
\caption{Visualisation of how the \gls{nef} ($\tilde{E}_\mathrm{req,free}$) as defined by the classical receiver reference ($A_e$ and $T_\mathrm{sys}$) leads onto the value of the free-space electric-field spectral density $\tilde{E}_\mathrm{free}$ for the Rydberg receiver. It is also shown how the Rydberg receiver is embedded within a field-shaping/resonant structure. This can lead to field enhancement and hence a larger electric-field spectral density local to the sensor $\tilde{E}_\mathrm{loc}$.}
\label{fig:free_to_loc}
\end{figure}

Translating Eq.~\eqref{eq:Eloc_placeholder} into the spectral density equivalent of the local field ($\tilde{E}_\mathrm{loc}$) and substituting in Eq.~\eqref{eq:Ereq_free_simple} (where $\tilde{E}_\mathrm{free}$ is taken to be a real incoming free-space field spectral density set by the \gls{nef} $\tilde{E}_\mathrm{req,free}$), the free-space requirement maps to the local requirement at the sensor as
\begin{equation}
  \tilde{E}_{\mathrm{loc}}
  = \beta_\mathrm{enh}\,\tilde{E}_{\mathrm{free}}
  = \beta_\mathrm{enh}\,
    \sqrt{\frac{k_\mathrm{B} T_\mathrm{sys}\,\eta_0}{\rho^2 A_e}}.
  \label{eq:Ereq_local_placeholder}
\end{equation}
For clarity, $\tilde{E}_\mathrm{req,free}$ is a sensitivity threshold (\gls{nef}) defined here from a classical receiver reference: an increase of aperture lowers that threshold and an increase of system temperature raises it. The \gls{nef} therefore specifies what the minimum equivalent free-space field (per $\sqrt{\mathrm{Hz}}$) can be, in order to achieve unity \gls{snr}. The reason why we consider the (aperture-plane) free-space field is to ensure compatibility when comparing a classical receiver and a Rydberg sensor. To reiterate, an increase of aperture does not increase the strength of a real incoming free-space field. Instead, it is the enhancement factor which determines whether a real incident free-space field can reach or exceed the \gls{nef} locally. A sensor meets or exceeds the classical reference when the following inequality is satisfied
\begin{equation*}
  \mathrm{NEF} \le \tilde{E}_{\mathrm{loc}}.
\end{equation*} The purpose of enhancing the local field is to counter the losses/inefficiency of the Rydberg sensor and thus to ensure a sufficient \gls{snr}. Figure~\ref{fig:free_to_loc} visualizes the definition of $\tilde{E}_\mathrm{req,free}$ from a classical receiver reference and the locations of the defined (incoming) electric fields for the Rydberg sensor.

\subsubsection{Example: Deep-space downlink, 34 m antenna, coherent case with \(Q\) matched to bandwidth}

A single receiving polarisation is assumed, with incident field-to-receiver polarisation coupling \(\rho^2=1\).
The parabolic antenna has diameter \(D=\SI{34}{m}\) and aperture efficiency \(\eta_\mathrm{ap}\approx0.65\), so the effective area is \(A_e=\eta_\mathrm{ap}\,\pi(D/2)^2\approx \SI{5.9e2}{m^2}\).
The receiver system noise temperature is \(T_\mathrm{sys}\).
A single cavity mode at frequency \(f_0\) has angular frequency \(\omega_0=2\pi f_0\), external quality factor \(Q_e\), internal quality factor \(Q_i\), and loaded quality factor \(Q_L\) with \(Q_L^{-1}=Q_e^{-1}+Q_i^{-1}\).
The cavity linewidth is \(\Delta f=f_0/Q_L\) and is chosen to admit the coherent signal bandwidth \(B_\mathrm{sig}\), i.e., \(Q_L\approx f_0/B_\mathrm{sig}\).
The \gls{rf} transfer efficiency from the antenna port into the cavity input is \(\eta_c\) (feeds, guides, match), and the probe samples the electric field in a mode with electric-energy volume \(V_\mathrm{eff}\). For critical coupling \(Q_e=Q_i\) the local \gls{rms} electric field requirement at the probe is~\cite{Pozar:ME:Ch6}:
\begin{equation*}
\beta_\mathrm{enh}
=\sqrt{\eta_c}\,
\sqrt{\frac{2Q_L}{\omega_0}}\,
\sqrt{\frac{A_e}{2\eta_0\,\varepsilon_0\,V_\mathrm{eff}}},
\end{equation*}
where the free-space impedance is \(\eta_0=\SI{377}{\mathrm{Ohm}}\) and the vacuum permittivity is \(\varepsilon_0=\SI{8.854e-12}{F\,m^{-1}}\).

\medskip
\noindent\textbf{X band (coherent case)}

\smallskip
\noindent
Centre frequency \(f_0=\SI{8.4}{GHz}\), signal bandwidth \(B_\mathrm{sig}=\SI{1}{MHz}\), giving
\(Q_L \approx f_0/B_\mathrm{sig} \approx 8.4\times10^{3}\).
With system noise temperature \(T_\mathrm{sys}=\SI{20}{K}\),
coupling efficiency \(\eta_c=0.8\),
and mode volume \(V_\mathrm{eff}=\SI{1.0e-5}{m^3}\),
the resulting sensitivities are
\begin{align*}
\tilde{E}_{\mathrm{free}}
&\approx 1.3\times10^{-11}
\ \mathrm{V\,m^{-1}\,Hz^{-1/2}},\\
\beta_\mathrm{enh}
&\approx 4.7\times10^{4},\\
\tilde{E}_{\mathrm{local}}
&\approx 6.2\times10^{-7}
\ \mathrm{V\,m^{-1}\,Hz^{-1/2}}.
\end{align*}

\medskip
\noindent\textbf{Ka band (coherent case)}

\smallskip
\noindent
Centre frequency \(f_0=\SI{32}{GHz}\), signal bandwidth \(B_\mathrm{sig}=\SI{2}{MHz}\), giving
\(Q_L \approx f_0/B_\mathrm{sig} \approx 1.6\times10^{4}\).
With system noise temperature \(T_\mathrm{sys}=\SI{70}{K}\),
coupling efficiency \(\eta_c=0.7\),
effective antenna area \(A_e\approx\SI{5.0e2}{m^2}\),
and mode volume \(V_\mathrm{eff}=\SI{2.0e-6}{m^3}\),
the resulting sensitivities are
\begin{align*}
\tilde{E}_{\mathrm{free}}
&\approx 2.7\times10^{-11}
\ \mathrm{V\,m^{-1}\,Hz^{-1/2}},\\
\beta_\mathrm{enh}
&\approx 6.5\times10^{4},\\
\tilde{E}_{\mathrm{local}}
&\approx 1.7\times10^{-6}
\ \mathrm{V\,m^{-1}\,Hz^{-1/2}}.
\end{align*}


\section{Methodology for Constructing Receiver Category Ranges}
\label{sec:methodology}

Representative instruments were selected from primary mission documentation
and technical references across the main operational classes:
deep-space communication, space \gls{vlbi}, passive radiometry, \gls{gnss} and satellite
communication, scatterometry and altimetry, sub-millimetre limb sounding,
and \gls{cmb} radiometry.
For each instrument, the following parameters are compiled:

\begin{itemize}
\item Center frequency $f_0$ [GHz],
\item Instantaneous \gls{rf} bandwidth [Hz],
\item Effective aperture $A_e$ [m$^2$],
\item System noise temperature $T_\mathrm{sys}$ [K],
\item Polarisation coupling factor $\rho^2$,
\item Derived free-space electric-field requirement $\tilde{E}_{\mathrm{free}}$ [V\,m$^{-1}$\,Hz$^{-1/2}$].
\end{itemize}

All values are in \gls{si} units. The instantaneous \gls{rf} bandwidth denotes the
noise-equivalent spectral width of the receiver front-end, not necessarily
the telemetry or measurement channel width used in operations.

\subsubsection{Construction of Category Ranges}

To synthesise representative ranges for each receiver class
(Table~\ref{tab:overview_ranges_propagated_verified}), the following procedure is applied:

\begin{enumerate}
\item \textbf{Categorisation:} Instruments are grouped by operational function
and by coherence type (coherent or incoherent).
\item \textbf{Extraction of extrema:} For each parameter
$(f_0, A_e, T_\mathrm{sys}, \text{bandwidth})$, the observed minimum and maximum
values across instruments in the same category are taken.
\item \textbf{Engineering margin:} Lower and upper bounds are expanded
by factors of 0.8 and 1.2, respectively, to reflect
typical uncertainty in realised performance and design margins.
Frequency ranges $f_0$ are \textit{not} expanded and retain their nominal values.
\item \textbf{Propagation of margins to field metric:}
The electric-field requirement is then recalculated using
the perturbed bounds of $A_e$ and $T_\mathrm{sys}$:
\begin{equation*}
\tilde{E}_{\mathrm{free,\,min}} = 
\sqrt{\frac{k_\mathrm{B} (0.8 T_\mathrm{sys,min})\, \eta_0}
      {\rho^2 (1.2 A_{e,\mathrm{max}})}}, \quad
\tilde{E}_{\mathrm{free,\,max}} =
\sqrt{\frac{k_\mathrm{B} (1.2 T_\mathrm{sys,max})\, \eta_0}
      {\rho^2 (0.8 A_{e,\mathrm{min}})}}.
\end{equation*}
This preserves the physical coupling among system parameters. We take $\rho^2 = 1$ for coherent, polarisation-matched systems and
$\rho^2 = 0.5$ for single-channel, unpolarised incoherent systems. This
maintains consistency between communication receivers and radiometric instruments.
\item \textbf{Rounding:} All ranges are rounded to two significant digits
or to practical engineering precision.
\end{enumerate}

The resulting tables provide reproducible bounds reflecting
the diversity of current instruments while avoiding dominance by outliers.

\begin{landscape}
\thispagestyle{empty}
\begin{table*}[p]
\vspace*{-2cm} 
\begin{adjustwidth}{-3cm}{-2cm}
\footnotesize
\caption{Synthesised dataset of exemplary spaceborne receivers including antenna temperature $T_A$ and receiver noise temperature $T_\mathrm{Rx}$, with $T_\mathrm{sys}=T_A+T_\mathrm{Rx}$. 
For each instrument we list centre frequency $f_0$, instantaneous \gls{rf} bandwidth, effective aperture $A_e$, 
polarisation factor $\rho^2$, and an input-referred equivalent free-space electric-field requirement per $\sqrt{\mathrm{Hz}}$, 
$\tilde{E}_\mathrm{free}=\sqrt{k_B T_\mathrm{sys}\eta_0/(\rho^2 A_e)}$, which expresses the system noise power spectral density $k_B T_{\mathrm{sys}}$ as the plane-wave electric-field spectral density that would deliver the same power through $A_e$ (with polarisation coupling $\rho^2$).
 A superscript indicates how a value was obtained; see notes below.}
\setlength{\tabcolsep}{4pt}
\renewcommand{\arraystretch}{1.15}
\begin{tabular}{@{}l l l c c c c c c c c l@{}}
\toprule
Instrument & Mission & Category & $f_0$ [GHz] & Bandwidth [Hz] & $A_e$ [m$^2$] & $T_A$ [K] & $T_\mathrm{Rx}$ [K] & $T_\mathrm{sys}$ [K] & $\rho^2$ & $\tilde{E}_\mathrm{free}$ [$\mathrm{V\,m^{-1}\,Hz^{-1/2}}$] & Ref. \\
\midrule
\multicolumn{12}{l}{\textbf{Deep-space communication, coherent}}\\
DSN 70 m BWG & Goldstone DSS-14 & Deep-space comm. & 8.4  & $4.0\times10^{8}$\textsuperscript{$\text{RF}$} & 2660\textsuperscript{$\physical$} & 5 & 18 & 23 & 1   & $6.7\times10^{-12}$ & \cite{dsn70m_interfaces_2015} \\
\gls{esa} DSA-3 35 m & Malarg\"ue & Deep-space comm. & 8.45 & $2.0\times10^{7}$\textsuperscript{$\text{RF}$} & 625\textsuperscript{$\physical$}  & 5 & 18 & 23 & 1   & $1.4\times10^{-11}$ & \cite{esa_dsa3_2013} \\
\midrule
\multicolumn{12}{l}{\textbf{Space \gls{vlbi}, incoherent}}\\
HALCA 8 m & VSOP & Space \gls{vlbi} & 1.6  & $1.6\times10^{7}$\textsuperscript{$\text{RF}$} & 12.6\textsuperscript{$\physical$} & 51 & 35 & 86   & 0.5 & $2.7\times10^{-10}$ & \cite{tokumaru_2000_halca_pasj} \\
Spektr-R 10 m & RadioAstron & Space \gls{vlbi} & 4.8  & $1.1\times10^{8}$\textsuperscript{$\text{RF}$} & 40\textsuperscript{$\physical$} & 40 & 30 & 70   & 0.5 & $1.4\times10^{-10}$ & \cite{kardashev_2013_radioastron_ar} \\
\midrule
\multicolumn{12}{l}{\textbf{Radiometry, L band, incoherent}}\\
SMAP radiometer & SMAP & L-band radiometer & 1.413 & $2.4\times10^{7}$\textsuperscript{$\text{noise}$} & 18.4\textsuperscript{$\physical$} & - & - & 500  & 0.5 & $5.3\times10^{-10}$ & \cite{piepmeier_2017_smap_tgrs} \\
Aquarius radiometer & Aquarius & L-band radiometer & 1.413 & $2.5\times10^{7}$\textsuperscript{$\text{noise}$} & 3.19\textsuperscript{$\physical$} & - & - & 960\textsuperscript{$\text{NEDT}$} & 0.5 & $1.8\times10^{-9}$ & \cite{Piepmeier2015} \\
\midrule
\multicolumn{12}{l}{\textbf{Interferometric radiometry, L band, incoherent}}\\
SMOS MIRAS element (single LICEF) & SMOS & L-band interferometer & 1.413 & $2.7\times10^{7}$\textsuperscript{$\text{noise}$} & 0.016\textsuperscript{$\text{gain}$} & 250 (assumed) & 210 & 460  & 0.5 & $1.5\times10^{-8}$ & \cite{MartnNeira2014} \\
\midrule
\multicolumn{12}{l}{\textbf{\gls{gnssro}, coherent}}\\
MetOp-SG GRAS-2 & MetOp-SG & \gls{gnssro} & 1.575 & $2.0\times10^{7}$\textsuperscript{$\text{RF}$} & 0.072\textsuperscript{$\text{gain}$} & 30\textsuperscript{$\text{coh-eq}$} & 129\textsuperscript{$\text{NF}$} & 159 & 1 & $3.4\times10^{-9}$ & \cite{Rasch2024} \\
\midrule
\multicolumn{12}{l}{\textbf{Altimeter, coherent}}\\
Jason-2 Poseidon-3 (Ku) & Jason-2 & Altimeter & 13.575 & $3.2\times10^{8}$\textsuperscript{$\text{chirp}$} & 0.74\textsuperscript{$\physical$} & 290\textsuperscript{$\text{coh-eq}$} & 316\textsuperscript{$\text{NF}$} & 606 & 1 & $2.4\times10^{-9}$ & \cite{eoportal_poseidon3b} \\
Jason-2 Poseidon-3 (C) & Jason-2 & Altimeter & 5.30 & $3.2\times10^{8}$\textsuperscript{$\text{chirp}$}  & 0.74\textsuperscript{$\physical$} & 290\textsuperscript{$\text{coh-eq}$} &  83\textsuperscript{$\text{NF}$} & 373  & 1   & $1.7\times10^{-9}$ & \cite{eoportal_poseidon3b} \\
\midrule
\multicolumn{12}{l}{\textbf{\gls{sar}, coherent}}\\
Sentinel-1 C-SAR (IW) & Sentinel-1A/B & \gls{sar} & 5.405 & $1.0\times10^{8}$\textsuperscript{$\text{chirp}$} & 6.56\textsuperscript{$\physical$} & 290\textsuperscript{$\text{coh-eq}$} & 316\textsuperscript{$\text{NF}$} & 606 & 1 & $6.9\times10^{-10}$ & \cite{Torres2012}\\
TerraSAR-X & TerraSAR-X & \gls{sar} & 9.65 & $3.0\times10^{8}$\textsuperscript{$\text{chirp}$} & 2.17\textsuperscript{$\physical$} & 290\textsuperscript{$\text{coh-eq}$} & 627\textsuperscript{$\text{NF}$} & 917 & 1 & $1.5\times10^{-9}$ & \cite{Airbus2015_TerraSARX_ProductGuide, eoportal_terrasarx}\\
\midrule
\multicolumn{12}{l}{\textbf{Radiometry, 60 GHz O$_2$ band, incoherent}}\\
NOAA-19 AMSU-A ch.9 & NOAA-19 & 60 GHz radiometer & 57.29 & $3.1\times10^{8}$\textsuperscript{$\text{noise}$} & 0.0079\textsuperscript{$\text{gain}$} & - & - & 1200\textsuperscript{$\text{NEDT}$} & 0.5 & $5.9\times10^{-8}$ & \cite{mo_2000_nesdis98, atovs_calval} \\
Suomi NPP ATMS ch.14 & Suomi NPP & 60 GHz radiometer & 57.29 & $8.0\times10^{6}$\textsuperscript{$\text{noise}$} & 0.0178\textsuperscript{$\text{gain}$} & - & - & $11000$\textsuperscript{$\text{NEDT}$} & 0.5 & $8.2\times10^{-8}$ & \cite{kim_2014_atms_jgr} \\
\midrule
\multicolumn{12}{l}{\textbf{Radiometry, 183 GHz H$_2$O band, incoherent}}\\
NOAA-17 AMSU-B ch.18 & NOAA-17 & 183 GHz radiometer & 183.3 & $1.0\times10^{9}$\textsuperscript{$\text{noise}$} & 0.006\textsuperscript{$\text{gain}$} & - & - & 4160\textsuperscript{$\text{NEDT}$} & 0.5 & $8.2\times10^{-8}$ & \cite{Saunders1995} \\
\midrule
\multicolumn{12}{l}{\textbf{Limb sounding, sub-millimetre, incoherent}}\\
JEM/SMILES 640 GHz & ISS (SMILES) & Limb sounder & 640 & $2.5\times10^{6}$\textsuperscript{$\text{noise}$} & 0.072\textsuperscript{$\physical$} & 250 (assumed) & 500 & 750 & 0.5 & $1.0\times10^{-8}$ & \cite{smiles_missionplan_ch3_2002}\\
Odin-SMR 557 GHz & Odin & Limb sounder & 557 & $1.0\times10^{6}$\textsuperscript{$\text{noise}$} & 0.66\textsuperscript{$\physical$} & - & - & 3300 & 0.5 & $2.9\times10^{-8}$ & \cite{frisk_2003_odin_aa}\\
2.1~THz heterodyne spectrometer & Keystone (concept) & Limb sounder & 2060 & $1.0\times10^{6}$\textsuperscript{$\text{noise}$} & 0.082\textsuperscript{$\physical$} & 250 (assumed) & 11000 & 11250 & 0.5 & $3.1\times10^{-8}$ & \cite{egusphere-2024-3648}\\
4.7~THz heterodyne spectrometer & Keystone (concept) & Limb sounder & 4745 & $1.0\times10^{6}$\textsuperscript{$\text{noise}$} & 0.082\textsuperscript{$\physical$} & 250 (assumed) & 25000 & 25250 & 0.5 & $4.6\times10^{-8}$ & \cite{egusphere-2024-3648}\\
\midrule
\multicolumn{12}{l}{\textbf{\gls{cmb} radiometer and polarimeter, incoherent}}\\
WMAP K-band & WMAP & \gls{cmb} radiometer & 22.8  & $4.0\times10^{9}$\textsuperscript{$\text{noise}$} & 0.72\textsuperscript{gain} & 2.7 & 26.3 & 29 & 0.5 & $6.5\times10^{-10}$ & \cite{limon_2006_wmap_3yr_supplement} \\
Planck-LFI 70 GHz & Planck & \gls{cmb} radiometer & 70  & $1.4\times10^{10}$\textsuperscript{$\text{noise}$} & 1.13\textsuperscript{gain} & 2.7 & 29.2 & 31.9  & 0.5 & $5.4\times10^{-10}$ & \cite{bersanelli_2010_planck_lfi_aa} \\
\bottomrule
\end{tabular}

\vspace{0.5em}
\begin{minipage}{0.97\linewidth}\footnotesize
\textbf{Notes on methods.} 
$\physical$: from the physical antenna/reflector size $A_{\text{phys}}$ and aperture efficiency $\eta_\mathrm{ap}$ (assumed as 0.65 if not provided in the reference), $A_e=\eta_\mathrm{ap}A_{\text{phys}}$. 
$A_e^\text{gain}$: from gain, $A_e=G\lambda^2/(4\pi)$, $G=10^{\mathrm{dBi}/10}$.  
$\text{\gls{rf}}$: \gls{rf} bandwidth; $\text{noise}$: radiometer channel; $\text{chirp}$: altimeter; $\text{coh-eq}$ is a bookkeeping term used only to express equivalent system noise in a radiometric form; it is not an observed antenna temperature. $\text{NF}$ means calculated based on Noise Figure $\text{NF}$ with assumed $T_0$=290 K, $T_{Rx}=(10^{NF/10}-1)T_0$. $\text{\gls{nedt}}$: is the system noise temperature inferred from reported \gls{nedt} and integration time via the radiometer equation,
$\mathrm{NEDT}=T_\mathrm{sys}/\sqrt{B\tau}$.
\end{minipage}

\label{tab:synthesized_instruments_with_refs}
\end{adjustwidth}
\end{table*}

\end{landscape}
\begin{landscape}
\thispagestyle{empty}
\begin{table*}[p]
\footnotesize
\centering
\caption{Exemplary parameter ranges for major spaceborne receiver classes,
derived from the verified instrument dataset with propagated $\pm20\%$ margins on 
$A_e$ and $T_\mathrm{sys}$. Frequency ranges are not expanded.
$T_\mathrm{sys}$ includes antenna and receiver contributions.}
\setlength{\tabcolsep}{4pt}
\renewcommand{\arraystretch}{1.2}
\begin{tabular}{@{}l c c c c c@{}}
\toprule
Application & $f_0$ [GHz] & $A_e$ [m$^2$] & $T_\mathrm{sys}$ [K] & Bandwidth [Hz] & $\tilde{E}_{\mathrm{free}}$ [$\mathrm{V/m/\sqrt{Hz}}$]\\
\midrule
Deep-space comm. (coh.)  & 8.4--8.45     & 500--3200              & 18--28                 & $1.6\times10^{7}$--$4.8\times10^{8}$  & $5.5\times10^{-12}$--$1.7\times10^{-11}$\\
Space \gls{vlbi} (inc.)        & 1.6--4.8      & 10--48                 & 56--100                & $1.3\times10^{7}$--$1.3\times10^{8}$  & $1.1\times10^{-10}$--$3.3\times10^{-10}$\\
Radiometry, L band (inc.)& 1.413         & 2.6--22                & 400--1200              & $1.9\times10^{7}$--$3.0\times10^{7}$  & $4.3\times10^{-10}$--$2.2\times10^{-9}$\\
Interferom. L band (inc.)& 1.413         & 0.013--0.019           & 370--550               & $2.2\times10^{7}$--$3.2\times10^{7}$  & $1.4\times10^{-8}$--$2.1\times10^{-8}$\\
\gls{gnssro}  & 1.575         & 0.058--0.086           & 130--190               & $1.6\times10^{7}$--$2.4\times10^{7}$  & $2.8\times10^{-9}$--$4.2\times10^{-9}$\\
Altimeter (coh.) & 5.3--13.575 & 0.59--0.89           & 300--730               & $2.6\times10^{8}$--$3.8\times10^{8}$  & $1.3\times10^{-9}$--$2.5\times10^{-9}$\\
\gls{sar} (coh.)               & 5.405--9.65   & 1.7--7.9               & 480--1100              & $8.0\times10^{7}$--$3.6\times10^{8}$  & $5.7\times10^{-10}$--$1.8\times10^{-9}$\\
Radiometry, 60 GHz (inc.)& 57.29         & 0.0063--0.021          & 960--13000 & $6.4\times10^{6}$--$3.7\times10^{8}$  & $2.2\times10^{-8}$--$1.5\times10^{-7}$\\
Radiometry, 183 GHz (inc.)& 183.3        & 0.0048--0.0072         & 3300--5000             & $8.0\times10^{8}$--$1.2\times10^{9}$  & $6.9\times10^{-8}$--$1.0\times10^{-7}$\\
Limb sounding, sub-mm (inc.) & 557--4745  & 0.058--0.79            & 600--30000 & $8.0\times10^{5}$--$3.0\times10^{6}$  & $2.8\times10^{-9}$--$7.4\times10^{-8}$\\
\gls{cmb} radiometer/pol. (inc.) & 22.8--70    & 0.58--1.4              & 23--38                 & $3.2\times10^{9}$--$1.7\times10^{10}$  & $4.2\times10^{-10}$--$8.3\times10^{-10}$\\

\bottomrule
\end{tabular}
\label{tab:overview_ranges_propagated_verified}
\end{table*}

\end{landscape}

\bibliography{sn-bibliography}

@inbook{Pozar:ME:Ch6,
  author    = {Pozar, David M.},
  title     = {Microwave Resonators},
  booktitle = {Microwave Engineering},
  edition   = {4},
  chapter   = {6},
  pages     = {272--307},
  year      = {2012},
  publisher = {Wiley},
  address   = {Hoboken, NJ}
}

@article{raimond2001manipulating,
  title = {Manipulating quantum entanglement with atoms and photons in a cavity},
  volume = {73},
  ISSN = {1539-0756},
  url = {http://dx.doi.org/10.1103/RevModPhys.73.565},
  DOI = {10.1103/revmodphys.73.565},
  number = {3},
  journal = {Reviews of Modern Physics},
  publisher = {American Physical Society (APS)},
  author = {Raimond,  J. M. and Brune,  M. and Haroche,  S.},
  year = {2001},
  month = aug,
  pages = {565–582}
}

@inbook{gallagher2006rydberg,
  title = {Rydberg Atoms},
  ISBN = {9783030738938},
  ISSN = {2522-8706},
  DOI = {10.1007/978-3-030-73893-8\_15},
  booktitle = {Springer Handbook of Atomic,  Molecular,  and Optical Physics},
  publisher = {Springer International Publishing},
  author = {Gallagher,  Thomas F.},
  year = {2023},
  pages = {231–240}
}

@article{li2003millimeter,
  title = {Millimeter-wave spectroscopy of cold Rb Rydberg atoms in a magneto-optical trap: Quantum defects of the ns, np, and nd series},
  volume = {67},
  ISSN = {1094-1622},
  url = {http://dx.doi.org/10.1103/PhysRevA.67.052502},
  DOI = {10.1103/physreva.67.052502},
  number = {5},
  journal = {Physical Review A},
  publisher = {American Physical Society (APS)},
  author = {Li,  Wenhui and Mourachko,  I. and Noel,  M. W. and Gallagher,  T. F.},
  year = {2003},
  month = may,
 pages={052502},
}

@article{jones2020probing,
  title = {Probing new physics using Rydberg states of atomic hydrogen},
  volume = {2},
  ISSN = {2643-1564},
  url = {http://dx.doi.org/10.1103/PhysRevResearch.2.013244},
  DOI = {10.1103/physrevresearch.2.013244},
  number = {1},
  journal = {Physical Review Research},
  publisher = {American Physical Society (APS)},
  author = {Jones,  Matthew P. A. and Potvliege,  Robert M. and Spannowsky,  Michael},
  year = {2020},
  pages={013244},
  month = mar 
}

@article{saffman2010quantum,
  title = {Quantum information with Rydberg atoms},
  volume = {82},
  ISSN = {1539-0756},
  url = {http://dx.doi.org/10.1103/RevModPhys.82.2313},
  DOI = {10.1103/revmodphys.82.2313},
  number = {3},
  journal = {Reviews of Modern Physics},
  publisher = {American Physical Society (APS)},
  author = {Saffman,  M. and Walker,  T. G. and Mølmer,  K.},
  year = {2010},
  month = aug,
  pages = {2313–2363}
}

@article{elgee2025electrically,
  title = {Electrically small Rydberg sensor for three-dimensional determination of radio-frequency k-vectors},
  volume = {23},
  ISSN = {2331-7019},
  url = {http://dx.doi.org/10.1103/pthj-gy98},
  DOI = {10.1103/pthj-gy98},
  number = {6},
  journal = {Physical Review Applied},
  publisher = {American Physical Society (APS)},
  author = {Elgee,  Peter K. and Cox,  Kevin C. and Hill,  Joshua C. and Kunz,  Paul D. and Meyer,  David H.},
  year = {2025},
  pages={064022},
  month = jun 
}

@article{meyer2021waveguide,
  title = {Waveguide-Coupled Rydberg Spectrum Analyzer from 0 to 20 GHz},
  volume = {15},
  ISSN = {2331-7019},
  url = {http://dx.doi.org/10.1103/PhysRevApplied.15.014053},
  DOI = {10.1103/physrevapplied.15.014053},
  number = {1},
  journal = {Physical Review Applied},
  publisher = {American Physical Society (APS)},
  author = {Meyer,  David H. and Kunz,  Paul D. and Cox,  Kevin C.},
  year = {2021},
  pages={014053},
  month = jan 
}

@article{meyer2020assessment,
  title = {Assessment of Rydberg atoms for wideband electric field sensing},
  volume = {53},
  ISSN = {1361-6455},
  url = {http://dx.doi.org/10.1088/1361-6455/ab6051},
  DOI = {10.1088/1361-6455/ab6051},
  number = {3},
  journal = {Journal of Physics B: Atomic,  Molecular and Optical Physics},
  publisher = {IOP Publishing},
  author = {Meyer,  David H and Castillo,  Zachary A and Cox,  Kevin C and Kunz,  Paul D},
  year = {2020},
  month = jan,
  pages = {034001}
}

@article{robinson2021determining,
  title = {Determining the angle-of-arrival of a radio-frequency source with a Rydberg atom-based sensor},
  volume = {118},
  ISSN = {1077-3118},
  url = {http://dx.doi.org/10.1063/5.0045601},
  DOI = {10.1063/5.0045601},
  number = {11},
  journal = {Applied Physics Letters},
  publisher = {AIP Publishing},
  author = {Robinson,  Amy K. and Prajapati,  Nikunjkumar and Senic,  Damir and Simons,  Matthew T. and Holloway,  Christopher L.},
  year = {2021},
  pages = {114001},
  month = mar 
}

@article{richardson2025study,
  title = {Study of angle of arrival estimation with linear arrays of simulated Rydberg atom receivers},
  volume = {2},
  ISSN = {2835-0103},
  url = {http://dx.doi.org/10.1063/5.0240787},
  DOI = {10.1063/5.0240787},
  number = {1},
  journal = {APL Quantum},
  publisher = {AIP Publishing},
  author = {Richardson,  D. and Dee,  J. and Kayim,  B. N. and Sawyer,  B. C. and Wyllie,  R. and Lee,  R. T. and Westafer,  R. S.},
  year = {2025},
  pages ={016123},
  month = feb 
}

@article{schlossberger2025angle,
  title = {Angle-of-arrival detection of radio-frequency waves via Rydberg-atom fluorescence imaging of standing waves in a glass vapor cell},
  volume = {24},
  ISSN = {2331-7019},
  url = {http://dx.doi.org/10.1103/6dl6-754w},
  DOI = {10.1103/6dl6-754w},
  number = {2},
  journal = {Physical Review Applied},
  publisher = {American Physical Society (APS)},
  author = {Schlossberger,  Noah and Talashila,  Rajavardhan and Prajapati,  Nikunjkumar and Holloway,  Christopher L.},
  year = {2025},
  month = aug,
  pages = {024056}
}

@article{yan2023three,
  title = {Three-dimensional location system based on an L-shaped array of Rydberg atomic receivers},
  volume = {48},
  ISSN = {1539-4794},
  url = {http://dx.doi.org/10.1364/OL.496057},
  DOI = {10.1364/ol.496057},
  number = {15},
  journal = {Optics Letters},
  publisher = {Optica Publishing Group},
  author = {Yan,  Yang and Yuan,  Jinpeng and Zhang,  Linjie and Xiao,  Liantuan and Jia,  Suotang and Wang,  Lirong},
  year = {2023},
  month = jul,
  pages = {3945--3948}
}

@article{meschede1985one,
  title = {One-Atom Maser},
  volume = {54},
  ISSN = {0031-9007},
  url = {http://dx.doi.org/10.1103/PhysRevLett.54.551},
  DOI = {10.1103/physrevlett.54.551},
  number = {6},
  journal = {Physical Review Letters},
  publisher = {American Physical Society (APS)},
  author = {Meschede,  D. and Walther,  H. and M\"{u}ller,  G.},
  year = {1985},
  month = feb,
  pages = {551–554}
}

@article{gleyzes2007quantum,
  title = {Quantum jumps of light recording the birth and death of a photon in a cavity},
  volume = {446},
  ISSN = {1476-4687},
  url = {http://dx.doi.org/10.1038/nature05589},
  DOI = {10.1038/nature05589},
  number = {7133},
  journal = {Nature},
  publisher = {Springer Science and Business Media LLC},
  author = {Gleyzes,  Sébastien and Kuhr,  Stefan and Guerlin,  Christine and Bernu,  Julien and Deléglise,  Samuel and Busk Hoff,  Ulrich and Brune,  Michel and Raimond,  Jean-Michel and Haroche,  Serge},
  year = {2007},
  month = mar,
  pages = {297–300}
}

@article{browaeys2020many,
  title = {Many-body physics with individually controlled Rydberg atoms},
  volume = {16},
  ISSN = {1745-2481},
  url = {http://dx.doi.org/10.1038/s41567-019-0733-z},
  DOI = {10.1038/s41567-019-0733-z},
  number = {2},
  journal = {Nature Physics},
  publisher = {Springer Science and Business Media LLC},
  author = {Browaeys,  Antoine and Lahaye,  Thierry},
  year = {2020},
  month = jan,
  pages = {132–142}
}

@article{zeiher2016many,
  title = {Many-body interferometry of a Rydberg-dressed spin lattice},
  volume = {12},
  ISSN = {1745-2481},
  url = {http://dx.doi.org/10.1038/nphys3835},
  DOI = {10.1038/nphys3835},
  number = {12},
  journal = {Nature Physics},
  publisher = {Springer Science and Business Media LLC},
  author = {Zeiher,  Johannes and van Bijnen,  Rick and Schauß,  Peter and Hild,  Sebastian and Choi,  Jae-yoon and Pohl,  Thomas and Bloch,  Immanuel and Gross,  Christian},
  year = {2016},
  month = aug,
  pages = {1095–1099}
}

@article{adams2019rydberg,
  title = {Rydberg atom quantum technologies},
  volume = {53},
  ISSN = {1361-6455},
  url = {http://dx.doi.org/10.1088/1361-6455/ab52ef},
  DOI = {10.1088/1361-6455/ab52ef},
  number = {1},
  journal = {Journal of Physics B: Atomic,  Molecular and Optical Physics},
  publisher = {IOP Publishing},
  author = {Adams,  C S and Pritchard,  J D and Shaffer,  J P},
  year = {2019},
  month = dec,
  pages = {012002}
}

@article{dicke1946thermal,
  title = {The Measurement of Thermal Radiation at Microwave Frequencies},
  volume = {17},
  ISSN = {1089-7623},
  url = {http://dx.doi.org/10.1063/1.1770483},
  DOI = {10.1063/1.1770483},
  number = {7},
  journal = {Review of Scientific Instruments},
  publisher = {AIP Publishing},
  author = {Dicke,  R. H.},
  year = {1946},
  month = jul,
  pages = {268–275}
}

@article{goggins1967feedback,
  title = {A Microwave Feedback Radiometer},
  volume = {AES-3},
  ISSN = {0018-9251},
  url = {http://dx.doi.org/10.1109/TAES.1967.5408717},
  DOI = {10.1109/taes.1967.5408717},
  number = {1},
  journal = {IEEE Transactions on Aerospace and Electronic Systems},
  publisher = {Institute of Electrical and Electronics Engineers (IEEE)},
  author = {Goggins,  William B.},
  year = {1967},
  month = jan,
  pages = {83–90}
}

@inproceedings{mennella2003plancklfi,
  author = {Mennella,  A. and Bersanelli,  M. and Butler,  R. C. and Maino,  D. and Mandolesi,  N. and Morgante,  G. and Valenziano,  L. and Villa,  F. and Gaier,  T. and Seiffert,  M. and Levin,  S. and Lawrence,  C. and Meinhold,  P. and Lubin,  P. and Tuovinen,  J. and Varis,  J. and Karttaavi,  T. and Hughes,  N. and Jukkala,  P. and Sjman,  P. and Kangaslahti,  P. and Roddis,  N. and Kettle,  D. and Winder,  F. and Blackhurst,  E. and Davis,  R. and Wilkinson,  A. and Castelli,  C. and Aja,  B. and Artal,  E. and de la Fuente,  L. and Mediavilla,  A. and Pascual,  J. P. and Gallegos,  J. and Martinez-Gonzalez,  E. and de Paco,  P. and Pradell,  L.},
  title = {Advanced pseudo-correlation radiometers for the Planck-LFI instrument},
  publisher = {arXiv preprint},
  booktitle = {Proc. 3rd ESA Workshop on Millimetre Wave Technology and Applications},
  pages = {69},
  doi = {10.48550/arxiv.astro-ph/0307116},
  year = {2003},
  address = {Espoo, Finland}
}

@article{hansen2025spectrometer,
  title = {Feasibility of a space-borne terahertz heterodyne spectrometer for atomic oxygen and temperature in the mesosphere and lower thermosphere},
  url = {http://dx.doi.org/10.5194/egusphere-2024-3648},
  DOI = {10.5194/egusphere-2024-3648},
  publisher = {Copernicus GmbH},
  journal = {Copernicus GmbH preprint},
  author = {Hansen,  Peder Bagge and Wienold,  Martin and H\"{u}bers,  Heinz-Wilhelm},
  year = {2025},
  month = mar 
}

@inproceedings{pradhan2024radiometer,
  title = {Hyperspectral Microwave Radiometer for Airborne Atmospheric Sounding},
  DOI = {10.1109/igarss53475.2024.10640807},
  booktitle = {IGARSS 2024 - 2024 IEEE International Geoscience and Remote Sensing Symposium},
  publisher = {IEEE},
  author = {Pradhan,  Omkar and Tanner,  Alan and Babenko,  Akim and Mohamed,  Ahmed and Brown,  Shannon and Shah,  Niyati and Kangaslahti,  Pekka and Bosch-Lluis,  Javier and Munoz-Martin,  Joan},
  year = {2024},
  month = jul,
  pages = {6218–6221},
  address = {Athens, Greece}
}

@inproceedings{pett2018radiometer,
  title = {Photonics-based Microwave Radiometer for Hyperspectral Earth Remote Sensing},
  DOI = {10.1109/mwp.2018.8552836},
  booktitle = {2018 International Topical Meeting on Microwave Photonics (MWP)},
  publisher = {IEEE},
  author = {Pett,  Todd and Lee,  Jennifer H. and Ehrlichman,  Yossef and Gevorgyan,  Hayk and Khilo,  Anatol and Popovic,  Milos},
  year = {2018},
  month = oct,
  pages = {1–4},
  address = {Toulouse, France}
}

@article{brown67sar,
  title = {Synthetic Aperture Radar},
  volume = {AES-3},
  ISSN = {0018-9251},
  url = {http://dx.doi.org/10.1109/TAES.1967.5408745},
  DOI = {10.1109/taes.1967.5408745},
  number = {2},
  journal = {IEEE Transactions on Aerospace and Electronic Systems},
  publisher = {Institute of Electrical and Electronics Engineers (IEEE)},
  author = {Brown,  William M.},
  year = {1967},
  month = mar,
  pages = {217–229}
}

@inbook{stewart17sar,
  title = {Synthetic aperture radar algorithms},
  ISBN = {9781420046076},
  ISSN = {1097-9409},
  DOI = {10.1201/9781420046076},
  journal = {Electrical Engineering Handbook},
  booktitle = {Digital Signal Processing Fundamentals},
  publisher = {CRC Press},
  author = {Stewart, Clay and Larson, Vic},
  year = {2009},
  pages = {773--786},
  month = nov,
  address = {Boca Raton, FL, USA}
}

@article{griffiths95isar,
  title = {Interferometric synthetic aperture radar},
  volume = {7},
  ISSN = {2051-218X},
  url = {http://dx.doi.org/10.1049/ecej:19950605},
  DOI = {10.1049/ecej:19950605},
  number = {6},
  journal = {Electronics \& Communication Engineering Journal},
  publisher = {Institution of Engineering and Technology (IET)},
  author = {Griffiths,  H.},
  year = {1995},
  month = dec,
  pages = {247–256}
}

@inproceedings{verbout92sar,
  title = {Polarimetric techniques for enhancing SAR imagery},
  volume = {1630},
  ISSN = {0277-786X},
  DOI = {10.1117/12.59015},
  booktitle = {Synthetic Aperture Radar},
  publisher = {SPIE},
  author = {Verbout,  Shawn M. and Netishen,  Christine M. and Novak,  Leslie M.},
  editor = {McCoy,  Richard D. and Tanenhaus,  Martin E.},
  year = {1992},
  month = may,
  pages = {141–173},
  address = {Los Angeles, CA, United States}
}

@article{torres12sentinel,
  title = {GMES Sentinel-1 mission},
  volume = {120},
  ISSN = {0034-4257},
  url = {http://dx.doi.org/10.1016/j.rse.2011.05.028},
  DOI = {10.1016/j.rse.2011.05.028},
  journal = {Remote Sensing of Environment},
  publisher = {Elsevier BV},
  author = {Torres,  Ramon and Snoeij,  Paul and Geudtner,  Dirk and Bibby,  David and Davidson,  Malcolm and Attema,  Evert and Potin,  Pierre and Rommen,  Bj\"{O}rn and Floury,  Nicolas and Brown,  Mike and Traver,  Ignacio Navas and Deghaye,  Patrick and Duesmann,  Berthyl and Rosich,  Betlem and Miranda,  Nuno and Bruno,  Claudio and L’Abbate,  Michelangelo and Croci,  Renato and Pietropaolo,  Andrea and Huchler,  Markus and Rostan,  Friedhelm},
  year = {2012},
  month = may,
  pages = {9–24}
}

@article{abbott2016gravitation,
  title = {Observation of Gravitational Waves from a Binary Black Hole Merger},
  volume = {116},
  ISSN = {1079-7114},
  url = {http://dx.doi.org/10.1103/PhysRevLett.116.061102},
  DOI = {10.1103/physrevlett.116.061102},
  number = {6},
  journal = {Physical Review Letters},
  publisher = {American Physical Society (APS)},
  author = {B.P. Abbott and R. Abbott and T.D. Abbott and others},
  year = {2016},
  pages   = {061102},
  month = feb 
}

@article{abbott2018gravitation,
  title = {Prospects for observing and localizing gravitational-wave transients with Advanced LIGO,  Advanced Virgo and KAGRA},
  volume = {23},
  ISSN = {1433-8351},
  url = {http://dx.doi.org/10.1007/s41114-020-00026-9},
  DOI = {10.1007/s41114-020-00026-9},
  number = {1},
  journal = {Living Reviews in Relativity},
  publisher = {Springer Science and Business Media LLC},
  author = {B.P. Abbott and R. Abbott and T.D. Abbott and others},
  year = {2020},
  pages = {3},
  month = sep 
}

@article{amaro2017laserantenna,
  title = {Astrophysics with the Laser Interferometer Space Antenna},
  volume = {26},
  ISSN = {1433-8351},
  url = {http://dx.doi.org/10.1007/s41114-022-00041-y},
  DOI = {10.1007/s41114-022-00041-y},
  number = {1},
  journal = {Living Reviews in Relativity},
  publisher = {Springer Science and Business Media LLC},
  author = {Amaro-Seoane,  Pau and Andrews,  Jeff and Arca Sedda,  Manuel and others},
  year = {2023},
  pages = {2},
  month = mar 
}

@article{yagi2011neutron,
  title = {Detector configuration of DECIGO/BBO and identification of cosmological neutron-star binaries},
  volume = {83},
  ISSN = {1550-2368},
  url = {http://dx.doi.org/10.1103/PhysRevD.83.044011},
  DOI = {10.1103/physrevd.83.044011},
  number = {4},
  journal = {Physical Review D},
  publisher = {American Physical Society (APS)},
  author = {Yagi,  Kent and Seto,  Naoki},
  year = {2011},
  pages   = {044011},
  note    = {Erratum: Phys.\ Rev.\ D\ 95, 109901 (2017)},
  month = feb 
}

@article{agazie2023nanograv,
  title = {The NANOGrav 15 yr Data Set: Evidence for a Gravitational-wave Background},
  volume = {951},
  ISSN = {2041-8213},
  url = {http://dx.doi.org/10.3847/2041-8213/acdac6},
  DOI = {10.3847/2041-8213/acdac6},
  number = {1},
  journal = {The Astrophysical Journal Letters},
  publisher = {American Astronomical Society},
  author = {Agazie,  Gabriella and Anumarlapudi,  Akash and Archibald,  Anne M. and others},
  year = {2023},
  month = jun,
  pages = {L8}
}

@article{reardon2023gravitation,
  title = {Search for an Isotropic Gravitational-wave Background with the Parkes Pulsar Timing Array},
  volume = {951},
  ISSN = {2041-8213},
  url = {http://dx.doi.org/10.3847/2041-8213/acdd02},
  DOI = {10.3847/2041-8213/acdd02},
  number = {1},
  journal = {The Astrophysical Journal Letters},
  publisher = {American Astronomical Society},
  author = {Reardon,  Daniel J. and Zic,  Andrew and Shannon,  Ryan M. and Hobbs,  George B. and Bailes,  Matthew and Di Marco,  Valentina and Kapur,  Agastya and Rogers,  Axl F. and Thrane,  Eric and Askew,  Jacob and Bhat,  N. D. Ramesh and Cameron,  Andrew and Curyło,  Małgorzata and Coles,  William A. and Dai,  Shi and Goncharov,  Boris and Kerr,  Matthew and Kulkarni,  Atharva and Levin,  Yuri and Lower,  Marcus E. and Manchester,  Richard N. and Mandow,  Rami and Miles,  Matthew T. and Nathan,  Rowina S. and Osłowski,  Stefan and Russell,  Christopher J. and Spiewak,  Renée and Zhang,  Songbo and Zhu,  Xing-Jiang},
  year = {2023},
  month = jun,
  pages = {L6}
}

@article{jing2020superhet,
  title = {Atomic superheterodyne receiver based on microwave-dressed Rydberg spectroscopy},
  volume = {16},
  ISSN = {1745-2481},
  url = {http://dx.doi.org/10.1038/s41567-020-0918-5},
  DOI = {10.1038/s41567-020-0918-5},
  number = {9},
  journal = {Nature Physics},
  publisher = {Springer Science and Business Media LLC},
  author = {Jing,  Mingyong and Hu,  Ying and Ma,  Jie and Zhang,  Hao and Zhang,  Linjie and Xiao,  Liantuan and Jia,  Suotang},
  year = {2020},
  month = jun,
  pages = {911–915}
}

@article{kanno2025gravitation,
  title = {Search for high-frequency gravitational waves with Rydberg atoms},
  volume = {85},
  ISSN = {1434-6052},
  url = {http://dx.doi.org/10.1140/epjc/s10052-024-13736-z},
  DOI = {10.1140/epjc/s10052-024-13736-z},
  number = {1},
  journal = {The European Physical Journal C},
  publisher = {Springer Science and Business Media LLC},
  author = {Kanno,  Sugumi and Soda,  Jiro and Taniguchi,  Akira},
  year = {2025},
  pages = {31},
  month = jan 
}

@article{low2012guide,
  title = {An experimental and theoretical guide to strongly interacting Rydberg gases},
  volume = {45},
  ISSN = {1361-6455},
  url = {http://dx.doi.org/10.1088/0953-4075/45/11/113001},
  DOI = {10.1088/0953-4075/45/11/113001},
  number = {11},
  journal = {Journal of Physics B: Atomic,  Molecular and Optical Physics},
  publisher = {IOP Publishing},
  author = {L\"{o}w,  Robert and Weimer,  Hendrik and Nipper,  Johannes and Balewski,  Jonathan B and Butscher,  Bj\"{o}rn and B\"{u}chler,  Hans Peter and Pfau,  Tilman},
  year = {2012},
  month = may,
  pages = {113001}
}

@article{gallagher1979interactions,
  title = {Interactions of Blackbody Radiation with Atoms},
  volume = {42},
  ISSN = {0031-9007},
  url = {http://dx.doi.org/10.1103/PhysRevLett.42.835},
  DOI = {10.1103/physrevlett.42.835},
  number = {13},
  journal = {Physical Review Letters},
  publisher = {American Physical Society (APS)},
  author = {Gallagher,  T. F. and Cooke,  W. E.},
  year = {1979},
  month = mar,
  pages = {835–839}
}

@article{figger1980photon,
  title = {A photon detector for submillimetre wavelengths using Rydberg atoms},
  volume = {33},
  ISSN = {0030-4018},
  url = {http://dx.doi.org/10.1016/0030-4018(80)90088-7},
  DOI = {10.1016/0030-4018(80)90088-7},
  number = {1},
  journal = {Optics Communications},
  publisher = {Elsevier BV},
  author = {Figger,  H. and Leuchs,  G. and Straubinger,  R. and Walther,  H.},
  year = {1980},
  month = apr,
  pages = {37–41}
}

@article{agarwal1984vacuum,
  title = {Vacuum-Field Rabi Splittings in Microwave Absorption by Rydberg Atoms in a Cavity},
  volume = {53},
  ISSN = {0031-9007},
  url = {http://dx.doi.org/10.1103/PhysRevLett.53.1732},
  DOI = {10.1103/physrevlett.53.1732},
  number = {18},
  journal = {Physical Review Letters},
  publisher = {American Physical Society (APS)},
  author = {Agarwal,  G. S.},
  year = {1984},
  month = oct,
  pages = {1732–1734}
}

@article{haroche2013nobel,
  title = {Nobel Lecture: Controlling photons in a box and exploring the quantum to classical boundary},
  author = {Haroche, Serge},
  journal = {Rev. Mod. Phys.},
  volume = {85},
  issue = {3},
  pages = {1083--1102},
  numpages = {0},
  year = {2013},
  month = {Jul},
  publisher = {American Physical Society},
  doi = {10.1103/RevModPhys.85.1083},
  url = {https://link.aps.org/doi/10.1103/RevModPhys.85.1083}
}

@misc{romalis2024rades,
  doi = {10.48550/ARXIV.2406.05106},
  author = {Romalis,  Michael V. and Wiedemann,  Joe and Zhang,  Shaobo and Dural,  Nezih},
  title = {Vapor cell Rydberg atom electrometry with time-separated fields},
  publisher = {arXiv preprint},
  year = {2024},
  copyright = {Creative Commons Attribution Non Commercial No Derivatives 4.0 International}
}

@article{wade2016imaging,
  title = {Real-time near-field terahertz imaging with atomic optical fluorescence},
  volume = {11},
  ISSN = {1749-4893},
  url = {http://dx.doi.org/10.1038/nphoton.2016.214},
  DOI = {10.1038/nphoton.2016.214},
  number = {1},
  journal = {Nature Photonics},
  publisher = {Springer Science and Business Media LLC},
  author = {Wade,  C. G. and Šibalić,  N. and de Melo,  N. R. and Kondo,  J. M. and Adams,  C. S. and Weatherill,  K. J.},
  year = {2016},
  month = nov,
  pages = {40–43}
}

@article{holloway2022rydberg,
  title = {Rydberg atom-based field sensing enhancement using a split-ring resonator},
  volume = {120},
  ISSN = {1077-3118},
  url = {http://dx.doi.org/10.1063/5.0088532},
  DOI = {10.1063/5.0088532},
  number = {20},
  journal = {Applied Physics Letters},
  publisher = {AIP Publishing},
  author = {Holloway,  Christopher L. and Prajapati,  Nikunjkumar and Artusio-Glimpse,  Alexandra B. and Berweger,  Samuel and Simons,  Matthew T. and Kasahara,  Yoshiaki and Alù,  Andrea and Ziolkowski,  Richard W.},
  year = {2022},
  pages = {204001},
  month = may 
}

@article{dowling2003quantum,
  title = {Quantum technology: the second quantum revolution},
  volume = {361},
  ISSN = {1471-2962},
  url = {http://dx.doi.org/10.1098/rsta.2003.1227},
  DOI = {10.1098/rsta.2003.1227},
  number = {1809},
  journal = {Philosophical Transactions of the Royal Society of London. Series A: Mathematical,  Physical and Engineering Sciences},
  publisher = {The Royal Society},
  author = {Dowling,  Jonathan P. and Milburn,  Gerard J.},
  editor = {MacFarlane,  A. G. J.},
  year = {2003},
  month = jun,
  pages = {1655–1674}
}

@article{acin2018quantum,
  title = {The quantum technologies roadmap: a European community view},
  volume = {20},
  ISSN = {1367-2630},
  url = {http://dx.doi.org/10.1088/1367-2630/aad1ea},
  DOI = {10.1088/1367-2630/aad1ea},
  number = {8},
  journal = {New Journal of Physics},
  publisher = {IOP Publishing},
  author = {Acín,  Antonio and Bloch,  Immanuel and Buhrman,  Harry and Calarco,  Tommaso and Eichler,  Christopher and Eisert,  Jens and Esteve,  Daniel and Gisin,  Nicolas and Glaser,  Steffen J and Jelezko,  Fedor and Kuhr,  Stefan and Lewenstein,  Maciej and Riedel,  Max F and Schmidt,  Piet O and Thew,  Rob and Wallraff,  Andreas and Walmsley,  Ian and Wilhelm,  Frank K},
  year = {2018},
  month = aug,
  pages = {080201}
}

@article{krelina2021quantum,
  title = {Quantum technology for military applications},
  volume = {8},
  ISSN = {2196-0763},
  url = {http://dx.doi.org/10.1140/epjqt/s40507-021-00113-y},
  DOI = {10.1140/epjqt/s40507-021-00113-y},
  number = {1},
  journal = {EPJ Quantum Technology},
  publisher = {Springer Science and Business Media LLC},
  author = {Krelina,  Michal},
  year = {2021},
  month = nov, 
pages={24}
}

@article{downes2022rapid,
  title = {Rapid readout of terahertz orbital angular momentum beams using atom-based imaging},
  volume = {47},
  ISSN = {1539-4794},
  url = {http://dx.doi.org/10.1364/OL.476945},
  DOI = {10.1364/ol.476945},
  number = {22},
  journal = {Optics Letters},
  publisher = {Optica Publishing Group},
  author = {Downes,  Lucy A. and Whiting,  Daniel J. and Adams,  C. Stuart and Weatherill,  Kevin J.},
  year = {2022},
  month = nov,
  pages = {6001--6004}
}

@article{li2024dual,
  title = {Dual-cameras terahertz imaging with multi-kilohertz frame rates and high sensitivity via Rydberg-atom vapor},
  volume = {58},
  ISSN = {1361-6463},
  url = {http://dx.doi.org/10.1088/1361-6463/ad9c8e},
  DOI = {10.1088/1361-6463/ad9c8e},
  number = {8},
  journal = {Journal of Physics D: Applied Physics},
  publisher = {IOP Publishing},
  author = {Li,  Xianzhe and Li,  Tao and Wan,  Jun and Zhang,  Bin and Huang,  Qirong and Yang,  Xinyu and Feng,  Lie and Zhang,  Kaiqing and Huang,  Wei and Deng,  Haixiao},
  year = {2024},
  month = dec,
  pages = {085109}
}

@article{borowka2024automotive,
  title = {Rydberg-atom-based system for benchmarking millimeter-wave automotive radar chips},
  volume = {22},
  ISSN = {2331-7019},
  url = {http://dx.doi.org/10.1103/PhysRevApplied.22.034067},
  DOI = {10.1103/physrevapplied.22.034067},
  number = {3},
  journal = {Physical Review Applied},
  publisher = {American Physical Society (APS)},
  author = {Borówka,  Sebastian and Krokosz,  Wiktor and Mazelanik,  Mateusz and Wasilewski,  Wojciech and Parniak,  Michał},
  year = {2024},
  pages   = {034067},
  month = sep 
}

@article{anderson2014multiphoton,
  title = {Two-photon microwave transitions and strong-field effects in a room-temperature Rydberg-atom gas},
  volume = {90},
  ISSN = {1094-1622},
  url = {http://dx.doi.org/10.1103/PhysRevA.90.043419},
  DOI = {10.1103/physreva.90.043419},
  number = {4},
  journal = {Physical Review A},
  publisher = {American Physical Society (APS)},
  author = {Anderson,  D. A. and Schwarzkopf,  A. and Miller,  S. A. and Thaicharoen,  N. and Raithel,  G. and Gordon,  J. A. and Holloway,  C. L.},
  year = {2014},
  pages   = {043419},
  month = oct 
}

@article{xue2021twophoton,
  title = {Microwave two-photon spectroscopy of cesium Rydberg atoms},
  volume = {29},
  ISSN = {1094-4087},
  url = {http://dx.doi.org/10.1364/OE.442703},
  DOI = {10.1364/oe.442703},
  number = {26},
  journal = {Optics Express},
  publisher = {Optica Publishing Group},
  author = {Xue,  Yongmei and Jiao,  Yuechun and Hao,  Liping and Zhao,  Jianming},
  year = {2021},
  month = dec,
  pages = {43827}
}

@article{anderson2017highintensity,
  title = {Continuous-frequency measurements of high-intensity microwave electric fields with atomic vapor cells},
  volume = {111},
  ISSN = {1077-3118},
  url = {http://dx.doi.org/10.1063/1.4996234},
  DOI = {10.1063/1.4996234},
  number = {5},
  journal = {Applied Physics Letters},
  publisher = {AIP Publishing},
  author = {Anderson,  D. A. and Raithel,  G.},
  year = {2017},
  pages = {053504},
  month = jul 
}

@article{kumar2017lockin,
  title = {Rydberg-atom based radio-frequency electrometry using frequency modulation spectroscopy in room temperature vapor cells},
  volume = {25},
  ISSN = {1094-4087},
  url = {http://dx.doi.org/10.1364/OE.25.008625},
  DOI = {10.1364/oe.25.008625},
  number = {8},
  journal = {Optics Express},
  publisher = {Optica Publishing Group},
  author = {Kumar,  Santosh and Fan,  Haoquan and K\"{u}bler,  Harald and Jahangiri,  Akbar J. and Shaffer,  James P.},
  year = {2017},
  month = apr,
  pages = {8625--8635}
}

@article{borowka2022mod,
  title = {Sensitivity of a Rydberg-atom receiver to frequency and amplitude
				modulation of microwaves},
  volume = {61},
  ISSN = {2155-3165},
  url = {http://dx.doi.org/10.1364/AO.472295},
  DOI = {10.1364/ao.472295},
  number = {29},
  journal = {Applied Optics},
  publisher = {Optica Publishing Group},
  author = {Borówka,  Sebastian and Pylypenko,  Uliana and Mazelanik,  Mateusz and Parniak,  Michał},
  year = {2022},
  month = oct,
  pages = {8806--8812}
}

@article{cai2022sensitivity,
  title = {Sensitivity Improvement and Determination of Rydberg Atom-Based Microwave Sensor},
  volume = {9},
  ISSN = {2304-6732},
  url = {http://dx.doi.org/10.3390/photonics9040250},
  DOI = {10.3390/photonics9040250},
  number = {4},
  journal = {Photonics},
  publisher = {MDPI AG},
  author = {Cai,  Minghao and Xu,  Zishan and You,  Shuhang and Liu,  Hongping},
  year = {2022},
  month = apr,
  pages = {250}
}

@article{sedlacek2013polarization,
  title = {Atom-Based Vector Microwave Electrometry Using Rubidium Rydberg Atoms in a Vapor Cell},
  volume = {111},
  ISSN = {1079-7114},
  url = {http://dx.doi.org/10.1103/PhysRevLett.111.063001},
  DOI = {10.1103/physrevlett.111.063001},
  number = {6},
  journal = {Physical Review Letters},
  publisher = {American Physical Society (APS)},
  author = {Sedlacek,  J. A. and Schwettmann,  A. and K\"{u}bler,  H. and Shaffer,  J. P.},
  year = {2013},
  pages   = {063001},
  month = aug 
}

@article{anderson2018resonant,
  title = {A vapor-cell atomic sensor for radio-frequency field detection using a polarization-selective field enhancement resonator},
  volume = {113},
  ISSN = {1077-3118},
  url = {http://dx.doi.org/10.1063/1.5038550},
  DOI = {10.1063/1.5038550},
  number = {7},
  journal = {Applied Physics Letters},
  publisher = {AIP Publishing},
  author = {Anderson,  D. A. and Paradis,  E. G. and Raithel,  G.},
  year = {2018},
  pages   = {073501},
  month = aug 
}

@article{mohapatra2008giant,
  title = {A giant electro-optic effect using polarizable dark states},
  volume = {4},
  ISSN = {1745-2481},
  url = {http://dx.doi.org/10.1038/nphys1091},
  DOI = {10.1038/nphys1091},
  number = {11},
  journal = {Nature Physics},
  publisher = {Springer Science and Business Media LLC},
  author = {Mohapatra,  Ashok K. and Bason,  Mark G. and Butscher,  Bj\"{o}rn and Weatherill,  Kevin J. and Adams,  Charles S.},
  year = {2008},
  month = sep,
  pages = {890–894}
}

@article{meyer2023simultaneous,
  title = {Simultaneous Multiband Demodulation Using a Rydberg Atomic Sensor},
  volume = {19},
  ISSN = {2331-7019},
  url = {http://dx.doi.org/10.1103/PhysRevApplied.19.014025},
  DOI = {10.1103/physrevapplied.19.014025},
  number = {1},
  journal = {Physical Review Applied},
  publisher = {American Physical Society (APS)},
  author = {Meyer,  David H. and Hill,  Joshua C. and Kunz,  Paul D. and Cox,  Kevin C.},
  year = {2023},
  pages={014025},
  month = jan 
}

@article{bason2010enhanced,
  title = {Enhanced electric field sensitivity of rf-dressed Rydberg dark states},
  volume = {12},
  ISSN = {1367-2630},
  url = {http://dx.doi.org/10.1088/1367-2630/12/6/065015},
  DOI = {10.1088/1367-2630/12/6/065015},
  number = {6},
  journal = {New Journal of Physics},
  publisher = {IOP Publishing},
  author = {Bason,  M G and Tanasittikosol,  M and Sargsyan,  A and Mohapatra,  A K and Sarkisyan,  D and Potvliege,  R M and Adams,  C S},
  year = {2010},
  month = jun,
  pages = {065015}
}

@article{paradis2019atomic,
  title = {Atomic measurements of high-intensity VHF-band radio-frequency fields with a Rydberg vapor-cell detector},
  volume = {100},
  ISSN = {2469-9934},
  url = {http://dx.doi.org/10.1103/PhysRevA.100.013420},
  DOI = {10.1103/physreva.100.013420},
  number = {1},
  journal = {Physical Review A},
  publisher = {American Physical Society (APS)},
  author = {Paradis,  Eric and Raithel,  Georg and Anderson,  David A.},
  year = {2019},
  pages = {013420},
  month = jul 
}

@article{fan2014subwavelength,
  title = {Subwavelength microwave electric-field imaging using Rydberg atoms inside atomic vapor cells},
  volume = {39},
  ISSN = {1539-4794},
  url = {http://dx.doi.org/10.1364/OL.39.003030},
  DOI = {10.1364/ol.39.003030},
  number = {10},
  journal = {Optics Letters},
  publisher = {Optica Publishing Group},
  author = {Fan,  H. Q. and Kumar,  S. and Daschner,  R. and K\"{u}bler,  H. and Shaffer,  J. P.},
  year = {2014},
  month = may,
  pages = {3030}
}

@article{holloway2014sub,
  title = {Sub-wavelength imaging and field mapping via electromagnetically induced transparency and Autler-Townes splitting in Rydberg atoms},
  volume = {104},
  ISSN = {1077-3118},
  url = {http://dx.doi.org/10.1063/1.4883635},
  DOI = {10.1063/1.4883635},
  number = {24},
  journal = {Applied Physics Letters},
  publisher = {AIP Publishing},
  author = {Holloway,  Christopher L. and Gordon,  Joshua A. and Schwarzkopf,  Andrew and Anderson,  David A. and Miller,  Stephanie A. and Thaicharoen,  Nithiwadee and Raithel,  Georg},
  year = {2014},
  pages = {244102},
  month = jun 
}

@article{wang2023precise,
  title = {Precise measurement of microwave polarization using a Rydberg atom-based mixer},
  volume = {31},
  ISSN = {1094-4087},
  url = {http://dx.doi.org/10.1364/OE.485662},
  DOI = {10.1364/oe.485662},
  number = {6},
  journal = {Optics Express},
  publisher = {Optica Publishing Group},
  author = {Wang,  Yuhan and Jia,  Fengdong and Hao,  Jianhai and Cui,  Yue and Zhou,  Fei and Liu,  Xiubin and Mei,  Jiong and Yu,  Yonghong and Liu,  Ya and Zhang,  Jian and Xie,  Feng and Zhong,  Zhiping},
  year = {2023},
  month = mar,
  pages = {10449--10457}
}

@article{anderson2020rydberg,
  title = {Rydberg Atoms for Radio-Frequency Communications and Sensing: Atomic Receivers for Pulsed RF Field and Phase Detection},
  volume = {35},
  ISSN = {1557-959X},
  url = {http://dx.doi.org/10.1109/MAES.2019.2960922},
  DOI = {10.1109/maes.2019.2960922},
  number = {4},
  journal = {IEEE Aerospace and Electronic Systems Magazine},
  publisher = {Institute of Electrical and Electronics Engineers (IEEE)},
  author = {Anderson,  David A. and Sapiro,  Rachel E. and Raithel,  Georg},
  year = {2020},
  month = apr,
  pages = {48–56}
}

@article{elgee2024complete,
  title = {Complete three-dimensional vector polarimetry with a Rydberg-atom rf electrometer},
  volume = {22},
  ISSN = {2331-7019},
  url = {http://dx.doi.org/10.1103/PhysRevApplied.22.064012},
  DOI = {10.1103/physrevapplied.22.064012},
  number = {6},
  journal = {Physical Review Applied},
  publisher = {American Physical Society (APS)},
  author = {Elgee,  Peter K. and Cox,  Kevin C. and Hill,  Joshua C. and Kunz,  Paul D. and Meyer,  David H.},
  year = {2024},
  pages = {064012},
  month = dec 
}

@article{schlossberger2024rydberg,
  title = {Rydberg states of alkali atoms in atomic vapour as SI-traceable field probes and communications receivers},
  volume = {6},
  ISSN = {2522-5820},
  url = {http://dx.doi.org/10.1038/s42254-024-00756-7},
  DOI = {10.1038/s42254-024-00756-7},
  number = {10},
  journal = {Nature Reviews Physics},
  publisher = {Springer Science and Business Media LLC},
  author = {Schlossberger,  Noah and Prajapati,  Nikunjkumar and Berweger,  Samuel and Rotunno,  Andrew P. and Artusio-Glimpse,  Alexandra B. and Simons,  Matthew T. and Sheikh,  Abrar A. and Norrgard,  Eric B. and Eckel,  Stephen P. and Holloway,  Christopher L.},
  year = {2024},
  month = sep,
  pages = {606–620}
}

@article{meyer2018digital,
  title = {Digital communication with Rydberg atoms and amplitude-modulated microwave fields},
  volume = {112},
  ISSN = {1077-3118},
  url = {http://dx.doi.org/10.1063/1.5028357},
  DOI = {10.1063/1.5028357},
  number = {21},
  journal = {Applied Physics Letters},
  publisher = {AIP Publishing},
  author = {Meyer,  David H. and Cox,  Kevin C. and Fatemi,  Fredrik K. and Kunz,  Paul D.},
  year = {2018},
  pages   = {093201},
  month = may 
}

@article{han2018coherent,
  title = {Coherent Microwave-to-Optical Conversion via Six-Wave Mixing in Rydberg Atoms},
  volume = {120},
  ISSN = {1079-7114},
  url = {http://dx.doi.org/10.1103/PhysRevLett.120.093201},
  DOI = {10.1103/physrevlett.120.093201},
  number = {9},
  journal = {Physical Review Letters},
  publisher = {American Physical Society (APS)},
  author = {Han,  Jingshan and Vogt,  Thibault and Gross,  Christian and Jaksch,  Dieter and Kiffner,  Martin and Li,  Wenhui},
  year = {2018},
  pages={093201},
  month = mar 
}

@article{vogt2019efficient,
  title = {Efficient microwave-to-optical conversion using Rydberg atoms},
  volume = {99},
  ISSN = {2469-9934},
  url = {http://dx.doi.org/10.1103/PhysRevA.99.023832},
  DOI = {10.1103/physreva.99.023832},
  number = {2},
  journal = {Physical Review A},
  publisher = {American Physical Society (APS)},
  author = {Vogt,  Thibault and Gross,  Christian and Han,  Jingshan and Pal,  Sambit B. and Lam,  Mark and Kiffner,  Martin and Li,  Wenhui},
  year = {2019},
  pages = {023832},
  month = feb 
}

@article{kumar2023quantum,
  title = {Quantum-enabled millimetre wave to optical transduction using neutral atoms},
  volume = {615},
  ISSN = {1476-4687},
  url = {http://dx.doi.org/10.1038/s41586-023-05740-2},
  DOI = {10.1038/s41586-023-05740-2},
  number = {7953},
  journal = {Nature},
  publisher = {Springer Science and Business Media LLC},
  author = {Kumar,  Aishwarya and Suleymanzade,  Aziza and Stone,  Mark and Taneja,  Lavanya and Anferov,  Alexander and Schuster,  David I. and Simon,  Jonathan},
  year = {2023},
  month = mar,
  pages = {614–619}
}

@article{schlossberger2024zeeman,
  title = {Zeeman-resolved Autler-Townes splitting in Rydberg atoms with tunable resonances and a single transition dipole moment},
  author = {Schlossberger, Noah and Rotunno, Andrew P. and Artusio-Glimpse, Alexandra B. and Prajapati, Nikunjkumar and Berweger, Samuel and Shylla, Dangka and Simons, Matthew T. and Holloway, Christopher L.},
  journal = {Phys. Rev. A},
  volume = {109},
  issue = {2},
  pages = {L021702},
  numpages = {6},
  year = {2024},
  month = {Feb},
  publisher = {American Physical Society},
  doi = {10.1103/PhysRevA.109.L021702},
  url = {https://link.aps.org/doi/10.1103/PhysRevA.109.L021702}
}

@article{liu2024stable,
  title = {Stable,  narrow-linewidth laser system with a broad frequency tunability and a fast switching time},
  volume = {49},
  ISSN = {1539-4794},
  url = {http://dx.doi.org/10.1364/OL.510825},
  DOI = {10.1364/ol.510825},
  number = {2},
  journal = {Optics Letters},
  publisher = {Optica Publishing Group},
  author = {Liu,  C. and Nickerson,  K. and Booth,  D. W. and Frechem,  J. and Tai,  H. and Miladi,  H. and Moore,  K. and Shaffer,  J. P.},
  year = {2024},
  month = jan,
  pages = {399--402}
}

@article{berweger2023engineering,
  title = {Rydberg-State Engineering: Investigations of Tuning Schemes for Continuous Frequency Sensing},
  author = {Berweger, Samuel and Prajapati, Nikunjkumar and Artusio-Glimpse, Alexandra B. and Rotunno, Andrew P. and Brown, Roger and Holloway, Christopher L. and Simons, Matthew T. and Imhof, Eric and Jefferts, Steven R. and Kayim, Baran N. and Viray, Michael A. and Wyllie, Robert and Sawyer, Brian C. and Walker, Thad G.},
  journal = {Phys. Rev. Appl.},
  volume = {19},
  issue = {4},
  pages = {044049},
  numpages = {13},
  year = {2023},
  month = {Apr},
  publisher = {American Physical Society},
  doi = {10.1103/PhysRevApplied.19.044049},
  url = {https://link.aps.org/doi/10.1103/PhysRevApplied.19.044049}
}

@article{li2024conversion,
  title = {Room temperature single-photon terahertz detection with thermal Rydberg atoms},
  volume = {11},
  ISSN = {1931-9401},
  url = {http://dx.doi.org/10.1063/5.0219879},
  DOI = {10.1063/5.0219879},
  number = {4},
  journal = {Applied Physics Reviews},
  publisher = {AIP Publishing},
  author = {Li,  Danyang and Bai,  Zhengyang and Zuo,  Xiaoliang and Wu,  Yuelong and Sheng,  Jiteng and Wu,  Haibin},
  year = {2024},
  pages = {041420},
  month = nov 
}

@article{gordon2019weak,
  title = {Weak electric-field detection with sub-1 Hz resolution at radio frequencies using a Rydberg atom-based mixer},
  volume = {9},
  ISSN = {2158-3226},
  url = {http://dx.doi.org/10.1063/1.5095633},
  DOI = {10.1063/1.5095633},
  number = {4},
  journal = {AIP Advances},
  publisher = {AIP Publishing},
  author = {Gordon,  Joshua A. and Simons,  Matthew T. and Haddab,  Abdulaziz H. and Holloway,  Christopher L.},
  year = {2019},
  pages = {045030},
  month = apr 
}

@article{weichman2024doppler,
  title = {Doppler sensitivity and resonant tuning of Rydberg atom-based antennas},
  volume = {57},
  ISSN = {1361-6455},
  url = {http://dx.doi.org/10.1088/1361-6455/ad6385},
  DOI = {10.1088/1361-6455/ad6385},
  number = {16},
  journal = {Journal of Physics B: Atomic,  Molecular and Optical Physics},
  publisher = {IOP Publishing},
  author = {Weichman,  Peter B},
  year = {2024},
  month = jul,
  pages = {165501}
}

@article{bussey2022shotnoise,
  title = {Quantum Shot Noise Limit in a Rydberg RF Receiver Compared to Thermal Noise Limit in a Conventional Receiver},
  volume = {6},
  ISSN = {2475-1472},
  url = {http://dx.doi.org/10.1109/LSENS.2022.3203465},
  DOI = {10.1109/lsens.2022.3203465},
  number = {9},
  journal = {IEEE Sensors Letters},
  publisher = {Institute of Electrical and Electronics Engineers (IEEE)},
  author = {Bussey,  Liam W. and Burton,  Fraser A. and Bongs,  Kai and Goldwin,  Jonathan and Whitley,  Tim},
  year = {2022},
  month = sep,
  pages = {1–4}
}

@article{backes2024performance,
  title = {Performance of antenna-based and Rydberg quantum RF sensors in the electrically small regime},
  volume = {125},
  ISSN = {1077-3118},
  url = {http://dx.doi.org/10.1063/5.0222827},
  DOI = {10.1063/5.0222827},
  number = {14},
  journal = {Applied Physics Letters},
  publisher = {AIP Publishing},
  author = {Backes,  K. M. and Elgee,  P. K. and LeBlanc,  K.-J. and Fancher,  C. T. and Meyer,  D. H. and Kunz,  P. D. and Malvania,  N. and Nicolich,  K. L. and Hill,  J. C. and Marlow,  B. L. Schmittberger and Cox,  K. C.},
  year = {2024},
  pages = {144002},
  month = sep 
}

@article{jau2020vapor,
  title = {Vapor-Cell-Based Atomic Electrometry for Detection Frequencies below 1 kHz},
  volume = {13},
  ISSN = {2331-7019},
  url = {http://dx.doi.org/10.1103/PhysRevApplied.13.054034},
  DOI = {10.1103/physrevapplied.13.054034},
  number = {5},
  journal = {Physical Review Applied},
  publisher = {American Physical Society (APS)},
  author = {Jau,  Yuan-Yu and Carter,  Tony},
  year = {2020},
  pages={054034},
  month = may 
}

@article{wang2023noise,
  title = {Noise analysis of the atomic superheterodyne receiver based on flat-top laser beams},
  volume = {31},
  ISSN = {1094-4087},
  url = {http://dx.doi.org/10.1364/OE.491718},
  DOI = {10.1364/oe.491718},
  number = {12},
  journal = {Optics Express},
  publisher = {Optica Publishing Group},
  author = {Wang,  Zheng and Jing,  Mingyong and Zhang,  Peng and Yuan,  Shaoxin and Zhang,  Hao and Zhang,  Linjie and Xiao,  Liantuan and Jia,  Suotang},
  year = {2023},
  month = may,
  pages = {19909--19917}
}

@article{mohapatra2007coherent,
  title = {Coherent Optical Detection of Highly Excited Rydberg States Using Electromagnetically Induced Transparency},
  volume = {98},
  ISSN = {1079-7114},
  url = {http://dx.doi.org/10.1103/PhysRevLett.98.113003},
  DOI = {10.1103/physrevlett.98.113003},
  number = {11},
  journal = {Physical Review Letters},
  publisher = {American Physical Society (APS)},
  author = {Mohapatra,  A. K. and Jackson,  T. R. and Adams,  C. S.},
  year = {2007},
  pages={113003},
  month = mar 
}

@article{schlossberger2025thermometry,
  title = {Primary quantum thermometry of mm-wave blackbody radiation via induced state transfer in Rydberg states of cold atoms},
  volume = {7},
  ISSN = {2643-1564},
  url = {http://dx.doi.org/10.1103/PhysRevResearch.7.L012020},
  DOI = {10.1103/physrevresearch.7.l012020},
  number = {1},
  journal = {Physical Review Research},
  publisher = {American Physical Society (APS)},
  author = {Schlossberger,  Noah and Rotunno,  Andrew P. and Eckel,  Stephen P. and Norrgard,  Eric B. and Manchaiah,  Dixith and Prajapati,  Nikunjkumar and Artusio-Glimpse,  Alexandra B. and Berweger,  Samuel and Simons,  Matthew T. and Shylla,  Dangka and Watterson,  William J. and Patrick,  Charles and Meraki,  Adil and Talashila,  Rajavardhan and Younes,  Amanda and La Mantia,  David S. and Holloway,  Christopher L.},
  year = {2025},
  pages={L012020},
  month = jan 
}

@article{lamantia2025sensor,
  title = {Compact blackbody-radiation atomic sensor: Measuring temperature using optically excited atoms in vapor cells},
  volume = {23},
  ISSN = {2331-7019},
  url = {http://dx.doi.org/10.1103/PhysRevApplied.23.044037},
  DOI = {10.1103/physrevapplied.23.044037},
  number = {4},
  journal = {Physical Review Applied},
  publisher = {American Physical Society (APS)},
  author = {La Mantia,  David S. and Lei,  Mingxin and Prajapati,  Nikunjkumar and Schlossberger,  Noah and Simons,  Matthew T. and Holloway,  Christopher L. and Scherschligt,  Julia and Eckel,  Stephen P. and Norrgard,  Eric B.},
  year = {2025},
  pages={044037},
  month = apr 
}

@article{zhou2023improving,
  title = {Improving the spectral resolution and measurement range of quantum microwave electrometry by cold Rydberg atoms},
  volume = {56},
  ISSN = {1361-6455},
  url = {http://dx.doi.org/10.1088/1361-6455/acae4f},
  DOI = {10.1088/1361-6455/acae4f},
  number = {2},
  journal = {Journal of Physics B: Atomic,  Molecular and Optical Physics},
  publisher = {IOP Publishing},
  author = {Zhou,  Fei and Jia,  Fengdong and Liu,  Xiubin and Yu,  Yonghong and Mei,  Jiong and Zhang,  Jian and Xie,  Feng and Zhong,  Zhiping},
  year = {2023},
  month = jan,
  pages = {025501}
}

@article{zhou2022effect,
  title = {The effect of the Doppler mismatch in microwave electrometry using Rydberg electromagnetically induced transparency and Autler–Townes splitting},
  volume = {55},
  ISSN = {1361-6455},
  url = {http://dx.doi.org/10.1088/1361-6455/ac5d8d},
  DOI = {10.1088/1361-6455/ac5d8d},
  number = {7},
  journal = {Journal of Physics B: Atomic,  Molecular and Optical Physics},
  publisher = {IOP Publishing},
  author = {Zhou,  Fei and Jia,  Feng-Dong and Mei,  Jiong and Liu,  Xiu-Bin and Zhang,  Huai-Yu and Yu,  Yong-Hong and Liang,  Wei-Chen and Qin,  Jian-Wei and Zhang,  Jian and Xie,  Feng and Zhong,  Zhi-Ping},
  year = {2022},
  month = apr,
  pages = {075501}
}

@article{jamieson2025uhf,
  title = {Continuous-time ultrahigh-frequency sensing using cold Rydberg atoms},
  volume = {24},
  ISSN = {2331-7019},
  url = {http://dx.doi.org/10.1103/2vxb-czvz},
  DOI = {10.1103/2vxb-czvz},
  number = {3},
  journal = {Physical Review Applied},
  publisher = {American Physical Society (APS)},
  author = {Jamieson,  Matt J. and Adams,  C. Stuart and Weatherill,  Kevin J. and Hanley,  Ryan K. and Alves,  Natalia and Keaveney,  James},
  year = {2025},
  month = sep,
  pages = {034022}
}

@article{duverger2024metrology,
  title = {Metrology of microwave fields based on trap-loss spectroscopy with cold Rydberg atoms},
  volume = {22},
  ISSN = {2331-7019},
  url = {http://dx.doi.org/10.1103/PhysRevApplied.22.044039},
  DOI = {10.1103/physrevapplied.22.044039},
  number = {4},
  journal = {Physical Review Applied},
  publisher = {American Physical Society (APS)},
  author = {Duverger,  Romain and Bonnin,  Alexis and Granier,  Romain and Marolleau,  Quentin and Blanchard,  Cédric and Zahzam,  Nassim and Bidel,  Yannick and Cadoret,  Malo and Bresson,  Alexandre and Schwartz,  Sylvain},
  year = {2024},
  pages = {044039},
  month = oct 
}

@article{liao2020electrometry,
  title = {Microwave electrometry via electromagnetically induced absorption in cold Rydberg atoms},
  volume = {101},
  ISSN = {2469-9934},
  url = {http://dx.doi.org/10.1103/PhysRevA.101.053432},
  DOI = {10.1103/physreva.101.053432},
  number = {5},
  journal = {Physical Review A},
  publisher = {American Physical Society (APS)},
  author = {Liao,  Kai-Yu and Tu,  Hai-Tao and Yang,  Shu-Zhe and Chen,  Chang-Jun and Liu,  Xiao-Hong and Liang,  Jie and Zhang,  Xin-Ding and Yan,  Hui and Zhu,  Shi-Liang},
  year = {2020},
  pages = {053432},
  month = may 
}

@article{tu2024approaching,
  title = {Approaching the standard quantum limit of a Rydberg-atom microwave electrometer},
  volume = {10},
  ISSN = {2375-2548},
  url = {http://dx.doi.org/10.1126/sciadv.ads0683},
  DOI = {10.1126/sciadv.ads0683},
  number = {51},
  journal = {Science Advances},
  publisher = {American Association for the Advancement of Science (AAAS)},
  author = {Tu,  Hai-Tao and Liao,  Kai-Yu and Wang,  Hong-Lei and Zhu,  Yi-Fei and Qiu,  Si-Yuan and Jiang,  Hao and Huang,  Wei and Bian,  Wu and Yan,  Hui and Zhu,  Shi-Liang},
  year = {2024},
  pages = {eads0683},
  month = dec 
}

@misc{santamaria2022comparison,
  doi = {10.48550/ARXIV.2209.00908},
  author = {Santamaria-Botello,  Gabriel and Verploegh,  Shane and Bottomley,  Eric and Popovic,  Zoya},
  title = {Comparison of Noise Temperature of Rydberg-Atom and Electronic Microwave Receivers},
  publisher = {arXiv preprint},
  year = {2022},
  copyright = {arXiv.org perpetual,  non-exclusive license}
}

@misc{yan2025bandwidth,
  doi = {10.48550/ARXIV.2506.10541},
  author = {Yan,  Yuhan and Yang,  Bowen and Li,  Xuejie and Zhao,  Haojie and Yu,  Binghong and Deng,  Jianliao and Chen,  L. Q. and Cheng,  Huadong},
  title = {Multi-Dress-State Engineered Rydberg Electrometry: Achieving 100-MHz-level Instantaneous-Bandwidth},
  publisher = {arXiv preprint},
  year = {2025},
  copyright = {arXiv.org perpetual,  non-exclusive license}
}

@article{sedlacek2012microwave,
  title = {Microwave electrometry with Rydberg atoms in a vapour cell using bright atomic resonances},
  volume = {8},
  ISSN = {1745-2481},
  url = {http://dx.doi.org/10.1038/nphys2423},
  DOI = {10.1038/nphys2423},
  number = {11},
  journal = {Nature Physics},
  publisher = {Springer Science and Business Media LLC},
  author = {Sedlacek,  Jonathon A. and Schwettmann,  Arne and K\"{u}bler,  Harald and L\"{o}w,  Robert and Pfau,  Tilman and Shaffer,  James P.},
  year = {2012},
  month = sep,
  pages = {819–824}
}

@article{legaie2024mmwave,
  title = {A millimeter-wave atomic receiver},
  volume = {6},
  ISSN = {2639-0213},
  url = {http://dx.doi.org/10.1116/5.0173654},
  DOI = {10.1116/5.0173654},
  number = {2},
  journal = {AVS Quantum Science},
  publisher = {American Vacuum Society},
  author = {Legaie,  Remy and Raithel,  Georg and Anderson,  David A.},
  year = {2024},
  pages   = {025701},
  month = apr 
}

@article{cai2023sensitivity,
  title = {Sensitivity extension of atom-based amplitude-modulation microwave electrometry via high Rydberg states},
  volume = {122},
  ISSN = {1077-3118},
  url = {http://dx.doi.org/10.1063/5.0146768},
  DOI = {10.1063/5.0146768},
  number = {16},
  journal = {Applied Physics Letters},
  publisher = {AIP Publishing},
  author = {Cai,  Minghao and You,  Shuhang and Zhang,  Shanshan and Xu,  Zishan and Liu,  Hongping},
  year = {2023},
  pages = {161103},
  month = apr 
}

@article{yang2024noise,
  title = {Sensitivity of Rydberg microwave electrometry limited by laser frequency noise},
  volume = {109},
  ISSN = {2469-9934},
  url = {http://dx.doi.org/10.1103/PhysRevA.109.032609},
  DOI = {10.1103/physreva.109.032609},
  number = {3},
  journal = {Physical Review A},
  publisher = {American Physical Society (APS)},
  author = {Yang,  Bowen and Yan,  Yuhan and Li,  Xuejie and Zhao,  Haojie and Xiao,  Ling and Li,  Xiaolin and Deng,  Jianliao and Cheng,  Huadong},
  year = {2024},
  pages = {032609},
  month = mar 
}

@article{yang2025local,
  title = {Local Oscillator Port-Integrated Resonators for Sensitivity Enhancement of VHF Band Rydberg Atomic Heterodyne Receivers},
  volume = {73},
  ISSN = {1557-9670},
  url = {http://dx.doi.org/10.1109/TMTT.2025.3528720},
  DOI = {10.1109/tmtt.2025.3528720},
  number = {8},
  journal = {IEEE Transactions on Microwave Theory and Techniques},
  publisher = {Institute of Electrical and Electronics Engineers (IEEE)},
  author = {Yang,  Kai and Lin,  Yi and Ding,  Zhenke and An,  Qiang and Sun,  Zhanshan and Fu,  Yunqi},
  year = {2025},
  month = aug,
  pages = {5061–5070}
}

@article{yang2024highly,
  title = {Highly sensitive microwave electrometry with enhanced instantaneous bandwidth},
  volume = {21},
  ISSN = {2331-7019},
  url = {http://dx.doi.org/10.1103/PhysRevApplied.21.L031003},
  DOI = {10.1103/physrevapplied.21.l031003},
  number = {3},
  journal = {Physical Review Applied},
  publisher = {American Physical Society (APS)},
  author = {Yang,  Bowen and Yan,  Yuhan and Li,  Xuejie and Xiao,  Ling and Li,  Xiaolin and Chen,  L.Q. and Deng,  Jianliao and Cheng,  Huadong},
  year = {2024},
  pages={L031003},
  month = mar 
}

@article{aoki2016random,
  title = {Observing random walks of atoms in buffer gas through resonant light absorption},
  volume = {94},
  ISSN = {2469-9934},
  url = {http://dx.doi.org/10.1103/PhysRevA.94.012703},
  DOI = {10.1103/physreva.94.012703},
  number = {1},
  journal = {Physical Review A},
  publisher = {American Physical Society (APS)},
  author = {Aoki,  Kenichiro and Mitsui,  Takahisa},
  year = {2016},
  pages={012703},
  month = jul 
}

@article{micalizio2020brownian,
  title = {Brownian motion-induced amplitude noise in vapor-cell frequency standards},
  volume = {22},
  ISSN = {1367-2630},
  url = {http://dx.doi.org/10.1088/1367-2630/aba464},
  DOI = {10.1088/1367-2630/aba464},
  number = {8},
  journal = {New Journal of Physics},
  publisher = {IOP Publishing},
  author = {Micalizio,  S and Godone,  A and Gozzelino,  M and Levi,  F},
  year = {2020},
  month = aug,
  pages = {083050}
}

@article{mao2024resonator,
  title = {Shortwave Ultrahigh-Sensitivity Rydberg Atomic Electric Field Sensing Based on a Subminiature Resonator},
  volume = {72},
  ISSN = {1558-2221},
  url = {http://dx.doi.org/10.1109/TAP.2024.3439862},
  DOI = {10.1109/tap.2024.3439862},
  number = {11},
  journal = {IEEE Transactions on Antennas and Propagation},
  publisher = {Institute of Electrical and Electronics Engineers (IEEE)},
  author = {Mao,  Ruiqi and Lin,  Yi and Zhou,  Aojie and Yang,  Kai and Fu,  Yunqi},
  year = {2024},
  month = nov,
  pages = {8165–8172}
}

@article{prajapati2021enhancement,
  title = {Enhancement of electromagnetically induced transparency based Rydberg-atom electrometry through population repumping},
  volume = {119},
  ISSN = {1077-3118},
  url = {http://dx.doi.org/10.1063/5.0069195},
  DOI = {10.1063/5.0069195},
  number = {21},
  journal = {Applied Physics Letters},
  publisher = {AIP Publishing},
  author = {Prajapati,  Nikunjkumar and Robinson,  Amy K. and Berweger,  Samuel and Simons,  Matthew T. and Artusio-Glimpse,  Alexandra B. and Holloway,  Christopher L.},
  year = {2021},
  pages   = {214001},
  month = nov 
}

@article{liu2022highly,
  title = {Highly Sensitive Measurement of a Megahertz rf Electric Field with a Rydberg-Atom Sensor},
  volume = {18},
  ISSN = {2331-7019},
  url = {http://dx.doi.org/10.1103/PhysRevApplied.18.014045},
  DOI = {10.1103/physrevapplied.18.014045},
  number = {1},
  journal = {Physical Review Applied},
  publisher = {American Physical Society (APS)},
  author = {Liu,  Bang and Zhang,  Li-Hua and Liu,  Zong-Kai and Zhang,  Zheng-Yuan and Zhu,  Zhi-Han and Gao,  Wei and Guo,  Guang-Can and Ding,  Dong-Sheng and Shi,  Bao-Sen},
  year = {2022},
  pages={014045},
  month = jul 
}

@article{bohaichuk2022transient,
  title = {Origins of Rydberg-Atom Electrometer Transient Response and Its Impact on Radio-Frequency Pulse Sensing},
  volume = {18},
  ISSN = {2331-7019},
  url = {http://dx.doi.org/10.1103/PhysRevApplied.18.034030},
  DOI = {10.1103/physrevapplied.18.034030},
  number = {3},
  journal = {Physical Review Applied},
  publisher = {American Physical Society (APS)},
  author = {Bohaichuk,  Stephanie M. and Booth,  Donald and Nickerson,  Kent and Tai,  Harry and Shaffer,  James P.},
  year = {2022},
  pages={034030},
  month = sep 
}

@misc{glick2025warm,
  doi = {10.48550/ARXIV.2506.04504},
  author = {Glick,  Jeremy and Anderson,  Brielle E. and Nunley,  T. Nathan and Bingaman,  Josiah and Liu,  Jian Jun and Meyer,  David H. and Kunz,  Paul D.},
  title = {Warm Vapor Rydberg EIT spectra in Doppler-free configurations},
  publisher = {arXiv preprint},
  year = {2025},
  copyright = {arXiv.org perpetual,  non-exclusive license}
}

@article{kitching2011atomic,
  title = {Atomic Sensors – A Review},
  volume = {11},
  ISSN = {1558-1748},
  url = {http://dx.doi.org/10.1109/JSEN.2011.2157679},
  DOI = {10.1109/jsen.2011.2157679},
  number = {9},
  journal = {IEEE Sensors Journal},
  publisher = {Institute of Electrical and Electronics Engineers (IEEE)},
  author = {Kitching,  John and Knappe,  Svenja and Donley,  Elizabeth A.},
  year = {2011},
  month = sep,
  pages = {1749–1758}
}

@inproceedings{venu2025three,
  title = {Three-photon Rydberg atom electrometry with enhanced sensitivity},
  DOI = {10.1117/12.3053172},
  booktitle = {Quantum Sensing,  Imaging,  and Precision Metrology III},
  publisher = {SPIE},
  author = {Venu,  Vijin and Bohaichuk,  Stephanie and Christaller,  Florian and Schmidt,  Matthias and Kubler,  Harald and Shaffer,  James P.},
  editor = {Shahriar,  Selim M.},
  year = {2025},
  volume={13392},
  pages={255--260},
  month = mar,
  address = {San Francisco, CA, USA}
}

@article{bohaichuk2023three,
  title = {Three-photon Rydberg-atom-based radio-frequency sensing scheme with narrow linewidth},
  volume = {20},
  ISSN = {2331-7019},
  url = {http://dx.doi.org/10.1103/PhysRevApplied.20.L061004},
  DOI = {10.1103/physrevapplied.20.l061004},
  number = {6},
  journal = {Physical Review Applied},
  publisher = {American Physical Society (APS)},
  author = {Bohaichuk,  Stephanie M. and Ripka,  Fabian and Venu,  Vijin and Christaller,  Florian and Liu,  Chang and Schmidt,  Matthias and K\"{u}bler,  Harald and Shaffer,  James P.},
  year = {2023},
  pages={L061004},
  month = dec 
}

@article{fan2015cell,
  title = {Effect of Vapor-Cell Geometry on Rydberg-Atom-Based Measurements of Radio-Frequency Electric Fields},
  volume = {4},
  ISSN = {2331-7019},
  url = {http://dx.doi.org/10.1103/PhysRevApplied.4.044015},
  DOI = {10.1103/physrevapplied.4.044015},
  number = {4},
  journal = {Physical Review Applied},
  publisher = {American Physical Society (APS)},
  author = {Fan,  Haoquan and Kumar,  Santosh and Sheng,  Jiteng and Shaffer,  James P. and Holloway,  Christopher L. and Gordon,  Joshua A.},
  year = {2015},
  pages={044015},
  month = oct 
}

@article{weichman2025resonant,
  title = {Resonant optical bistability in support of enhanced Rydberg-atom sensors},
  volume = {112},
  ISSN = {2469-9934},
  url = {http://dx.doi.org/10.1103/k2n6-1xm3},
  DOI = {10.1103/k2n6-1xm3},
  number = {4},
  journal = {Physical Review A},
  publisher = {American Physical Society (APS)},
  author = {Weichman,  Peter B.},
  year = {2025},
  month = oct,
  pages = {042805}
}

@BOOK{hulst1981light,
  title     = "Light scattering by small particles",
  author    = "van de Hulst, H C",
  publisher = "Dover Publications",
  series    = "Dover Books on Physics",
  month     =  dec,
  year      =  1981,
  address   = "Mineola, NY",
  language  = "en"
}

@book{westphal1972dielectric,
  title={Dielectric constant and loss data},
  author={Westphal, William B and Sils, Aina},
  year={1972},
  publisher={Air Force Materials Laboratory, Air Force Systems Command},
  address = {Wright-Patterson Air Force Base, Ohio}
}

@article{strom1977temperature,
  title = {Temperature and frequency dependences of the far-infrared and microwave optical absorption in amorphous materials},
  volume = {16},
  ISSN = {0556-2805},
  url = {http://dx.doi.org/10.1103/PhysRevB.16.5512},
  DOI = {10.1103/physrevb.16.5512},
  number = {12},
  journal = {Physical Review B},
  publisher = {American Physical Society (APS)},
  author = {Strom,  U. and Taylor,  P. C.},
  year = {1977},
  month = dec,
  pages = {5512–5522}
}

@article{carr2013nonequilibrium,
  title = {Nonequilibrium Phase Transition in a Dilute Rydberg Ensemble},
  volume = {111},
  ISSN = {1079-7114},
  url = {http://dx.doi.org/10.1103/PhysRevLett.111.113901},
  DOI = {10.1103/physrevlett.111.113901},
  number = {11},
  journal = {Physical Review Letters},
  publisher = {American Physical Society (APS)},
  author = {Carr,  C. and Ritter,  R. and Wade,  C. G. and Adams,  C. S. and Weatherill,  K. J.},
  year = {2013},
  pages={113901},
  month = sep 
}

@article{elgee2023satellite,
  title = {Satellite radio detection via dual-microwave Rydberg spectroscopy},
  volume = {123},
  ISSN = {1077-3118},
  url = {http://dx.doi.org/10.1063/5.0158150},
  DOI = {10.1063/5.0158150},
  number = {8},
  journal = {Applied Physics Letters},
  publisher = {AIP Publishing},
  author = {Elgee,  Peter K. and Hill,  Joshua C. and LeBlanc,  Kermit-James E. and Ko,  Gabriel D. and Kunz,  Paul D. and Meyer,  David H. and Cox,  Kevin C.},
  year = {2023},
  pages = {084001},
  month = aug 
}

@article{holloway2014broadband,
  title = {Broadband Rydberg Atom-Based Electric-Field Probe for SI-Traceable,  Self-Calibrated Measurements},
  volume = {62},
  ISSN = {1558-2221},
  url = {http://dx.doi.org/10.1109/TAP.2014.2360208},
  DOI = {10.1109/tap.2014.2360208},
  number = {12},
  journal = {IEEE Transactions on Antennas and Propagation},
  publisher = {Institute of Electrical and Electronics Engineers (IEEE)},
  author = {Holloway,  Christopher L. and Gordon,  Joshua A. and Jefferts,  Steven and Schwarzkopf,  Andrew and Anderson,  David A. and Miller,  Stephanie A. and Thaicharoen,  Nithiwadee and Raithel,  Georg},
  year = {2014},
  month = dec,
  pages = {6169–6182}
}

@article{sandidge2024structures,
  title = {Resonant Structures for Sensitivity Enhancement of Rydberg-Atom Microwave Receivers},
  volume = {72},
  ISSN = {1557-9670},
  url = {http://dx.doi.org/10.1109/TMTT.2024.3355763},
  DOI = {10.1109/tmtt.2024.3355763},
  number = {4},
  journal = {IEEE Transactions on Microwave Theory and Techniques},
  publisher = {Institute of Electrical and Electronics Engineers (IEEE)},
  author = {Sandidge,  Georgia and Santamaria-Botello,  Gabriel and Bottomley,  Eric and Fan,  Haoquan and Popović,  Zoya},
  year = {2024},
  month = apr,
  pages = {2057–2066}
}

@InProceedings{sandidge2025impact,
  author    = {Sandidge, Georgia and Popovic, Zoya},
  booktitle = {2025 IEEE Radio and Wireless Symposium (RWS)},
  title     = {Impact of Rb Condensation on Resonant Structures for Microwave Rydberg-Atom Electrometers},
  year      = {2025},
  address   = {San Juan, PR, USA},
  month     = jan,
  pages     = {45–47},
  publisher = {IEEE},
  doi       = {10.1109/rws62086.2025.10904838},
}

@article{ma2025study,
  title = {Study on electric field shielding in SiO2 and CaF2 vapor cell for Rydberg atom electric field sensors},
  volume = {58},
  ISSN = {1361-6463},
  url = {http://dx.doi.org/10.1088/1361-6463/adb04c},
  DOI = {10.1088/1361-6463/adb04c},
  number = {13},
  journal = {Journal of Physics D: Applied Physics},
  publisher = {IOP Publishing},
  author = {Ma,  Kai and Xiao,  Dongping and Zhang,  Huaiqing and Wang,  Xin and Wei,  Xutao},
  year = {2025},
  month = feb,
  pages = {135113}
}

@article{ma2020dc,
  title = {DC electric fields in electrode-free glass vapor cell by photoillumination},
  volume = {28},
  ISSN = {1094-4087},
  url = {http://dx.doi.org/10.1364/OE.380748},
  DOI = {10.1364/oe.380748},
  number = {3},
  journal = {Optics Express},
  publisher = {Optica Publishing Group},
  author = {Ma,  L. and Paradis,  E. and Raithel,  G.},
  year = {2020},
  month = jan,
  pages = {3676--3685}
}

@article{prajapati2023comparison,
  title = {Sensitivity comparison of two-photon vs three-photon Rydberg electrometry},
  volume = {134},
  ISSN = {1089-7550},
  url = {http://dx.doi.org/10.1063/5.0147827},
  DOI = {10.1063/5.0147827},
  number = {2},
  journal = {Journal of Applied Physics},
  publisher = {AIP Publishing},
  author = {Prajapati,  Nikunjkumar and Bhusal,  Narayan and Rotunno,  Andrew P. and Berweger,  Samuel and Simons,  Matthew T. and Artusio-Glimpse,  Alexandra B. and Ju Wang,  Ying and Bottomley,  Eric and Fan,  Haoquan and Holloway,  Christopher L.},
  year = {2023},
  pages = {023101},
  month = jul 
}

@article{prajapati2024fluorescence,
  title = {Investigation of fluorescence versus transmission readout for three-photon Rydberg excitation used in electrometry},
  volume = {6},
  ISSN = {2639-0213},
  url = {http://dx.doi.org/10.1116/5.0201928},
  DOI = {10.1116/5.0201928},
  number = {3},
  journal = {AVS Quantum Science},
  publisher = {American Vacuum Society},
  author = {Prajapati,  Nikunjkumar and Berweger,  Samuel and Rotunno,  Andrew P. and Artusio-Glimpse,  Alexandra B. and Schlossberger,  Noah and Shylla,  Dangka and Watterson,  William J. and Simons,  Matthew T. and LaMantia,  David and Norrgard,  Eric B. and Eckel,  Stephen P. and Holloway,  Christopher L.},
  year = {2024},
  pages = {034401},
  month = jul 
}

@article{prajapati2022video,
  title = {TV and video game streaming with a quantum receiver: A study on a Rydberg atom-based receiver’s bandwidth and reception clarity},
  volume = {4},
  ISSN = {2639-0213},
  url = {http://dx.doi.org/10.1116/5.0098057},
  DOI = {10.1116/5.0098057},
  number = {3},
  journal = {AVS Quantum Science},
  publisher = {American Vacuum Society},
  author = {Prajapati,  Nikunjkumar and Rotunno,  Andrew P. and Berweger,  Samuel and Simons,  Matthew T. and Artusio-Glimpse,  Alexandra B. and Voran,  Stephen D. and Holloway,  Christopher L.},
  year = {2022},
  pages = {035001},
  month = aug 
}

@article{downes2020imaging,
  title = {Full-Field Terahertz Imaging at Kilohertz Frame Rates Using Atomic Vapor},
  volume = {10},
  ISSN = {2160-3308},
  url = {http://dx.doi.org/10.1103/PhysRevX.10.011027},
  DOI = {10.1103/physrevx.10.011027},
  number = {1},
  journal = {Physical Review X},
  publisher = {American Physical Society (APS)},
  author = {Downes,  Lucy A. and MacKellar,  Andrew R. and Whiting,  Daniel J. and Bourgenot,  Cyril and Adams,  Charles S. and Weatherill,  Kevin J.},
  year = {2020},
  pages     = {011027},
  month = feb 
}

@article{chen2022terahertz,
  title = {Terahertz electrometry via infrared spectroscopy of atomic vapor},
  volume = {9},
  ISSN = {2334-2536},
  url = {http://dx.doi.org/10.1364/OPTICA.456761},
  DOI = {10.1364/optica.456761},
  number = {5},
  journal = {Optica},
  publisher = {Optica Publishing Group},
  author = {Chen,  Shuying and Reed,  Dominic J. and MacKellar,  Andrew R. and Downes,  Lucy A. and Almuhawish,  Nourah F. A. and Jamieson,  Matthew J. and Adams,  Charles S. and Weatherill,  Kevin J.},
  year = {2022},
  month = apr,
  pages = {485--491}
}

@inproceedings{prajapati2024rydberg,
  title = {Rydberg Atom Electrometry: Recent Sensitivity and Bandwidth Improvements},
  DOI = {10.1109/emceurope59828.2024.10722127},
  booktitle = {2024 International Symposium on Electromagnetic Compatibility – EMC Europe},
  publisher = {IEEE},
  author = {Prajapati,  Nikunjkumar and Artusio-Glimpse,  Alexandra and Berweger,  Samuel and Simons,  Matthew T. and Schlossberger,  Noah and Shylla,  Dangka and Watterson,  William and Manchaiah,  Dixith and Holloway,  Christopher L.},
  year = {2024},
  month = sep,
  pages = {323–328},
  address = {Brugge, Belgium}
}

@article{allinson2024multiband,
  title = {Simultaneous multiband radio-frequency detection using high-orbital-angular-momentum states in a Rydberg-atom receiver},
  volume = {6},
  ISSN = {2643-1564},
  url = {http://dx.doi.org/10.1103/PhysRevResearch.6.023317},
  DOI = {10.1103/physrevresearch.6.023317},
  number = {2},
  journal = {Physical Review Research},
  publisher = {American Physical Society (APS)},
  author = {Allinson,  Gianluca and Jamieson,  Matthew J. and Mackellar,  Andrew R. and Downes,  Lucy and Adams,  C. Stuart and Weatherill,  Kevin J.},
  year = {2024},
  pages     = {023317},
  month = jun 
}

@article{nowosielski2024qam,
  title = {Warm Rydberg atom-based quadrature amplitude-modulated receiver},
  volume = {32},
  ISSN = {1094-4087},
  url = {http://dx.doi.org/10.1364/OE.529977},
  DOI = {10.1364/oe.529977},
  number = {17},
  journal = {Optics Express},
  publisher = {Optica Publishing Group},
  author = {Nowosielski,  Jan and Jastrzębski,  Marcin and Halavach,  Pavel and Łukanowski,  Karol and Jarzyna,  Marcin and Mazelanik,  Mateusz and Wasilewski,  Wojciech and Parniak,  Michał},
  year = {2024},
  month = aug,
  pages = {30027--30039}
}

@article{borowka2024conversion,
  title = {Continuous wideband microwave-to-optical converter based on room-temperature Rydberg atoms},
  volume = {18},
  ISSN = {1749-4893},
  url = {http://dx.doi.org/10.1038/s41566-023-01295-w},
  DOI = {10.1038/s41566-023-01295-w},
  number = {1},
  journal = {Nature Photonics},
  publisher = {Springer Science and Business Media LLC},
  author = {Borówka,  Sebastian and Pylypenko,  Uliana and Mazelanik,  Mateusz and Parniak,  Michał},
  year = {2023},
  month = oct,
  pages = {32–38}
}

@article{amarloo2025photonic,
  title = {A photonic crystal receiver for Rydberg atom-based sensing},
  volume = {4},
  ISSN = {2731-3395},
  url = {http://dx.doi.org/10.1038/s44172-025-00408-3},
  DOI = {10.1038/s44172-025-00408-3},
  number = {1},
  journal = {Communications Engineering},
  publisher = {Springer Science and Business Media LLC},
  author = {Amarloo,  Hadi and Noaman,  Mohammad and Yu,  Su-Peng and Booth,  Donald and Mirzaee,  Somayeh and Pandiyan,  Rajesh and Christaller,  Florian and Shaffer,  James P.},
  year = {2025},
  pages={70},
  month = apr 
}

@article{simons2018fiber,
  title = {Fiber-coupled vapor cell for a portable Rydberg atom-based radio frequency electric field sensor},
  volume = {57},
  ISSN = {2155-3165},
  url = {http://dx.doi.org/10.1364/AO.57.006456},
  DOI = {10.1364/ao.57.006456},
  number = {22},
  journal = {Applied Optics},
  publisher = {Optica Publishing Group},
  author = {Simons,  Matt T. and Gordon,  Joshua A. and Holloway,  Christopher L.},
  year = {2018},
  month = jul,
  pages = {6456--6460}
}

@article{anderson2021self,
  title = {A Self-Calibrated SI-Traceable Rydberg Atom-Based Radio Frequency Electric Field Probe and Measurement Instrument},
  volume = {69},
  ISSN = {1558-2221},
  url = {http://dx.doi.org/10.1109/TAP.2021.3060540},
  DOI = {10.1109/tap.2021.3060540},
  number = {9},
  journal = {IEEE Transactions on Antennas and Propagation},
  publisher = {Institute of Electrical and Electronics Engineers (IEEE)},
  author = {Anderson,  David Alexander and Sapiro,  Rachel Elizabeth and Raithel,  Georg},
  year = {2021},
  month = sep,
  pages = {5931–5941}
}

@article{simons2019rydberg,
  title = {A Rydberg atom-based mixer: Measuring the phase of a radio frequency wave},
  volume = {114},
  ISSN = {1077-3118},
  url = {http://dx.doi.org/10.1063/1.5088821},
  DOI = {10.1063/1.5088821},
  number = {11},
  journal = {Applied Physics Letters},
  publisher = {AIP Publishing},
  author = {Simons,  Matthew T. and Haddab,  Abdulaziz H. and Gordon,  Joshua A. and Holloway,  Christopher L.},
  year = {2019},
  pages = {114101},
  month = mar 
}

@article{holguinlerma2020laser,
  title = {480-nm distributed-feedback InGaN laser diode for 10.5-Gbit/s visible-light communication},
  volume = {45},
  ISSN = {1539-4794},
  url = {http://dx.doi.org/10.1364/OL.385954},
  DOI = {10.1364/ol.385954},
  number = {3},
  journal = {Optics Letters},
  publisher = {Optica Publishing Group},
  author = {Holguin-Lerma,  Jorge A. and Kong,  Meiwei and Alkhazragi,  Omar and Sun,  Xiaobin and Khee Ng,  Tien and Ooi,  Boon S.},
  year = {2020},
  month = jan,
  pages = {742--745}
}

@article{trageser2024blue,
  title = {Blue GaN-based DFB laser diode with sub-MHz linewidth},
  volume = {32},
  ISSN = {1094-4087},
  url = {http://dx.doi.org/10.1364/OE.525498},
  DOI = {10.1364/oe.525498},
  number = {13},
  journal = {Optics Express},
  publisher = {Optica Publishing Group},
  author = {Trageser,  Emily and Zhang,  Haojun and Palmer,  Sonya and Morin,  Theodore and Guo,  Joel and Zhang,  Jiaao and Geske,  Evan and Wang,  Heming and Boes,  Andreas and Nakamura,  Shuji and Bowers,  John E. and DenBaars,  Steven P.},
  year = {2024},
  month = jun,
  pages = {23372--23380}
}

@article{jia2024squeezing,
  title = {Squeezing the quantum noise of a gravitational-wave detector below the standard quantum limit},
  volume = {385},
  ISSN = {1095-9203},
  url = {http://dx.doi.org/10.1126/science.ado8069},
  DOI = {10.1126/science.ado8069},
  number = {6715},
  journal = {Science},
  publisher = {American Association for the Advancement of Science (AAAS)},
  author = {Jia,  Wenxuan and Xu,  Victoria and Kuns,  Kevin and others},
  year = {2024},
  month = sep,
  pages = {1318–1321}
}

@article{brown2023very,
  title = {Very-high- and ultrahigh-frequency electric-field detection using high angular momentum Rydberg states},
  volume = {107},
  ISSN = {2469-9934},
  url = {http://dx.doi.org/10.1103/PhysRevA.107.052605},
  DOI = {10.1103/physreva.107.052605},
  number = {5},
  journal = {Physical Review A},
  publisher = {American Physical Society (APS)},
  author = {Brown,  Roger C. and Kayim,  Baran and Viray,  Michael A. and Perry,  Abigail R. and Sawyer,  Brian C. and Wyllie,  Robert},
  year = {2023},
  pages={052605},
  month = may 
}

@misc{zhu2025general,
  doi = {10.48550/ARXIV.2506.23455},
  author = {Zhu,  Jieao and Dai,  Linglong},
  title = {General Signal Model and Capacity Limit for Rydberg Quantum Information System},
  publisher = {arXiv preprint},
  year = {2025},
  copyright = {Creative Commons Zero v1.0 Universal}
}

@article{hu2022continuously,
  title = {Continuously tunable radio frequency electrometry with Rydberg atoms},
  volume = {121},
  ISSN = {1077-3118},
  url = {http://dx.doi.org/10.1063/5.0086357},
  DOI = {10.1063/5.0086357},
  number = {1},
  journal = {Applied Physics Letters},
  publisher = {AIP Publishing},
  author = {Hu,  Jinlian and Li,  Huaqiang and Song,  Rong and Bai,  Jingxu and Jiao,  Yuechun and Zhao,  Jianming and Jia,  Suotang},
  year = {2022},
  pages = {014002},
  month = jul 
}

@article{yuan2023quantum,
  title = {Quantum sensing of microwave electric fields based on Rydberg atoms},
  volume = {86},
  ISSN = {1361-6633},
  url = {http://dx.doi.org/10.1088/1361-6633/acf22f},
  DOI = {10.1088/1361-6633/acf22f},
  number = {10},
  journal = {Reports on Progress in Physics},
  publisher = {IOP Publishing},
  author = {Yuan,  Jinpeng and Yang,  Wenguang and Jing,  Mingyong and Zhang,  Hao and Jiao,  Yuechun and Li,  Weibin and Zhang,  Linjie and Xiao,  Liantuan and Jia,  Suotang},
  year = {2023},
  month = sep,
  pages = {106001}
}

@article{downes2023practical,
  title = {A practical guide to terahertz imaging using thermal atomic vapour},
  volume = {25},
  ISSN = {1367-2630},
  url = {http://dx.doi.org/10.1088/1367-2630/acb80c},
  DOI = {10.1088/1367-2630/acb80c},
  number = {3},
  journal = {New Journal of Physics},
  publisher = {IOP Publishing},
  author = {Downes,  Lucy A and Torralbo-Campo,  Lara and Weatherill,  Kevin J},
  year = {2023},
  month = mar,
  pages = {035002}
}

@article{fan2015rfsensing,
  title = {Atom based RF electric field sensing},
  volume = {48},
  ISSN = {1361-6455},
  url = {http://dx.doi.org/10.1088/0953-4075/48/20/202001},
  DOI = {10.1088/0953-4075/48/20/202001},
  number = {20},
  journal = {Journal of Physics B: Atomic,  Molecular and Optical Physics},
  publisher = {IOP Publishing},
  author = {Fan,  Haoquan and Kumar,  Santosh and Sedlacek,  Jonathon and K\"{u}bler,  Harald and Karimkashi,  Shaya and Shaffer,  James P},
  year = {2015},
  month = sep,
  pages = {202001}
}

@book{haus2000electromagnetic,
  title = {Electromagnetic Noise and Quantum Optical Measurements},
  ISBN = {9783662041901},
  ISSN = {1439-2674},
  DOI = {10.1007/978-3-662-04190-1},
  journal = {Advanced Texts in Physics},
  publisher = {Springer Berlin Heidelberg},
  author = {Haus,  Hermann A.},
  address = {Berlin Heidelberg New York},
  year = {2000}
}

@article{santamaria2018sensitivity,
  title = {Sensitivity limits of millimeter-wave photonic radiometers based on efficient electro-optic upconverters},
  volume = {5},
  ISSN = {2334-2536},
  url = {http://dx.doi.org/10.1364/OPTICA.5.001210},
  DOI = {10.1364/optica.5.001210},
  number = {10},
  journal = {Optica},
  publisher = {Optica Publishing Group},
  author = {Santamaría Botello,  Gabriel and Sedlmeir,  Florian and Rueda,  Alfredo and Abdalmalak,  Kerlos Atia and Brown,  Elliott R. and Leuchs,  Gerd and Preu,  Sascha and Segovia-Vargas,  Daniel and Strekalov,  Dmitry V. and García Muñoz,  Luis Enrique and Schwefel,  Harald G. L.},
  year = {2018},
  month = oct,
  pages = {1210--1219}
}

@article{fancher2021comm,
  title = {Rydberg Atom Electric Field Sensors for Communications and Sensing},
  volume = {2},
  ISSN = {2689-1808},
  url = {http://dx.doi.org/10.1109/TQE.2021.3065227},
  DOI = {10.1109/tqe.2021.3065227},
  journal = {IEEE Transactions on Quantum Engineering},
  publisher = {Institute of Electrical and Electronics Engineers (IEEE)},
  author = {Fancher,  Charles T. and Scherer,  David R. and John,  Marc C. St. and Marlow,  Bonnie L. Schmittberger},
  year = {2021},
  pages = {1–13}
}

@article{liu2023electric,
  title = {Electric Field Measurement and Application Based on Rydberg Atoms},
  volume = {1},
  ISSN = {2836-9440},
  url = {http://dx.doi.org/10.23919/emsci.2022.0015},
  DOI = {10.23919/emsci.2022.0015},
  number = {2},
  journal = {Electromagnetic Science},
  publisher = {Institute of Electrical and Electronics Engineers (IEEE)},
  author = {Liu,  Bang and Zhang,  Lihua and Liu,  Zongkai and Deng,  Zian and Ding,  Dongsheng and Shi,  Baosen and Guo,  Guangcan},
  year = {2023},
  month = jun,
  pages = {1–16}
}

@article{boller1991eit,
  title = {Observation of electromagnetically induced transparency},
  volume = {66},
  ISSN = {0031-9007},
  url = {http://dx.doi.org/10.1103/PHYSREVLETT.66.2593},
  DOI = {10.1103/physrevlett.66.2593},
  number = {20},
  journal = {Physical Review Letters},
  publisher = {American Physical Society (APS)},
  author = {Boller,  K.-J. and Imamoğlu,  A. and Harris,  S. E.},
  year = {1991},
  month = may,
  pages = {2593–2596}
}

@article{sibalic2017arc,
  title = {ARC: An open-source library for calculating properties of alkali Rydberg atoms},
  volume = {220},
  ISSN = {0010-4655},
  url = {http://dx.doi.org/10.1016/J.CPC.2017.06.015},
  DOI = {10.1016/j.cpc.2017.06.015},
  journal = {Computer Physics Communications},
  publisher = {Elsevier BV},
  author = {Šibalić,  N. and Pritchard,  J.D. and Adams,  C.S. and Weatherill,  K.J.},
  year = {2017},
  month = nov,
  pages = {319–331}
}

@article{cox2018electricallysmall,
  title = {Quantum-Limited Atomic Receiver in the Electrically Small Regime},
  volume = {121},
  ISSN = {1079-7114},
  url = {http://dx.doi.org/10.1103/PhysRevLett.121.110502},
  DOI = {10.1103/physrevlett.121.110502},
  number = {11},
  journal = {Physical Review Letters},
  publisher = {American Physical Society (APS)},
  author = {Cox,  Kevin C. and Meyer,  David H. and Fatemi,  Fredrik K. and Kunz,  Paul D.},
  year = {2018},
  pages = {110502},
  month = sep 
}

@article{zhang2024rydberg,
  title = {Rydberg atom electric field sensing for metrology,  communication and hybrid quantum systems},
  volume = {69},
  ISSN = {2095-9273},
  url = {http://dx.doi.org/10.1016/j.scib.2024.03.032},
  DOI = {10.1016/j.scib.2024.03.032},
  number = {10},
  journal = {Science Bulletin},
  publisher = {Elsevier BV},
  author = {Zhang,  Hao and Ma,  Yu and Liao,  Kaiyu and Yang,  Wenguang and Liu,  Zongkai and Ding,  Dongsheng and Yan,  Hui and Li,  Wenhui and Zhang,  Linjie},
  year = {2024},
  month = may,
  pages = {1515–1535}
}

@article{scholl2021simulation,
  title = {Quantum simulation of 2D antiferromagnets with hundreds of Rydberg atoms},
  volume = {595},
  ISSN = {1476-4687},
  url = {http://dx.doi.org/10.1038/s41586-021-03585-1},
  DOI = {10.1038/s41586-021-03585-1},
  number = {7866},
  journal = {Nature},
  publisher = {Springer Science and Business Media LLC},
  author = {Scholl,  Pascal and Schuler,  Michael and Williams,  Hannah J. and Eberharter,  Alexander A. and Barredo,  Daniel and Schymik,  Kai-Niklas and Lienhard,  Vincent and Henry,  Louis-Paul and Lang,  Thomas C. and Lahaye,  Thierry and L\"{a}uchli,  Andreas M. and Browaeys,  Antoine},
  year = {2021},
  month = jul,
  pages = {233–238}
}

@article{anderson2022interferometer,
  title = {Optical Radio-Frequency Phase Measurement With an Internal-State Rydberg Atom Interferometer},
  volume = {17},
  ISSN = {2331-7019},
  url = {http://dx.doi.org/10.1103/PhysRevApplied.17.044020},
  DOI = {10.1103/physrevapplied.17.044020},
  number = {4},
  journal = {Physical Review Applied},
  publisher = {American Physical Society (APS)},
  author = {Anderson,  D.A. and Sapiro,  R.E. and Gon\c{c}alves,  L.F. and Cardman,  R. and Raithel,  G.},
  year = {2022},
  pages = {044020},
  month = apr 
}

@article{berweger2023loop,
  title = {Closed-loop quantum interferometry for phase-resolved Rydberg-atom field sensing},
  volume = {20},
  ISSN = {2331-7019},
  url = {http://dx.doi.org/10.1103/PhysRevApplied.20.054009},
  DOI = {10.1103/physrevapplied.20.054009},
  number = {5},
  journal = {Physical Review Applied},
  publisher = {American Physical Society (APS)},
  author = {Berweger,  Samuel and Artusio-Glimpse,  Alexandra B. and Rotunno,  Andrew P. and Prajapati,  Nikunjkumar and Christesen,  Joseph D. and Moore,  Kaitlin R. and Simons,  Matthew T. and Holloway,  Christopher L.},
  year = {2023},
  pages = {054009},
  month = nov 
}

@article{borowka2024alloptical,
  title = {Optically-biased Rydberg microwave receiver enabled by hybrid nonlinear interferometry},
  volume = {16},
  ISSN = {2041-1723},
  url = {http://dx.doi.org/10.1038/s41467-025-63951-9},
  DOI = {10.1038/s41467-025-63951-9},
  number = {1},
  journal = {Nature Communications},
  publisher = {Springer Science and Business Media LLC},
  author = {Borówka,  Sebastian and Mazelanik,  Mateusz and Wasilewski,  Wojciech and Parniak,  Michał},
  year = {2025},
  pages = {8975},
  month = oct 
}

@article{wang2023polarisation,
  title = {Precise measurement of microwave polarization using a Rydberg atom-based mixer},
  volume = {31},
  ISSN = {1094-4087},
  url = {http://dx.doi.org/10.1364/OE.485662},
  DOI = {10.1364/oe.485662},
  number = {6},
  journal = {Optics Express},
  publisher = {Optica Publishing Group},
  author = {Wang,  Yuhan and Jia,  Fengdong and Hao,  Jianhai and Cui,  Yue and Zhou,  Fei and Liu,  Xiubin and Mei,  Jiong and Yu,  Yonghong and Liu,  Ya and Zhang,  Jian and Xie,  Feng and Zhong,  Zhiping},
  year = {2023},
  month = mar,
  pages = {10449--10457}
}

@article{holloway2019phase,
  title = {Detecting and Receiving Phase-Modulated Signals With a Rydberg Atom-Based Receiver},
  volume = {18},
  ISSN = {1548-5757},
  url = {http://dx.doi.org/10.1109/LAWP.2019.2931450},
  DOI = {10.1109/lawp.2019.2931450},
  number = {9},
  journal = {IEEE Antennas and Wireless Propagation Letters},
  publisher = {Institute of Electrical and Electronics Engineers (IEEE)},
  author = {Holloway,  Christopher L. and Simons,  Matthew T. and Gordon,  Joshua A. and Novotny,  David},
  year = {2019},
  month = sep,
  pages = {1853–1857}
}

@article{simons2021review,
  title = {Rydberg atom-based sensors for radio-frequency electric field metrology,  sensing,  and communications},
  volume = {18},
  ISSN = {2665-9174},
  url = {http://dx.doi.org/10.1016/j.measen.2021.100273},
  DOI = {10.1016/j.measen.2021.100273},
  journal = {Measurement: Sensors},
  publisher = {Elsevier BV},
  author = {Simons,  Matthew T. and Artusio-Glimpse,  Alexandra B. and Robinson,  Amy K. and Prajapati,  Nikunjkumar and Holloway,  Christopher L.},
  year = {2021},
  month = dec,
  pages = {100273}
}

@article{liu2022multifrequency,
  title = {Deep learning enhanced Rydberg multifrequency microwave recognition},
  volume = {13},
  ISSN = {2041-1723},
  url = {http://dx.doi.org/10.1038/s41467-022-29686-7},
  DOI = {10.1038/s41467-022-29686-7},
  number = {1},
  journal = {Nature Communications},
  publisher = {Springer Science and Business Media LLC},
  author = {Liu,  Zong-Kai and Zhang,  Li-Hua and Liu,  Bang and Zhang,  Zheng-Yuan and Guo,  Guang-Can and Ding,  Dong-Sheng and Shi,  Bao-Sen},
  year = {2022},
  pages = {1997},
  month = apr 
}

@article{arumugam2024soil,
  title = {Remote sensing of soil moisture using Rydberg atoms and satellite signals of opportunity},
  volume = {14},
  ISSN = {2045-2322},
  url = {http://dx.doi.org/10.1038/s41598-024-68914-6},
  DOI = {10.1038/s41598-024-68914-6},
  number = {1},
  journal = {Scientific Reports},
  publisher = {Springer Science and Business Media LLC},
  author = {Arumugam,  Darmindra and Park,  Jun-Hee and Feyissa,  Brook and Bush,  Jack and Mysore Nagaraja,  Srinivas Prasad},
  year = {2024},
  pages = {18025},
  month = aug 
}

@article{mao2024terahertz,
  title = {Terahertz-Band Near-Space Communications: From a Physical-Layer Perspective},
  volume = {62},
  ISSN = {1558-1896},
  url = {http://dx.doi.org/10.1109/MCOM.004.2200429},
  DOI = {10.1109/mcom.004.2200429},
  number = {2},
  journal = {IEEE Communications Magazine},
  publisher = {Institute of Electrical and Electronics Engineers (IEEE)},
  author = {Mao,  Tianqi and Zhang,  Leyi and Xiao,  Zhenyu and Han,  Zhu and Xia,  Xiang-Gen},
  year = {2024},
  month = feb,
  pages = {110–116}
}

@article{li2025terahertz,
  title = {Terahertz Science and Technology in Astronomy,  Telecommunications,  and Biophysics},
  volume = {8},
  ISSN = {2639-5274},
  url = {http://dx.doi.org/10.34133/research.0586},
  DOI = {10.34133/research.0586},
  journal = {Research},
  publisher = {American Association for the Advancement of Science (AAAS)},
  author = {Li,  Jing and Deng,  Xianjin and Li,  Yangmei and Hu,  Jie and Miao,  Wei and Lin,  Changxing and Jiang,  Jun and Shi,  Shengcai},
  year = {2025},
  pages = {0586},
  month = jan 
}

@inproceedings{kanhere2021position,
  title = {Outdoor sub-THz Position Location and Tracking using Field Measurements at 142 GHz},
  DOI = {10.1109/icc42927.2021.9500482},
  booktitle = {ICC 2021 - IEEE International Conference on Communications},
  publisher = {IEEE},
  author = {Kanhere,  Ojas and Rappaport,  Theodore S.},
  year = {2021},
  month = jun,
  pages = {1–6},
  address = {Montreal, Canada}
}

@inproceedings{lu2016thzradar,
  title = {Application of terahertz technology in cooperative detection of space targets},
  DOI = {10.1109/ucmmt.2016.7874011},
  booktitle = {2016 IEEE 9th UK-Europe-China Workshop on Millimetre Waves and Terahertz Technologies (UCMMT)},
  publisher = {IEEE},
  author = {Lu,  Zejian and Liu,  Xiao and Pan,  Yue and Huang,  Sheng and Chen,  Long and Wang,  Hai},
  year = {2016},
  month = sep,
  pages = {202–205},
  address = {Qingdao, China}
}

@article{gue2023darkmatter,
  title = {Search for vector dark matter in microwave cavities with Rydberg atoms},
  volume = {108},
  ISSN = {2470-0029},
  url = {http://dx.doi.org/10.1103/PhysRevD.108.035042},
  DOI = {10.1103/physrevd.108.035042},
  number = {3},
  journal = {Physical Review D},
  publisher = {American Physical Society (APS)},
  author = {Gué,  Jordan and Hees,  Aurélien and Lodewyck,  Jér\^ome and Le Targat,  Rodolphe and Wolf,  Peter},
  year = {2023},
  pages = {035042},
  month = aug 
}

@article{graham2024axion,
  title = {Rydberg-atom-based single-photon detection for haloscope axion searches},
  volume = {109},
  ISSN = {2470-0029},
  url = {http://dx.doi.org/10.1103/PhysRevD.109.032009},
  DOI = {10.1103/physrevd.109.032009},
  number = {3},
  journal = {Physical Review D},
  publisher = {American Physical Society (APS)},
  author = {Graham,  Eleanor and Ghosh,  Sumita and Zhu,  Yuqi and Bai,  Xiran and Cahn,  Sidney B. and Durcan,  Elsa and Jewell,  Michael J. and Speller,  Danielle H. and Zacarias,  Sabrina M. and Zhou,  Laura T. and Maruyama,  Reina H.},
  year = {2024},
  pages = {032009},
  month = feb 
}

@article{engelhardt2024axion,
  title = {Detecting axion dark matter with Rydberg atoms via induced electric dipole transitions},
  volume = {6},
  ISSN = {2643-1564},
  url = {http://dx.doi.org/10.1103/PhysRevResearch.6.023017},
  DOI = {10.1103/physrevresearch.6.023017},
  number = {2},
  journal = {Physical Review Research},
  publisher = {American Physical Society (APS)},
  author = {Engelhardt,  Georg and Bhoonah,  Amit and Liu,  W. Vincent},
  year = {2024},
  pages = {023017},
  month = apr 
}

@article{alonso2022coldatoms,
  title = {Cold atoms in space: community workshop summary and proposed road-map},
  volume = {9},
  ISSN = {2196-0763},
  url = {http://dx.doi.org/10.1140/epjqt/s40507-022-00147-w},
  DOI = {10.1140/epjqt/s40507-022-00147-w},
  number = {1},
  journal = {EPJ Quantum Technology},
  publisher = {Springer Science and Business Media LLC},
  author = {Alonso,  Iván and Alpigiani,  Cristiano and Altschul,  Brett and others},
  year = {2022},
  pages = {30},
  month = nov 
}

@article{otto2021data,
  title = {Data capacity scaling of a distributed Rydberg atomic receiver array},
  volume = {129},
  ISSN = {1089-7550},
  url = {http://dx.doi.org/10.1063/5.0048415},
  DOI = {10.1063/5.0048415},
  number = {15},
  journal = {Journal of Applied Physics},
  publisher = {AIP Publishing},
  author = {Otto,  J. Susanne and Hunter,  Marisol K. and Kjærgaard,  Niels and Deb,  Amita B.},
  year = {2021},
  pages = {154503},
  month = apr 
}

@article{gaborit2014probe,
  title = {Single Shot and Vectorial Characterization of Intense Electric Field in Various Environments With Pigtailed Electrooptic Probe},
  volume = {42},
  ISSN = {1939-9375},
  url = {http://dx.doi.org/10.1109/TPS.2014.2301023},
  DOI = {10.1109/tps.2014.2301023},
  number = {5},
  journal = {IEEE Transactions on Plasma Science},
  publisher = {Institute of Electrical and Electronics Engineers (IEEE)},
  author = {Gaborit,  Gwenael and Jarrige,  Pierre and Lecoche,  Frederic and Dahdah,  Jean and Duraz,  Eric and Volat,  Christophe and Duvillaret,  Lionel},
  year = {2014},
  month = may,
  pages = {1265–1273}
}

@article{schmidt2025alloptical,
  title = {All-Optical Radio-Frequency Phase Detection for Rydberg Atom Sensors Using Oscillatory Dynamics},
  volume = {135},
  ISSN = {1079-7114},
  url = {http://dx.doi.org/10.1103/23kb-7h7q},
  DOI = {10.1103/23kb-7h7q},
  number = {9},
  journal = {Physical Review Letters},
  publisher = {American Physical Society (APS)},
  author = {Schmidt,  Matthias and Bohaichuk,  Stephanie M. and Venu,  Vijin and Wang,  Ruoxi and K\"{u}bler,  Harald and Shaffer,  James P.},
  year = {2025},
  pages = {093602},
  month = aug 
}

@article{li2023super,
  title = {Super low-frequency electric field measurement based on Rydberg atoms},
  volume = {31},
  ISSN = {1094-4087},
  url = {http://dx.doi.org/10.1364/OE.499244},
  DOI = {10.1364/oe.499244},
  number = {18},
  journal = {Optics Express},
  publisher = {Optica Publishing Group},
  author = {Li,  Ling and Jiao,  Yuechun and Hu,  Jinlian and Li,  Huaqiang and Shi,  Meng and Zhao,  Jianming and Jia,  Suotang},
  year = {2023},
  month = aug,
  pages = {29228--29234}
}

@techreport{dsn70m_interfaces_2015,
  title        = {70-m Subnet Telecommunications Interfaces (DSN 810-005, Module 101 Rev. F)},
  author       = {Slobin, S.},
  institution  = {Jet Propulsion Laboratory, NASA},
  year         = {2015},
  url          = {https://deepspace.jpl.nasa.gov/dsndocs/810-005/101/101E.pdf},
  note         = {pp. 21–22}
}

@article{esa_dsa3_2013,
  title = {The European Space Agency’s Deep-Space Antennas},
  volume = {95},
  ISSN = {0018-9219},
  url = {http://dx.doi.org/10.1109/JPROC.2007.905189},
  DOI = {10.1109/jproc.2007.905189},
  number = {11},
  journal = {Proceedings of the IEEE},
  publisher = {Institute of Electrical and Electronics Engineers (IEEE)},
  author = {Vassallo,  Enrico and Martin,  Rolf and Madde,  Roberto and Lanucara,  Marco and Besso,  Piermario and Droll,  Peter and Galtie,  GÉrard and De Vicente,  Javier},
  year = {2007},
  month = nov,
  pages = {2111–2131}
}

@article{tokumaru_2000_halca_pasj,
  title = {HALCA’s Onboard VLBI Observing System},
  volume = {52},
  ISSN = {0004-6264},
  url = {http://dx.doi.org/10.1093/pasj/52.6.967},
  DOI = {10.1093/pasj/52.6.967},
  number = {6},
  journal = {Publications of the Astronomical Society of Japan},
  publisher = {Oxford University Press (OUP)},
  author = {Kobayashi,  Hideyuki and Wajima,  Kiyoaki and Hirabayashi,  Hisashi and Murata,  Yasuhiro and Kawaguchi,  Noriyuki and Kameno,  Seiji and Shibata,  Katsunori M. and Fujisawa,  Kenta and Inoue,  Makoto and Hirosawa,  Haruto},
  year = {2000},
  month = dec,
  pages = {967–973}
}

@article{kardashev_2013_radioastron_ar,
  title = {“RadioAstron”-A telescope with a size of 300 000 km: Main parameters and first observational results},
  volume = {57},
  ISSN = {1562-6881},
  url = {http://dx.doi.org/10.1134/S1063772913030025},
  DOI = {10.1134/s1063772913030025},
  number = {3},
  journal = {Astronomy Reports},
  publisher = {Pleiades Publishing Ltd},
  author = {Kardashev,  N. S. and Khartov,  V. V. and Abramov,  V. V. and others},
  year = {2013},
  month = mar,
  pages = {153–194}
}

@article{piepmeier_2017_smap_tgrs,
  title = {SMAP L-Band Microwave Radiometer: Instrument Design and First Year on Orbit},
  volume = {55},
  ISSN = {1558-0644},
  url = {http://dx.doi.org/10.1109/tgrs.2016.2631978},
  DOI = {10.1109/tgrs.2016.2631978},
  number = {4},
  journal = {IEEE Transactions on Geoscience and Remote Sensing},
  publisher = {Institute of Electrical and Electronics Engineers (IEEE)},
  author = {Piepmeier,  Jeffrey R. and Focardi,  Paolo and Horgan,  Kevin A. and Knuble,  Joseph and Ehsan,  Negar and Lucey,  Jared and Brambora,  Clifford and Brown,  Paula R. and Hoffman,  Pamela J. and French,  Richard T. and Mikhaylov,  Rebecca L. and Kwack,  Eug-Yun and Slimko,  Eric M. and Dawson,  Douglas E. and Hudson,  Derek and Peng,  Jinzheng and Mohammed,  Priscilla N. and De Amici,  Giovanni and Freedman,  Adam P. and Medeiros,  James and Sacks,  Fred and Estep,  Robert and Spencer,  Michael W. and Chen,  Curtis W. and Wheeler,  Kevin B. and Edelstein,  Wendy N. and O’Neill,  Peggy E. and Njoku,  Eni G.},
  year = {2017},
  month = apr,
  pages = {1954–1966}
}

@article{Piepmeier2015,
  title = {Aquarius L-Band Microwave Radiometer: 3 Years of Radiometric Performance and Systematic Effects},
  volume = {8},
  ISSN = {2151-1535},
  url = {http://dx.doi.org/10.1109/JSTARS.2015.2435493},
  DOI = {10.1109/jstars.2015.2435493},
  number = {12},
  journal = {IEEE Journal of Selected Topics in Applied Earth Observations and Remote Sensing},
  publisher = {Institute of Electrical and Electronics Engineers (IEEE)},
  author = {Piepmeier,  Jeffrey R. and Hong,  Liang and Pellerano,  Fernando A.},
  year = {2015},
  month = dec,
  pages = {5416–5423}
}

@article{MartnNeira2014,
  title = {Microwave interferometric radiometry in remote sensing: An invited historical review},
  volume = {49},
  ISSN = {1944-799X},
  url = {http://dx.doi.org/10.1002/2013RS005230},
  DOI = {10.1002/2013rs005230},
  number = {6},
  journal = {Radio Science},
  publisher = {American Geophysical Union (AGU)},
  author = {Martín‐Neira,  M. and LeVine,  D. M. and Kerr,  Y. and Skou,  N. and Peichl,  M. and Camps,  A. and Corbella,  I. and Hallikainen,  M. and Font,  J. and Wu,  J. and Mecklenburg,  S. and Drusch,  M.},
  year = {2014},
  month = jun,
  pages = {415–449}
}

@article{Rasch2024,
  title = {The GRAS-2 radio occultation mission},
  volume = {17},
  ISSN = {1867-8548},
  url = {http://dx.doi.org/10.5194/amt-17-6213-2024},
  DOI = {10.5194/amt-17-6213-2024},
  number = {20},
  journal = {Atmospheric Measurement Techniques},
  publisher = {Copernicus GmbH},
  author = {Rasch,  Joel and Carlstr\"{o}m,  Anders and Christensen,  Jacob and Liljegren,  Thomas},
  year = {2024},
  month = oct,
  pages = {6213–6222}
}

@misc{eoportal_poseidon3b,
  title        = {Jason-3 Mission: Poseidon-3B Altimeter},
  organization = {European Space Agency},
  howpublished = {\url{https://www.eoportal.org/satellite-missions/jason-3#poseidon-3b-altimeter}},
  note         = {EO Portal Directory},
  year         = {2024},
  urldate      = {2026-01-07}
}

@article{kim_2014_atms_jgr,
  title = {S‐NPP ATMS instrument prelaunch and on‐orbit performance evaluation},
  volume = {119},
  ISSN = {2169-8996},
  url = {http://dx.doi.org/10.1002/2013JD020483},
  DOI = {10.1002/2013jd020483},
  number = {9},
  journal = {Journal of Geophysical Research: Atmospheres},
  publisher = {American Geophysical Union (AGU)},
  author = {Kim,  Edward and Lyu,  Cheng‐Hsuan J. and Anderson,  Kent and Vincent Leslie,  R. and Blackwell,  William J.},
  year = {2014},
  month = may,
  pages = {5653–5670}
}

@article{frisk_2003_odin_aa,
  title = {The Odin satellite: I. Radiometer design and test},
  volume = {402},
  ISSN = {1432-0746},
  url = {http://dx.doi.org/10.1051/0004-6361:20030335},
  DOI = {10.1051/0004-6361:20030335},
  number = {3},
  journal = {Astronomy \& Astrophysics},
  publisher = {EDP Sciences},
  author = {Frisk,  U. and Hagström,  M. and Ala-Laurinaho,  J. and others},
  year = {2003},
  month = apr,
  pages = {L27–L34}
}

@article{bersanelli_2010_planck_lfi_aa,
  title = {Planck pre-launch status: Design and description of the Low Frequency Instrument},
  volume = {520},
  ISSN = {1432-0746},
  url = {http://dx.doi.org/10.1051/0004-6361/200912853},
  DOI = {10.1051/0004-6361/200912853},
  journal = {Astronomy and Astrophysics},
  publisher = {EDP Sciences},
  author = {Bersanelli,  M. and Mandolesi,  N. and Butler,  R. C. and others},
  year = {2010},
  month = sep,
  pages = {A4}
}

@article{egusphere-2024-3648,
  title = {Feasibility of a space-borne terahertz heterodyne spectrometer for atomic oxygen and temperature in the mesosphere and lower thermosphere},
  url = {http://dx.doi.org/10.5194/egusphere-2024-3648},
  DOI = {10.5194/egusphere-2024-3648},
  publisher = {Copernicus GmbH},
  author = {Hansen,  Peder Bagge and Wienold,  Martin and H\"{u}bers,  Heinz-Wilhelm},
  year = {2025},
  JOURNAL = {EGUsphere},
  month = mar 
}

@article{jmse13020355,
  title = {Combined Motion Compensation Method for Long Synthetic Aperture Radar Based on Subaperture Processing},
  volume = {13},
  ISSN = {2077-1312},
  url = {http://dx.doi.org/10.3390/jmse13020355},
  DOI = {10.3390/jmse13020355},
  number = {2},
  journal = {Journal of Marine Science and Engineering},
  publisher = {MDPI AG},
  author = {Zhang,  Yuan and Huang,  Limin and Xu,  Zhichao and Wang,  Zihao and Chen,  Biao},
  year = {2025},
  month = feb,
  pages = {355}
}

@article{ZHOU2025133513,
  title = {Sensitivity of multi-frequency and multi-polarization SAR to soil moisture at different depths in agricultural regions},
  volume = {660},
  ISSN = {0022-1694},
  url = {http://dx.doi.org/10.1016/j.jhydrol.2025.133513},
  DOI = {10.1016/j.jhydrol.2025.133513},
  journal = {Journal of Hydrology},
  publisher = {Elsevier BV},
  author = {Zhou,  Xin and Wang,  Jinfei and Shan,  Bo and He,  Yongjun and Xing,  Minfeng},
  year = {2025},
  month = oct,
  pages = {133513}
}

@article{rs15153742,
  title = {Radar Target Characterization and Deep Learning in Radar Automatic Target Recognition: A Review},
  volume = {15},
  ISSN = {2072-4292},
  url = {http://dx.doi.org/10.3390/rs15153742},
  DOI = {10.3390/rs15153742},
  number = {15},
  journal = {Remote Sensing},
  publisher = {MDPI AG},
  author = {Jiang,  Wen and Wang,  Yanping and Li,  Yang and Lin,  Yun and Shen,  Wenjie},
  year = {2023},
  month = jul,
  pages = {3742}
}

@article{Torres2012,
  title = {GMES Sentinel-1 mission},
  volume = {120},
  ISSN = {0034-4257},
  url = {http://dx.doi.org/10.1016/j.rse.2011.05.028},
  DOI = {10.1016/j.rse.2011.05.028},
  journal = {Remote Sensing of Environment},
  publisher = {Elsevier BV},
  author = {Torres,  Ramon and Snoeij,  Paul and Geudtner,  Dirk and Bibby,  David and Davidson,  Malcolm and Attema,  Evert and Potin,  Pierre and Rommen,  Bj\"{O}rn and Floury,  Nicolas and Brown,  Mike and Traver,  Ignacio Navas and Deghaye,  Patrick and Duesmann,  Berthyl and Rosich,  Betlem and Miranda,  Nuno and Bruno,  Claudio and L’Abbate,  Michelangelo and Croci,  Renato and Pietropaolo,  Andrea and Huchler,  Markus and Rostan,  Friedhelm},
  year = {2012},
  month = may,
  pages = {9–24}
}

@techreport{Airbus2015_TerraSARX_ProductGuide,
  author      = {{Airbus Defence and Space}},
  title       = {TerraSAR-X Image Product Guide, Issue 2.1},
  year        = {2015},
  institution = {Airbus Defence and Space},
  note        = {Antenna 4.8 m × 0.7 m; centre frequency 9.65 GHz; chirp bandwidth up to 300 MHz},
  url         = {https://airbusus.com/wp-content/uploads/2020/06/r459_9_20171004_tsxx-airbusds-ma-0009_tsx-productguide_i2.01.pdf}
}

@misc{eoportal_terrasarx,
  title        = {TerraSAR-X},
  author       = {{European Space Agency (ESA)}},
  year         = {2024},
  url          = {https://www.eoportal.org/satellite-missions/terrasar-x},
  note         = {Earth Observation Portal (eoPortal), accessed 2026-01-07}
}

@techreport{mo_2000_nesdis98,
  author       = {Mo, T.},
  title        = {NOAA-L and NOAA-M AMSU-A Antenna Pattern Corrections},
  institution  = {NOAA National Environmental Satellite, Data, and Information Service},
  number       = {NOAA Technical Report NESDIS 98},
  year         = {2000},
  address      = {Washington, D.C.}
}

@techreport{atovs_calval,
  title        = {ATOVS and AVHRR Calibration and Validation Plan},
  institution  = {NOAA National Environmental Satellite, Data, and Information Service},
  year         = {1998},
  note         = {Program calibration and validation document},
  author = {Klaes, Dieter and Ackermann, Jörg and Borde, Régis}
}

@article{Saunders1995,
  title = {The radiometric characterization of AMSU-B},
  volume = {43},
  ISSN = {0018-9480},
  url = {http://dx.doi.org/10.1109/22.375222},
  DOI = {10.1109/22.375222},
  number = {4},
  journal = {IEEE Transactions on Microwave Theory and Techniques},
  publisher = {Institute of Electrical and Electronics Engineers (IEEE)},
  author = {Saunders,  R.W. and Hewison,  T.J. and Stringer,  S.J. and Atkinson,  N.C.},
  year = {1995},
  month = apr,
  pages = {760–771}
}

@techreport{smiles_missionplan_ch3_2002,
  author      = {{NICT SMILES Project Team}},
  title       = {JEM/SMILES Mission Plan, Version 2.11, Chapter 3: Instrumental Capabilities},
  institution = {National Institute of Information and Communications Technology (NICT)},
  year        = {2002},
  url         = {https://smiles.nict.go.jp/Mission_Plan/version2.1/chap-3_ver2.11.pdf},
  note        = {Project-level technical mission plan; provides AOS noise bandwidth and antenna specifications}
}

@techreport{limon_2006_wmap_3yr_supplement,
  author      = {{WMAP Science Team}},
  title       = {Wilkinson Microwave Anisotropy Probe (WMAP): Three-Year Explanatory Supplement},
  institution = {NASA Goddard Space Flight Center},
  editor      = {Limon, M.},
  year        = {2006},
  note        = {Editor listed as M. Limon; official WMAP team explanatory supplement; last revised March 16, 2006},
  url         = {https://lambda.gsfc.nasa.gov/product/wmap/dr2/pub_papers/threeyear/supplement/wmap_3yr_supplement.pdf}
}

@article{simons2016detuning,
  title = {Using frequency detuning to improve the sensitivity of electric field measurements via electromagnetically induced transparency and Autler-Townes splitting in Rydberg atoms},
  volume = {108},
  ISSN = {1077-3118},
  url = {http://dx.doi.org/10.1063/1.4947231},
  DOI = {10.1063/1.4947231},
  number = {17},
  journal = {Applied Physics Letters},
  publisher = {AIP Publishing},
  author = {Simons,  Matt T. and Gordon,  Joshua A. and Holloway,  Christopher L. and Anderson,  David A. and Miller,  Stephanie A. and Raithel,  Georg},
  year = {2016},
  pages = {174101},
  month = apr 
}

@article{jiao2016spectroscopy,
  title = {Spectroscopy of cesium Rydberg atoms in strong radio-frequency fields},
  volume = {94},
  ISSN = {2469-9934},
  url = {http://dx.doi.org/10.1103/PhysRevA.94.023832},
  DOI = {10.1103/physreva.94.023832},
  number = {2},
  journal = {Physical Review A},
  publisher = {American Physical Society (APS)},
  author = {Jiao,  Yuechun and Han,  Xiaoxuan and Yang,  Zhiwei and Li,  Jingkui and Raithel,  Georg and Zhao,  Jianming and Jia,  Suotang},
  year = {2016},
  pages = {023832},
  month = aug 
}

@article{krokosz2025comb,
  title = {Electric-field metrology of a terahertz frequency comb using Rydberg atoms},
  volume = {12},
  ISSN = {2334-2536},
  url = {http://dx.doi.org/10.1364/OPTICA.578051},
  DOI = {10.1364/optica.578051},
  number = {11},
  journal = {Optica},
  publisher = {Optica Publishing Group},
  author = {Krokosz,  Wiktor and Nowosielski,  Jan and Kasza,  Bartosz and Borówka,  Sebastian and Mazelanik,  Mateusz and Wasilewski,  Wojciech and Parniak,  Michał},
  year = {2025},
  month = nov,
  pages = {1854}
}

@article{yao2022detuned,
  title = {Sensitivity enhancement of far-detuned RF field sensing based on Rydberg atoms dressed by a near-resonant RF field},
  volume = {47},
  ISSN = {1539-4794},
  url = {http://dx.doi.org/10.1364/OL.465048},
  DOI = {10.1364/ol.465048},
  number = {20},
  journal = {Optics Letters},
  publisher = {Optica Publishing Group},
  author = {Yao,  Jiawei and An,  Qiang and Zhou,  Yanli and Yang,  Kai and Wu,  Fengchuan and Fu,  Yunqi},
  year = {2022},
  month = oct,
  pages = {5256}
}

@article{zuo2025digital,
  title = {Digital terahertz communication with Rydberg-atom-based coherent photon conversion},
  volume = {126},
  ISSN = {1077-3118},
  url = {http://dx.doi.org/10.1063/5.0250550},
  DOI = {10.1063/5.0250550},
  number = {19},
  journal = {Applied Physics Letters},
  publisher = {AIP Publishing},
  author = {Zuo,  Xiaoliang and Li,  Qingbin and Li,  Danyang and Wu,  Haiteng and Sheng,  Jiteng and Wu,  Haibin},
  year = {2025},
  pages = {194002},
  month = may 
}

@article{anderson2021amfm,
  title = {An Atomic Receiver for AM and FM Radio Communication},
  volume = {69},
  ISSN = {1558-2221},
  url = {http://dx.doi.org/10.1109/TAP.2020.2987112},
  DOI = {10.1109/tap.2020.2987112},
  number = {5},
  journal = {IEEE Transactions on Antennas and Propagation},
  publisher = {Institute of Electrical and Electronics Engineers (IEEE)},
  author = {Anderson,  David Alexander and Sapiro,  Rachel Elizabeth and Raithel,  Georg},
  year = {2021},
  month = may,
  pages = {2455–2462}
}

@article{holloway2021amfm,
  title = {A Multiple-Band Rydberg Atom-Based Receiver: AM/FM Stereo Reception},
  volume = {63},
  ISSN = {1558-4143},
  url = {http://dx.doi.org/10.1109/MAP.2020.2976914},
  DOI = {10.1109/map.2020.2976914},
  number = {3},
  journal = {IEEE Antennas and Propagation Magazine},
  publisher = {Institute of Electrical and Electronics Engineers (IEEE)},
  author = {Holloway,  Christopher and Simons,  Mathew and Haddab,  Abdulaziz H. and Gordon,  Joshua A and Anderson,  David A and Raithel,  Georg and Voran,  Steven},
  year = {2021},
  month = jun,
  pages = {63–76}
}

@article{watterson2025radar,
  title = {An imaging radar using a Rydberg atom receiver},
  volume = {127},
  ISSN = {1077-3118},
  url = {http://dx.doi.org/10.1063/5.0287757},
  DOI = {10.1063/5.0287757},
  number = {16},
  journal = {Applied Physics Letters},
  publisher = {AIP Publishing},
  author = {Watterson,  William J. and Prajapati,  Nikunjkumar and Castillo-Garza,  Rodrigo and Berweger,  Samuel and Schlossberger,  Noah and Artusio-Glimpse,  Alexandra and Holloway,  Christopher L. and Simons,  Matthew T.},
  year = {2025},
  month = oct,
  pages = {161101}
}

@misc{chen2025radar,
  doi = {10.48550/ARXIV.2506.11833},
  author = {Chen,  Minze and Mao,  Tianqi and Zhu,  Zhiao and Feng,  Haonan and Gao,  Ge and Wu,  Zhonghuai and Xiao,  Wei and Li,  Zhongxiang and Zheng,  Dezhi},
  title = {High-Resolution Quantum Sensing with Rydberg Atomic Receiver: Principles,  Experiments and Future Prospects},
  publisher = {arXiv preprint},
  year = {2025},
  copyright = {Creative Commons Attribution 4.0 International}
}

@article{prajapati2024high,
  title = {High angular momentum coupling for enhanced Rydberg-atom sensing in the very-high frequency band},
  volume = {135},
  ISSN = {1089-7550},
  url = {http://dx.doi.org/10.1063/5.0179496},
  DOI = {10.1063/5.0179496},
  number = {7},
  journal = {Journal of Applied Physics},
  publisher = {AIP Publishing},
  author = {Prajapati,  Nikunjkumar and Kunzler,  Jakob W. and Artusio-Glimpse,  Alexandra B. and Rotunno,  Andrew P. and Berweger,  Samuel and Simons,  Matthew T. and Holloway,  Christopher L. and Gardner,  Chad M. and Mcbeth,  Michael S. and Younts,  Robert A.},
  year = {2024},
  pages = {074402},
  month = feb 
}

@article{Abend2023,
  title = {Technology roadmap for cold-atoms based quantum inertial sensor in space},
  volume = {5},
  ISSN = {2639-0213},
  url = {http://dx.doi.org/10.1116/5.0098119},
  DOI = {10.1116/5.0098119},
  number = {1},
  journal = {AVS Quantum Science},
  publisher = {American Vacuum Society},
  author = {Abend,  Sven and Allard,  Baptiste and Arnold,  Aidan S. and others},
  year = {2023},
  pages = {019201},
  month = mar 
}

@article{prost2025microwave,
  title = {Microwave field imaging inside an atomic cell by fluorescence thermography},
  volume = {127},
  ISSN = {1077-3118},
  url = {http://dx.doi.org/10.1063/5.0293816},
  DOI = {10.1063/5.0293816},
  number = {18},
  journal = {Applied Physics Letters},
  publisher = {AIP Publishing},
  author = {Prost,  D. and Bonnin,  A. and Schwartz,  S.},
  year = {2025},
  pages = {184002},
  month = nov 
}

@article{Trinh2024,
  title = {Modulation transfer protocol for Rydberg RF receivers},
  volume = {125},
  ISSN = {1077-3118},
  url = {http://dx.doi.org/10.1063/5.0216969},
  DOI = {10.1063/5.0216969},
  number = {15},
  journal = {Applied Physics Letters},
  publisher = {AIP Publishing},
  author = {Trinh,  Duc-Anh and Adwaith,  K. V. and Branco,  Mickael and Rouxel,  Aliénor and Welinski,  Sacha and Berger,  Perrine and Goldfarb,  Fabienne and Bretenaker,  Fabien},
  year = {2024},
  pages = {154001},
  month = oct 
}

@article{Wu2023,
  title = {Linear dynamic range of a Rydberg-atom microwave superheterodyne receiver},
  volume = {107},
  ISSN = {2469-9934},
  url = {http://dx.doi.org/10.1103/PhysRevA.107.043108},
  DOI = {10.1103/physreva.107.043108},
  number = {4},
  journal = {Physical Review A},
  publisher = {American Physical Society (APS)},
  author = {Wu,  Fengchuan and An,  Qiang and Sun,  Zhanshan and Fu,  Yunqi},
  year = {2023},
  pages = {043108},
  month = apr 
}

@misc{kaur2025impact,
  doi = {10.48550/ARXIV.2508.17506},
  author = {Kaur,  Channprit and Shen,  Pinrui and Booth,  Donald and Todd,  Andrew and Shaffer,  James P.},
  title = {The Impact of Thermal Fields on Rydberg Atom Radio Frequency Sensors},
  publisher = {arXiv preprint},
  year = {2025},
  copyright = {arXiv.org perpetual,  non-exclusive license}
}

\end{document}